\documentclass[preprint,showpacs,preprintnumbers,amsmath,amssymb,floatfix]{revtex4-1}
\usepackage{graphicx,color}% Include figure files
\usepackage{dcolumn}% Align table columns on decimal point
\usepackage{bm}% bold math
\usepackage{supertabular}

\begin{document}

\title{Atmospheric neutrino flux calculation using the NRLMSISE00 atmospheric model}
\author{M.~Honda}
\email[]{mhonda@icrr.u-tokyo.ac.jp}
\homepage[]{http://icrr.u-tokyo.ac.jp/~mhonda}
\affiliation{Institute for Cosmic Ray Research, the University of Tokyo, 5-1-5 Kashiwa-no-ha, Kashiwa, Chiba 277-8582, Japan}
\author{M. Sajjad \surname{Athar}}
\email{sajathar@gmail.com}
\affiliation{Department of Physics, Aligarh Muslim University, Aligarh-202002, India}
\author{T.~Kajita}
\email[]{kajita@icrr.u-tokyo.ac.jp}
\affiliation{Institute for Cosmic Ray Research, and
Institute for the Physics and Mathematical of the Universe, the University of Tokyo, 5-1-5 Kashiwa-no-ha, Kashiwa, Chiba 277-8582, Japan }
\author{K.~Kasahara}
\email[]{kasahara@icrc.u-tokyo.ac.jp}
\affiliation{Research Institute for Science and Engineering, Waseda University, 3-4-1 Okubo Shinjuku-ku, Tokyo, 169-8555, Japan.}
\author{S.~Midorikawa}
\email[]{midori@aomori-u.ac.jp}
\affiliation{Faculty of Software and Information Technology, Aomori University, Aomori, 030-0943 Japan.}
\date{\today}

\begin{abstract}

We extend our calculation of the atmospheric neutrino fluxes 
to polar and tropical regions. 
It is well known that the air density profile in the polar and the 
tropical regions are different from the mid-latitude region.
Also there are large seasonal variations in the polar region.
In this extension,
we use the
NRLMSISE-00 global atmospheric model~\cite{nrlmsise-00} 
replacing the US-standard '76 atmospheric model~\cite{us-standard}, 
which has no positional or seasonal variations.
With the NRLMSISE-00 atmospheric model, we study the atmospheric 
neutrino flux at the polar and tropical regions with seasonal
variations.
The geomagnetic model IGRF~\cite{igrf}  we have used in our calculations seems 
accurate enough in the polar regions also.
However, 
the polar and the equatorial regions are the 
two extremes in the IGRF model, and
the magnetic field configurations are largely different to
each other.
Note, the equatorial region is also the tropical region generally.
We study the effect of the geomagnetic field on the atmospheric 
neutrino flux in these extreme regions.

\end{abstract}

\maketitle

\section{introduction}

In this paper, we extend the calculation of the atmospheric 
neutrino flux~\cite{hkkm2004,hkkms2006,hkkm2011} 
to the sites in polar and tropical regions. 
In our earliest full 3D-calculation~\cite{hkkm2004}, 
we used DPMJET-III~\cite{dpm}
for the hadronic interaction model above 5~GeV, and NUCRIN~\cite{nucrin}
below 5~GeV.
We modified DPMJET-III as in Ref.~\cite{hkkms2006} 
to reproduce the experimental muon spectra better,
mainly using the data observed by BESS group~\cite{BESSTeV}.
In a recent work~\cite{hkkm2011}, 
we introduced JAM interaction model for the low energy
hadronic interactions.
JAM is a nuclear interaction model developed with 
PHITS (Particle and Heavy-Ion Transport code System)~\cite{phits}.
In Ref.~\cite{hkkm2011}, 
we could reproduce the observed muon flux at the low energies at 
balloon altitude~\cite{Abe:2003cd} with DPMJET-III above 32 GeV and JAM below  
better than with the combination of DPMJET-III above 5~GeV and 
NUCRIN below that.
Besides the interaction model, we have also improved the calculation 
scheme according to the increase of available computational power, such as
the ``virtual detector correction''
introduced in Ref.~\cite{hkkms2006}
and the optimization of it in Ref.~\cite{hkkm2011}.
The statistics of the Monte Carlo simulation is also improved at every 
step of the work.

We used the US-standard '76 atmospheric model in our earlier 
calculations.
The US-standard '76 atmospheric model had been used generally 
in the study of cosmic rays in the atmosphere for a long 
time~\cite{battis2003,barr2004}.
But, the air density profile in US-standard~'76 
is represented as a function of altitude only,
and has no time variation and no position dependence around the Earth. 
In Ref~\cite{shkkm2006}, we discussed the validity of 
using such a atmospheric model in the calculation of atmospheric 
neutrino flux, assuming small variations to the US-standard '76 
atmospheric model.
However,
the difference of air density profile in the polar regions and 
mid-latitude regions is lager than the considered variations in the study.
Also there is a large seasonal variation of the air density profile 
in the polar region.
We install the NRLMSISE-00 global atmospheric model~\cite{nrlmsise-00}
which represents proper  position dependence 
and the time variations on the Earth, to calculate the
atmospheric neutrino flux in the polar and tropical regions.

We have used the IGRF geomagnetic field model~\cite{igrf} 
in our calculation, 
and it is accurate enough in the polar and tropical
($\eqsim$ equatorial) regions.
However,
the geomagnetic field strongly affects the atmospheric neutrino flux,
and is largely different in the polar and equatorial regions.
The extension in this paper is also the study of atmospheric neutrino 
flux under these widely different geomagnetic field conditions.

 The models of primary cosmic rays spectra and interactions are the
 same as those in Ref.~\cite{hkkm2011}.
 The model of primary cosmic rays are constructed based on the
 AMS01~\cite{AMS01p1} and BESS~\cite{BESSphe,BESSTeV}
 observations.
 However, there are newer cosmic ray observation
 experiments~\cite{atic05,cream,pamela2011,AMS02p},
 and we will make a short comment on them
 and the error of our calculation in the summary.
 We note, the combination of the modified DPMJET-III
 above 32 GeV and JAM below that 
 reproduces the observed muon spectra
 best with the present cosmic ray spectra model. 

In our 3D-calculations of the atmospheric neutrino flux, 
we followed the motion of all the cosmic rays,
which penetrate the rigidity cutoff, and their secondaries.
Then we examine all the neutrinos produced during their propagation in 
the atmosphere,
and register the neutrinos which hit the virtual detector
assumed around the target neutrino observation site.
Therefore, we do not need a change in the calculation scheme other 
than the atmospheric model 
to calculate the atmospheric neutrino flux 
at a new site in the polar and tropical regions.

We study in detail the atmospheric neutrino flux at the 
India-based Neutrino Observatory (INO) site
(lat, lon)=($9^\circ59'', 77^\circ16''$)
for the tropical and equatorial region,
and the South Pole ($-90^\circ00'', 0^\circ00''$)  
and Pyh\"asalmi ($63^\circ40'', 6^\circ41''$) mine (Finland) for 
the polar regions in this paper.
Also we compare the atmospheric neutrino flux calculated with the 
NRLMSISE-00 atmospheric model and 
that calculated with the US-standard '76 atmospheric model at 
Super Kamiokande (SK) site ($36^\circ26'', 137^\circ10''$).

We note that the atmospheric neutrino production height is important for 
the analysis of neutrino oscillations as well as the flux.
As the production height is related to the atmospheric model,
we study the production height of atmospheric neutrinos 
with the NRLMSISE-00 atmospheric model in detail,
and compare the atmospheric neutrino production height with that
calculated with US-standard '76 atmospheric model at the SK site.

 The tables for the flux and production height calculated
 in this paper are available in the web page, 
 \url{http://www.icrr.u-tokyo.ac.jp/~mhonda} .

\section{\label{msise}NRLMSISE-00 atmospheric model}

\begin{figure*}[htb]
 \centering
  {
\includegraphics[width=6cm]{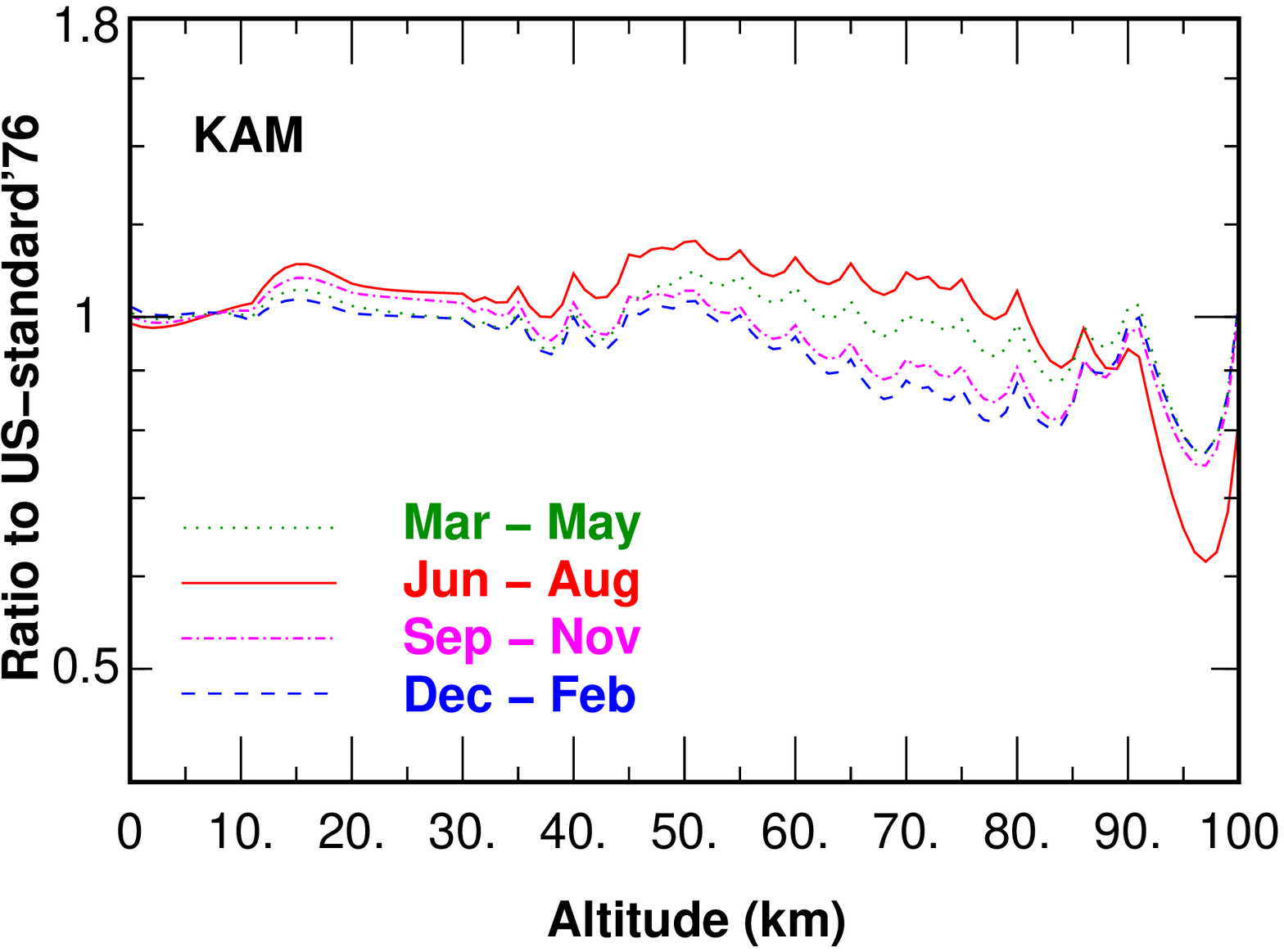} 
\hspace{5mm}
\includegraphics[width=6cm]{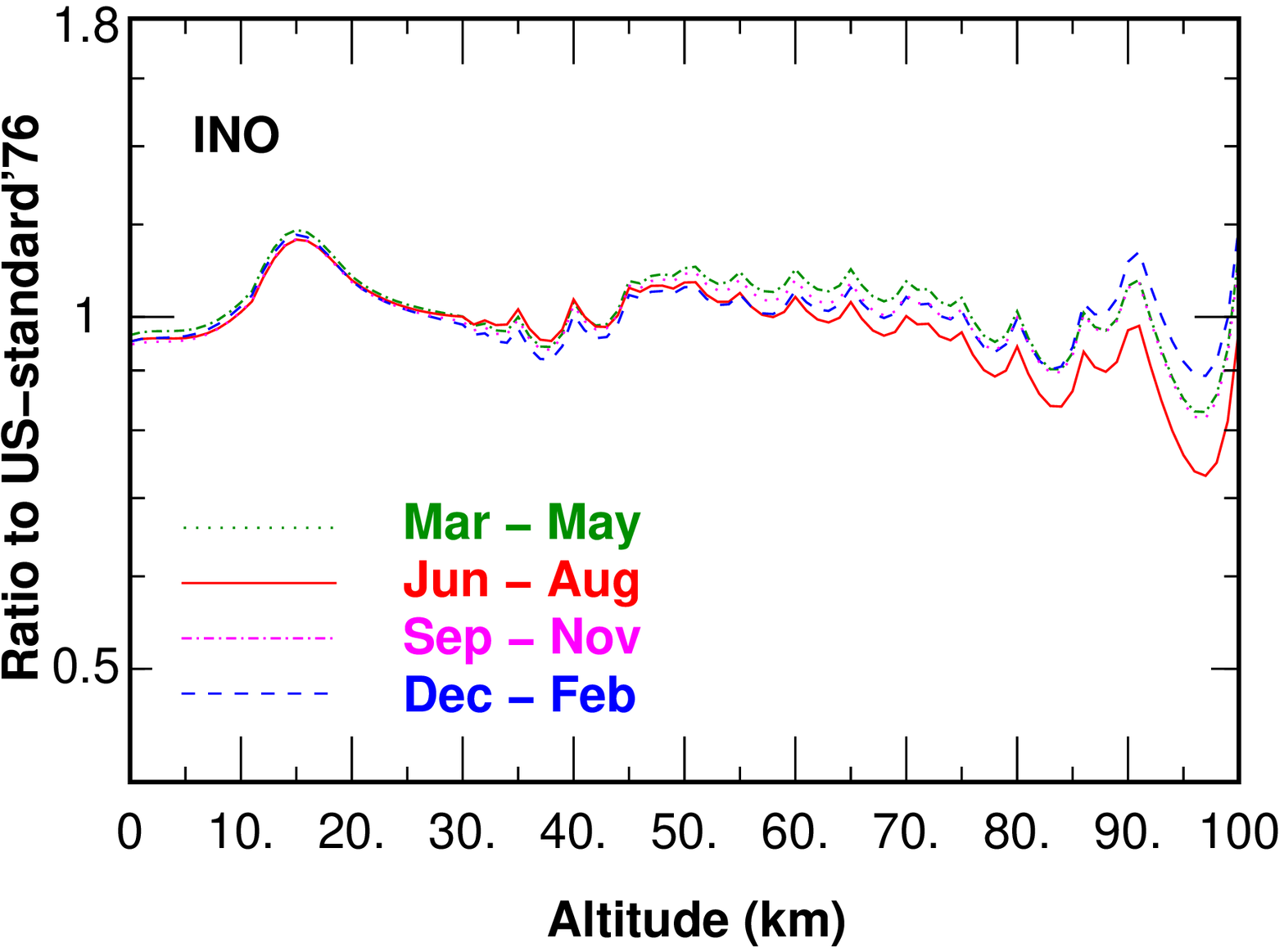}
\\
\includegraphics[width=6cm]{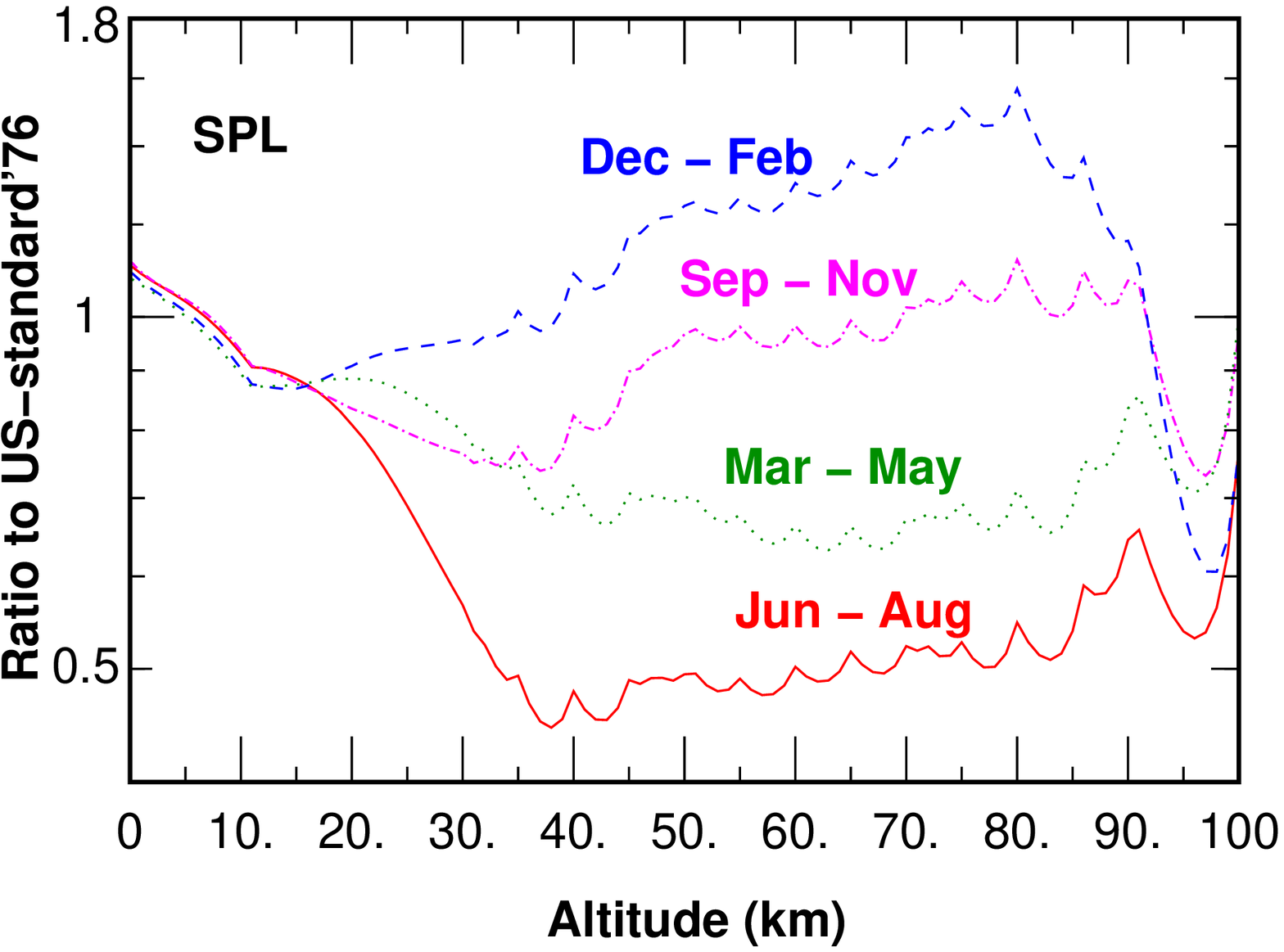}
\hspace{5mm}
\includegraphics[width=6cm]{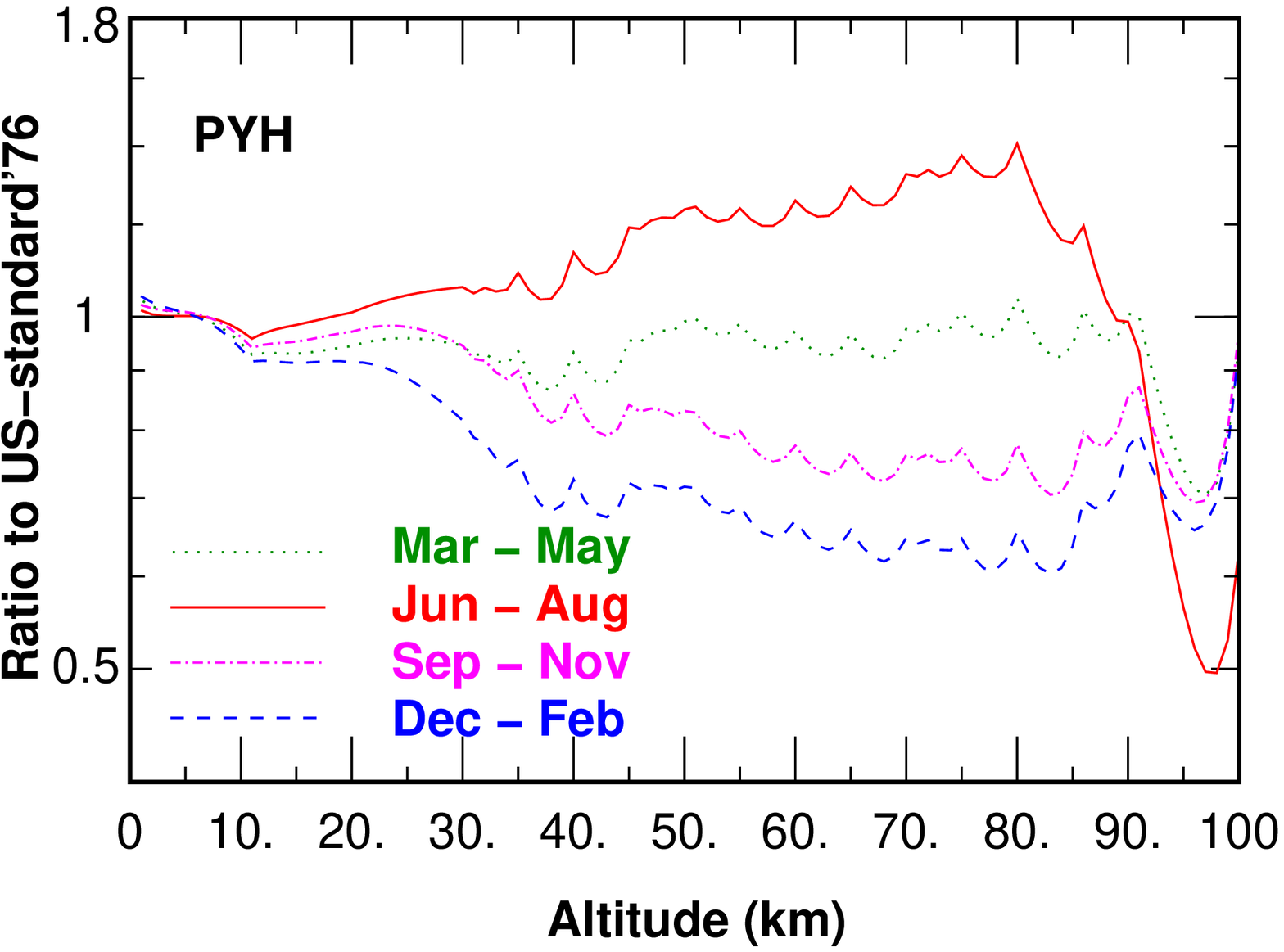}
  }
\caption{
The ratio of air density in the NRLMSISE-00 atmospheric model to 
that of US-standard '76 atmospheric model for the SK site (KAM), 
INO site (INO), South pole (SPL), and Pyh\"asalmi mine (PYH),
in the 4 seasons,
March -- May, June -- August, September -- November, and December -- February.
}
 \label{fig:ratio2us}
 \end{figure*}

NRLMSISE-00~\cite{nrlmsise-00} is an empirical, global model of the Earth's
atmosphere from ground to space. 
It models the temperatures and densities of the atmosphere's components. 
%
%A primary use of this model is to aid predictions of satellite orbital decay 
%due to atmospheric drag. 
%This model has also been used by astronomers to calculate the mass of air 
%between telescopes and laser beams in order to assess the impact of laser 
%guide stars on the non-lasing telescopes.~\cite{nrlmsise-00}
%
%Coulson, Dolores M. & Roth, Katherine C., 
%Adaptive Optics Systems II. Edited by Ellerbroek, Brent L.; Hart, Michael; Hubin, Norbert; Wizinowich, Peter L. 
%Proceedings of the SPIE, Volume 7736, pp. 773652-773652-9 (2010)
%
However,
the air density profile is the most important quantity 
in the calculation of atmospheric neutrino flux.
We calculate the ratio of the air density in 4 seasons, 
March -- May, June -- August, September -- November, and December -- February
at the SK site (KAM), INO site (INO), South pole (SPL), and Pyh\"asalmi mine
(PYH) by the NRLMSISE-00 atmospheric model to 
that by the US-standard '76 atmospheric model,
and show it in Fig.~\ref{fig:ratio2us} as a function of altitude.

In the tropical region (INO), 
we see a $\sim$~20~\% larger 
air density than the US-standard '76 atmospheric model
at the altitude of $\sim15$~km a.s.l 
(\underbar{a}bove \underbar{s}ea \underbar{l}evel).
However, except for that, the air density 
profile is similar to the US-standard '76 atmospheric model,
and there is almost no seasonal 
variation by the NRLMSISE-00 atmospheric model.
In the mid-latitude region (KAM),
we find seasonal variations but they are small,
and the air density profile 
is close to that of the US-standard '76 atmospheric model
through all the seasons below 40~km a.s.l.
On the other hand, we find large seasonal variations in the Polar region 
(SPL and PYH) above 10~km a.s.l.,
especially at the South Pole (SPL).
At the Pyh\"asalmi mine, the air density profile is similar to
that of the US-standard '76 atmospheric model below 10~km a.s.l,
but at the South Pole, the air density decreases quicker than that
even below 10~km a.s.l.

Thus, we expect some seasonal variations of atmospheric
neutrino flux except for the tropical region, and
study it some in detail with the NRLMSISE-00 atmospheric model in
the section~\ref{flux}.

\section{\label{geomagnetizm}Geomagnetic Field and Sites}

We have already reported a large effect of geomagnetic field on the
calculation of atmospheric neutrino flux for several sites in mid-latitude,
such as SK and SNO sites~\cite{hkkm2004,hkkm2011},
through the rigidity cutoff and muon bending.

As a naive illustration of the rigidity cutoff,
consider the gyro motion of cosmic rays guided by the horizontal 
component of the geomagnetic field near the Earth.
The cosmic rays with a very small radius of gyro motion
can not arrive at a very close point to the earth.
As the radius of the gyro motion becomes larger,
the cosmic rays can access the point, but the Earth works as a 
slant shield, limiting the access azimuth angle of cosmic rays.
When the radius is large enough, 
the Earth becomes a flat shield just limiting the upward going cosmic rays,
and the limitation on the access azimuth angle disappears.
This mechanism is called the rigidity cutoff.

The effect of the muon bending on the atmospheric neutrino
flux is often explained by the difference of the arrival 
zenith angle of neutrinos from that without the muon bending by 
the geomagnetic field.
As the atmospheric neutrino flux above a few GeV
has a large arrival zenith angle dependence,
a little difference of the zenith angle results 
in a large difference of the flux separately visible 
from the rigidity cutoff at these energies.
For the difference  of the arrival zenith angle,
the horizontal component of the geomagnetic field 
is responsible.

\begin{figure*}[htb]
 \centering
  {
\includegraphics[width=12cm]{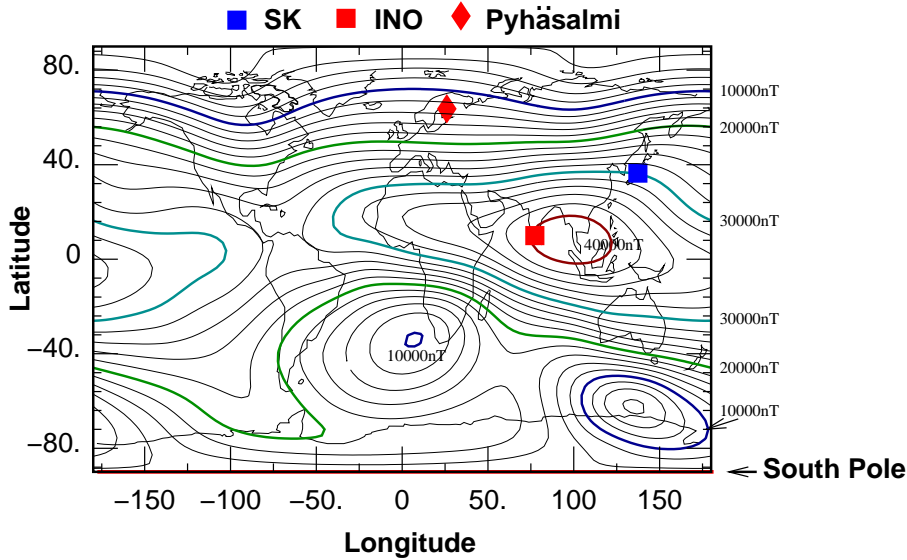} 
  }
\caption{
The horizontal component and the sites,
Kamioka (KAM), South pole (SPL),
INO site (INO), and Pyh\"asalmi mine (PYH).
where atmospheric neutrino flux
is calculated.
}
\label{fig:geomag}
\end{figure*}

Thus, the horizontal component of geomagnetic field ($B_h$) 
is  an important parameter 
to understand the effects of rigidity cutoff and muon bending.
In Fig.~\ref{fig:geomag}, we draw the strength of the horizontal 
component of the
geomagnetic field using the IGRF geomagnetic model for the year of 2010 
with the 
position of the SK site ($B_h\sim 30000$nT), INO site  ($B_h\sim 40000$nT),
South Pole ($B_h\sim 16000$nT), 
and Pyh\'asalmi mine ($B_h\sim 13000$nT)
for which we are going to calculate the atmospheric neutrino flux.

%Bh=16639.3 nT sp
%13302.0(pyh)
%18420.2 (sno)
%40186.6 ino
%30088.1 (kam)

\section{\label{calculation}Calculation of Atmospheric neutrino flux}

Except for the atmospheric model, the calculation scheme, including the
interaction model and the cosmic ray spectra model are the same 
as in the previous work~\cite{hkkm2011}.
We assume the surface of the Earth as a sphere with radius of
$R_e=6378.140$~km.
In addition, we assume 2 more spheres, the injection sphere
with a radius of $R_{inj} = R_e + 100$ km and escape sphere with radius  
$R_{esc}=10 \times R_e$.
Before Ref.~\cite{hkkms2006},
we assumed one more, the simulation sphere with radius of $R_{sim}$ 
($R_{inj} < R_{sim} < R_{esc}$).
We discarded the cosmic rays which go outside of this sphere
after the injection.
We took $R_{sim} = R_e + 3000$~km in Ref.~\cite{hkkms2006}.
However,  now 
we identify the simulation sphere and the escape sphere by taking
$R_{sim} = R_{esc}$ after Ref.~\cite{hkkm2011}.

For each cosmic ray event simulation,
we sample an energy and a chemical composition 
of  cosmic ray to simulate,
according to the cosmic ray spectra model. 
Then we sample the position
and the  initial direction of the cosmic ray on the injection sphere
to start the simulation.
For  each cosmic ray, we apply the rigidity cutoff test and check
if it reaches this position penetrating the rigidity cutoff.

This rigidity cutoff test is carried out by the back tracing method,
solving the equation of motion in the inverse time direction
in the geomagnetic field.
When the sampled cosmic ray reaches the escape sphere without 
touching the injection sphere again, we judge that the cosmic ray 
can reach the starting position from deep space,
and feed it into the simulator of the propagation in the atmosphere.

Generally, the transition from inhibited  to allowed rigidity
is not clear.
Most cosmic rays, which fail the rigidity cutoff test, hit the 
injection sphere very quickly before completing one cycle of  
gyro motion.
However, some of them with  rigidity near the transition,
travel a long distance before hitting the injection sphere. 
Also a cosmic ray with slightly higher rigidity than 
one  which passes the rigidity cutoff test
may fail the test even starting from the same position and direction.

A primary cosmic ray may pass through the injection sphere more 
than once.
In this case we have to select the first entrance to avoid double 
counting.
For this purpose, we require that the cosmic ray should not come back 
to the injection sphere again during back tracing.

The simulator of cosmic ray propagation in the atmosphere follows
the motion of all the cosmic rays, both primary cosmic rays which 
passed the rigidity cutoff test
and 
secondary cosmic rays produced in interactions of the cosmic rays
recursively, 
until they reach the escape sphere,
or hit the surface of the Earth,
or interact with an air nucleus, or decay.
Each neutrino produced in the cosmic ray interaction or decay 
is examined if its path will take it
through the virtual detector assumed around the neutrino observation
site, and when it goes through the virtual detector, it is registered
and the number is used to calculate the atmospheric neutrino flux.
We take a circle with radius of 1113.2~km as the virtual detector.
Note, the radius corresponds a change in longitude of
10~degrees on the equator. 
%the angle from the center of the Earth of 10~degrees,

 As the virtual detector is far larger than the real neutrino detector,
 we introduce the ``virtual detector correction''.
 Assuming a circle with a radius of $\theta_d$ around the real detector
 as the virtual detector,
 the flux $\Phi_{d}$ defined as the average flux in the
 virtual detector may be written in the form as
 \begin{equation}
 \label{eq:wide_flux}
 \Phi_{d} = \Phi_0 + \Phi_0^{(2)} \theta_d^2~ + \dots ,
 \end{equation} 
 with the flux $\Phi_0$ at the real detector~\cite{hkkms2006}.
 Then, we can cancel out the $\Phi_0^{(2)} \theta_d^2 $ term,
 using two fluxes $\Phi_1$ and $\Phi_2$
 determined in the
virtual detectors with radii $\theta_1$ and $\theta_2$
respectively  as
 \begin{equation}
 \label{eq:point_flux}
 \Phi_0 \simeq 
 \frac
 {\theta_1^2\Phi_2  - \theta_2^2\Phi_1}
 {\theta_1^2  - \theta_2^2}
 =
 \frac
 {\Phi_2  - r^2\Phi_1}
 {1 - r^2}    ,
 \end{equation}
 where $r=\theta_2$/$\theta_1$.
 We took $r=1/2$ in Ref.~\cite{hkkms2006}, but optimize
 it to $r=1/\sqrt{2}$ in Ref.~\cite{hkkm2011}
 to minimize the statistical error.

\section{\label{flux}Atmospheric neutrino flux at each site}

\begin{figure*}[htb]
 \centering
  {
\includegraphics[width=6cm]{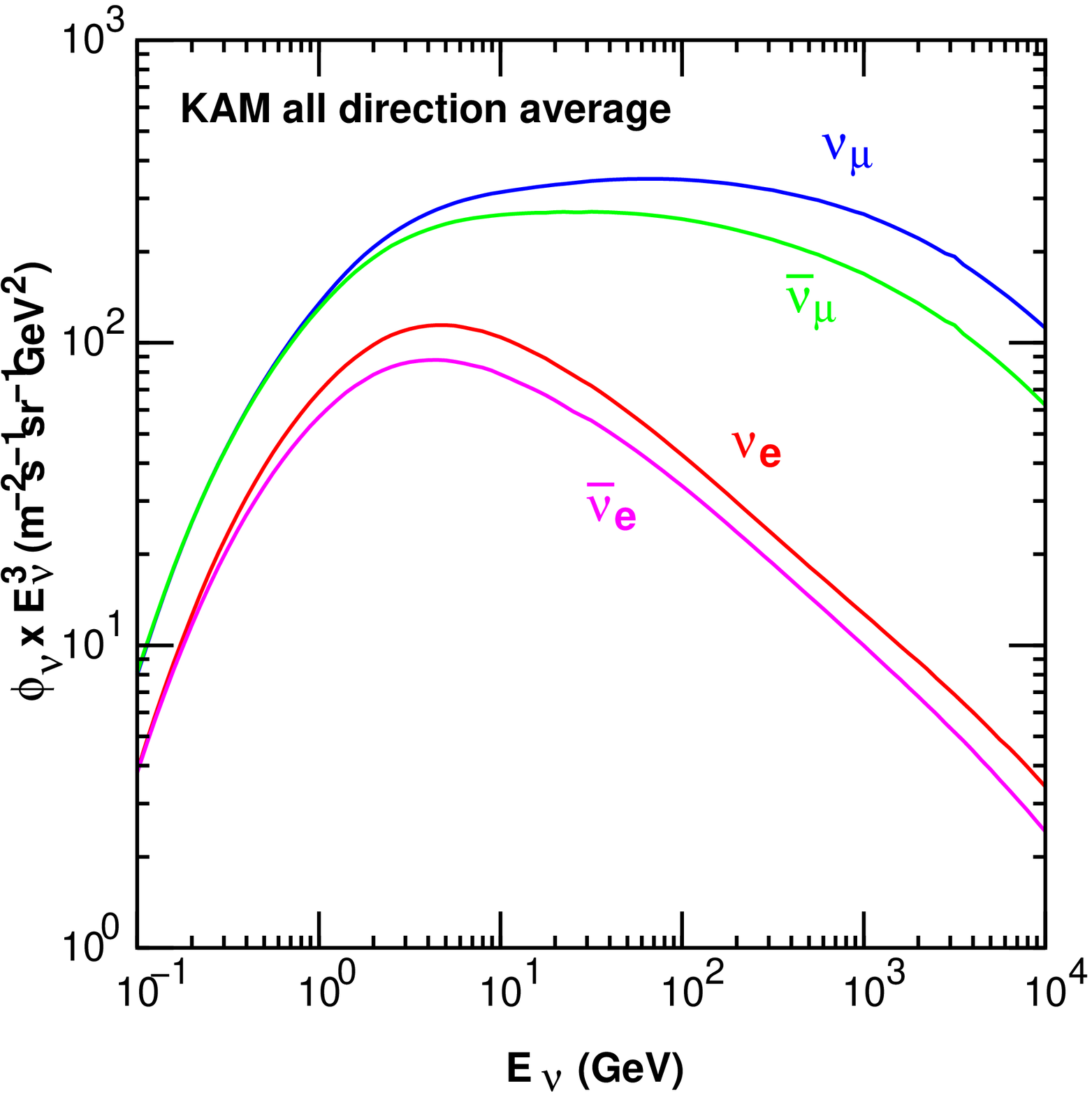} 
\hspace{5mm}
\includegraphics[width=6cm]{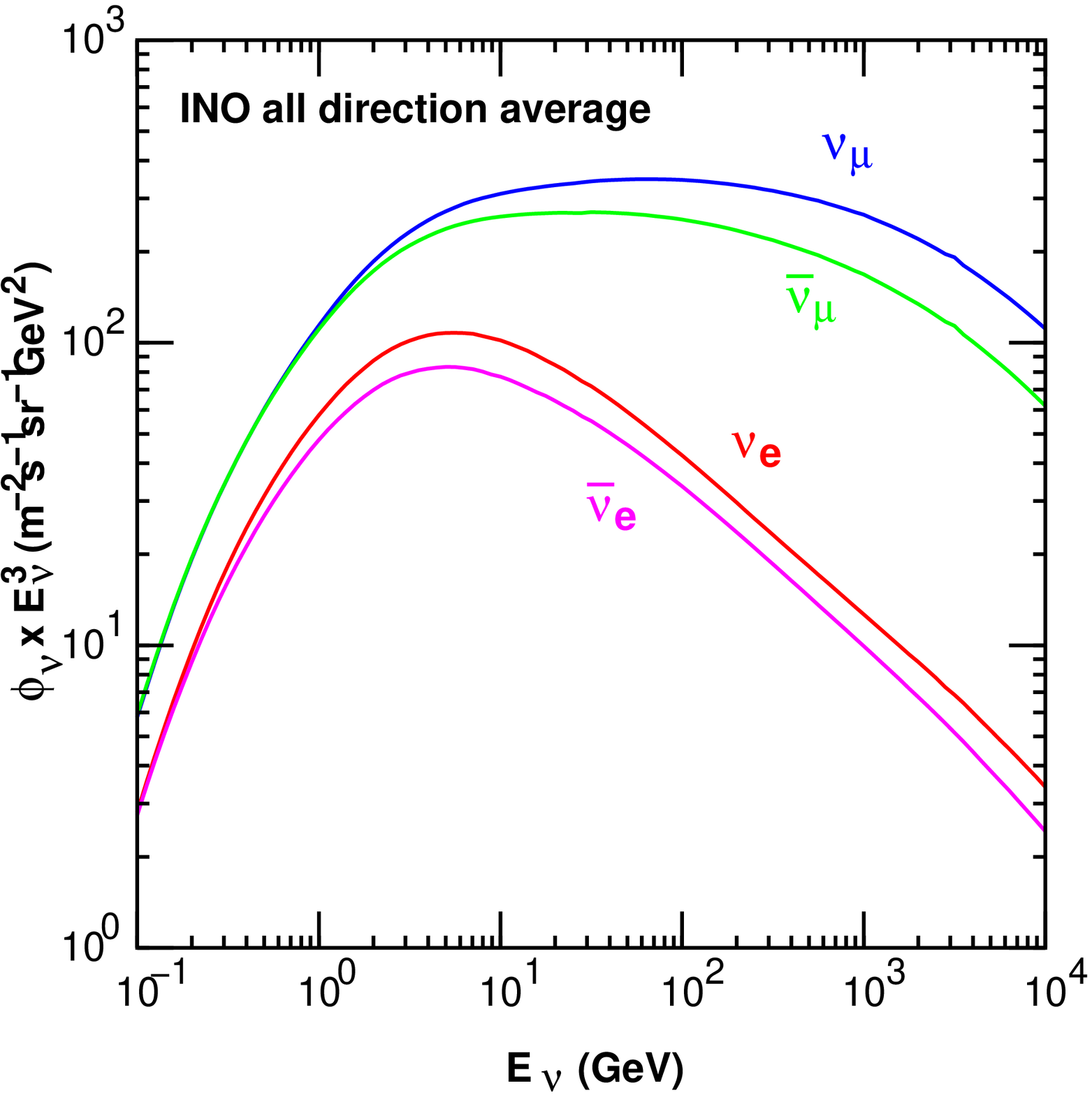}
\hspace{5mm}\\
\includegraphics[width=6cm]{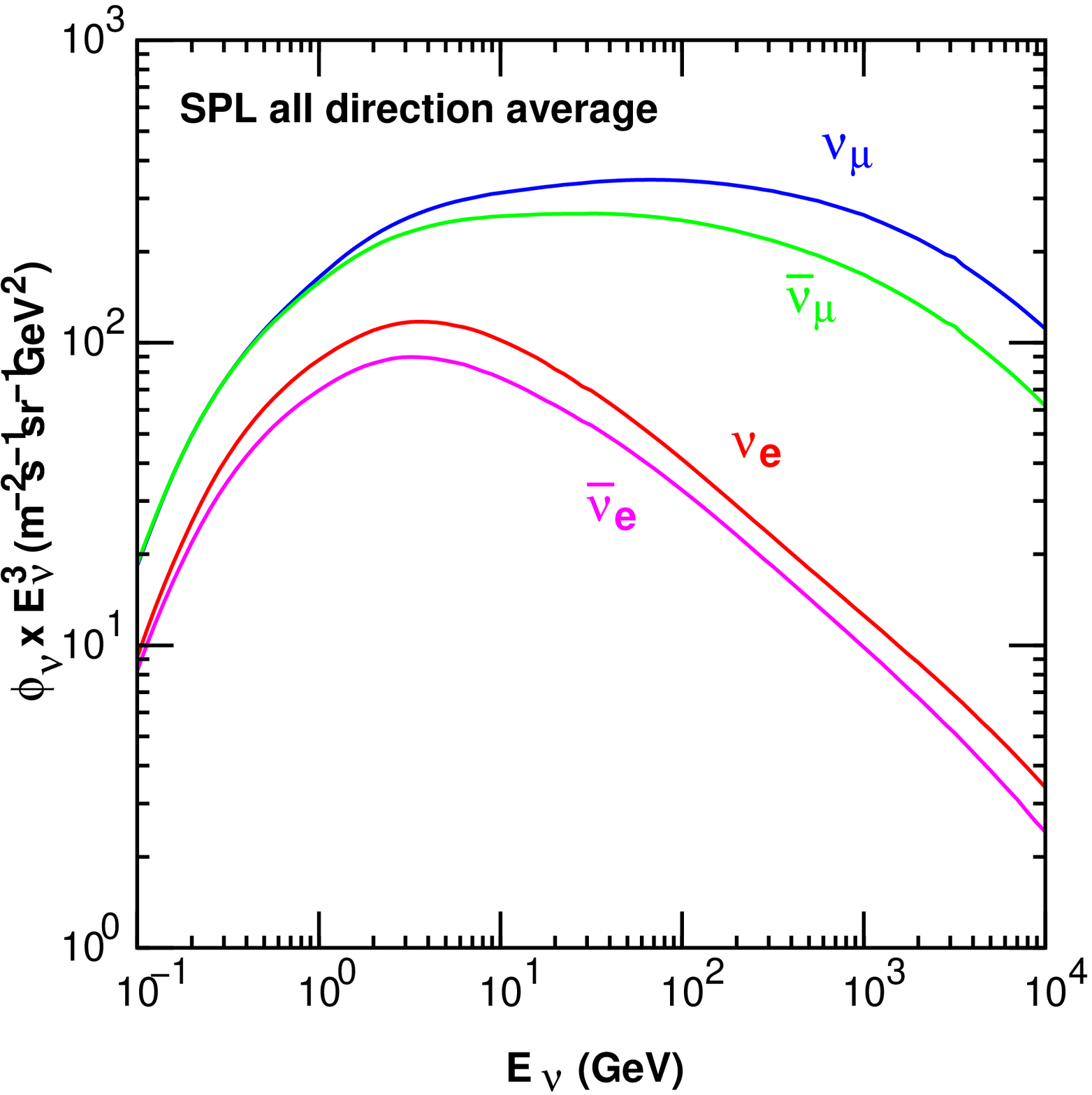}
\hspace{5mm}
\includegraphics[width=6cm]{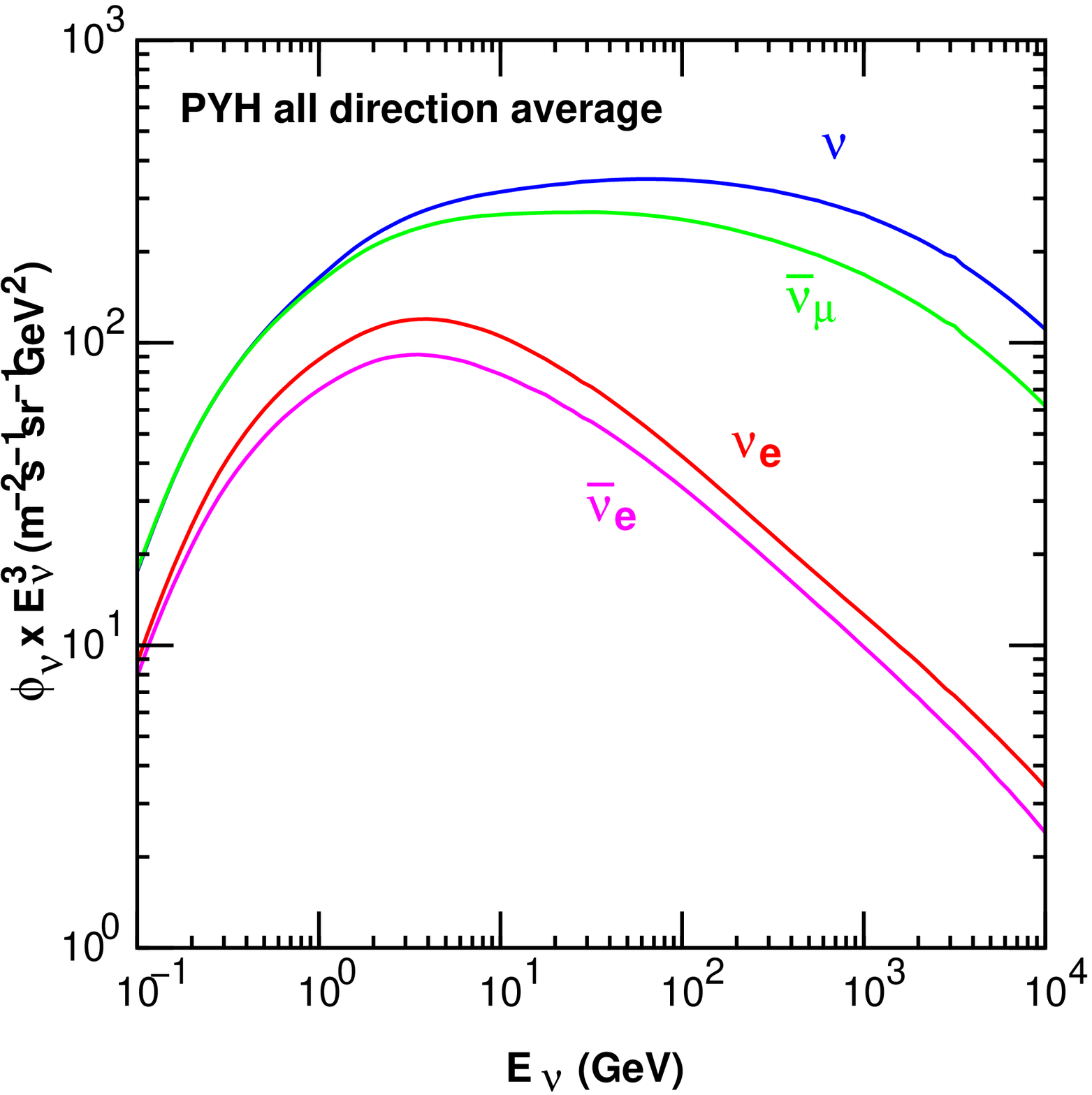}
  }
\caption{
All-direction averaged atmospheric neutrino flux for 4 sites 
averaging over one year. 
KAM stands for the SK site, INO for the INO site, SPL for the South Pole, and 
PYH for the Pyh\"asalmi mine.
}
 \label{fig:alldir}
 \end{figure*}

In Fig.~\ref{fig:alldir}, we show the one year average of atmospheric 
neutrino fluxes 
at the SK site, INO site, South Pole, and Pyh\"asalmi mine,
averaging over all the directions.
The qualitative features are the same at all the sites.
but we find a difference of flux among the sites 
by factor $\sim$ 3 at the low energy
end due to the large difference of the cutoff rigidity among these sites.
The differences of flux  among the sites above 10~GeV is small in the
figure. However we expect a large seasonal variations at 
South Pole and  Pyh\"asalmi mine, and we will study them
in the next subsection some in detail.

\subsection{\label{sec:seasonal} Seasonal variation of the atmospheric neutrino flux}

\begin{figure*}[htb]
 \centering
  {
    \includegraphics[width=4cm]{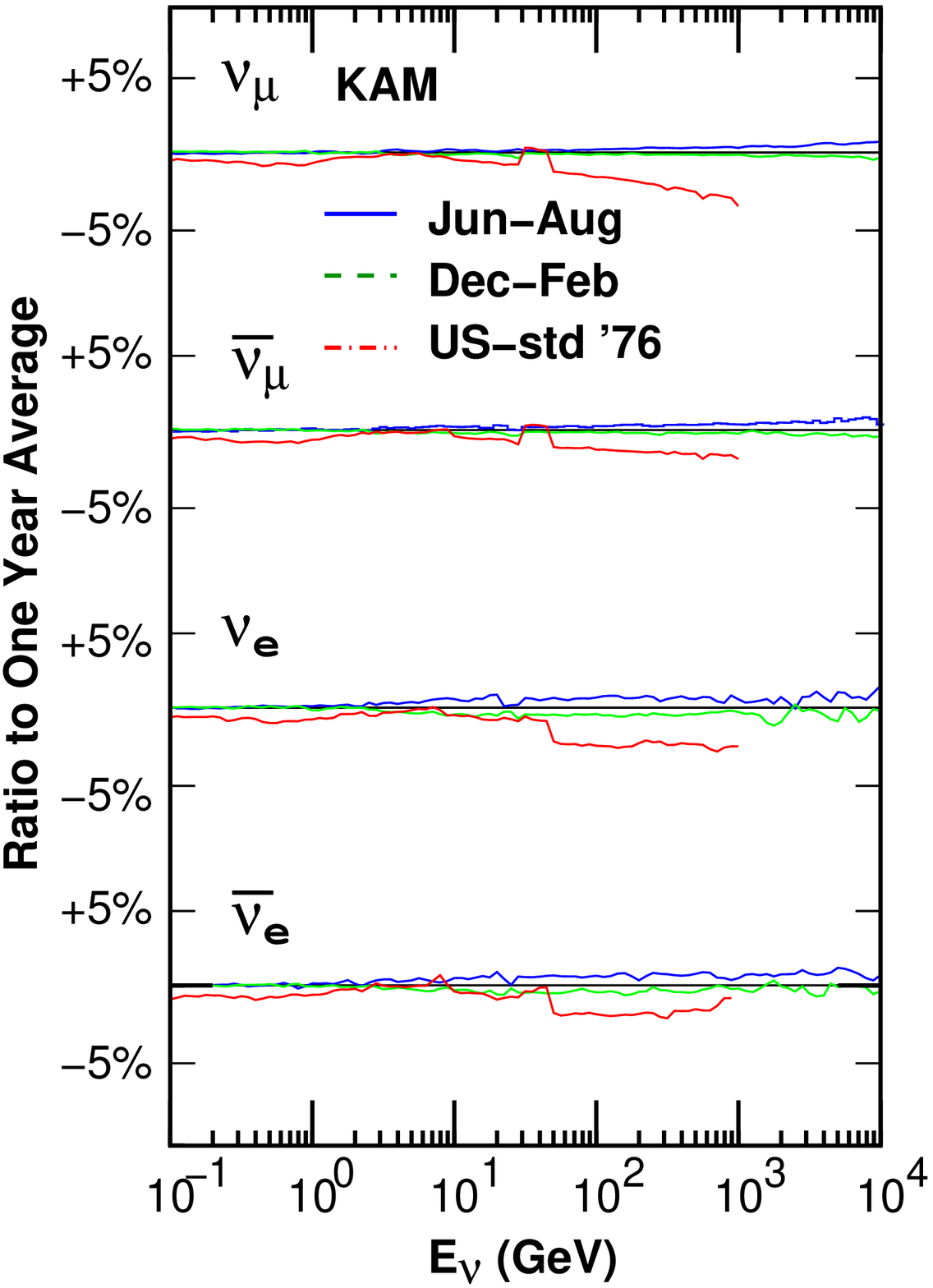} 
    \includegraphics[width=4cm]{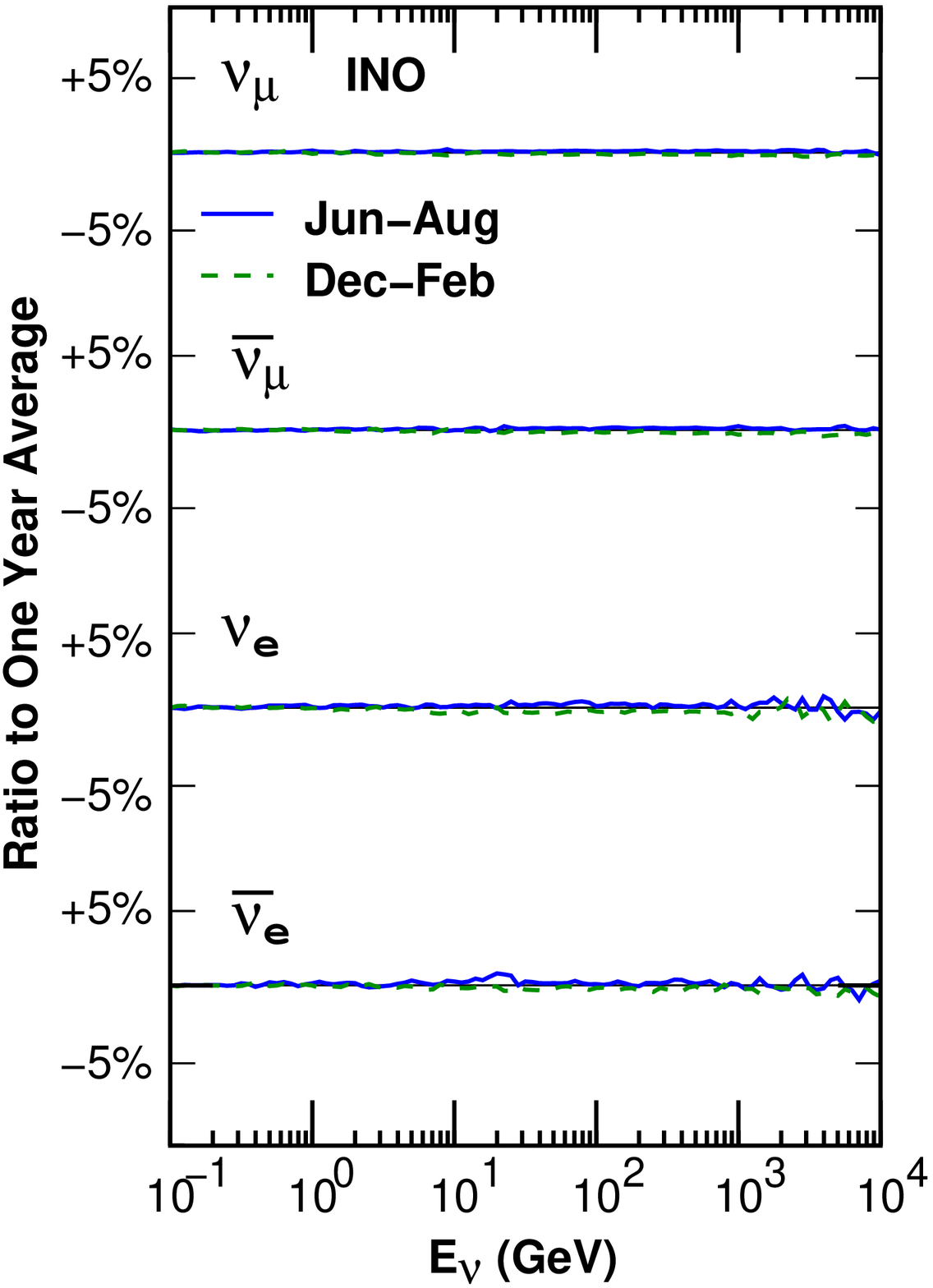}
    \includegraphics[width=4cm]{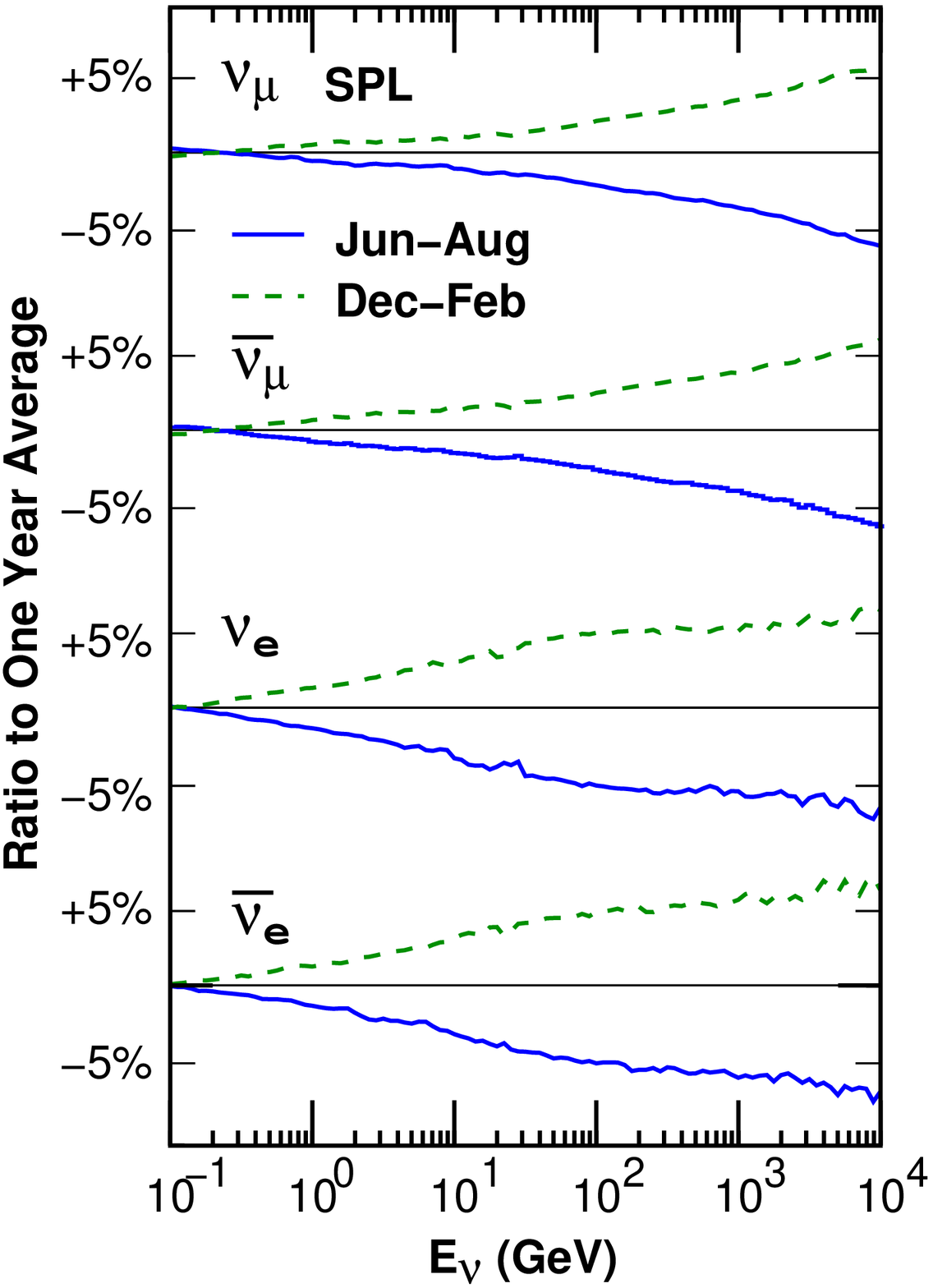}
    \includegraphics[width=4cm]{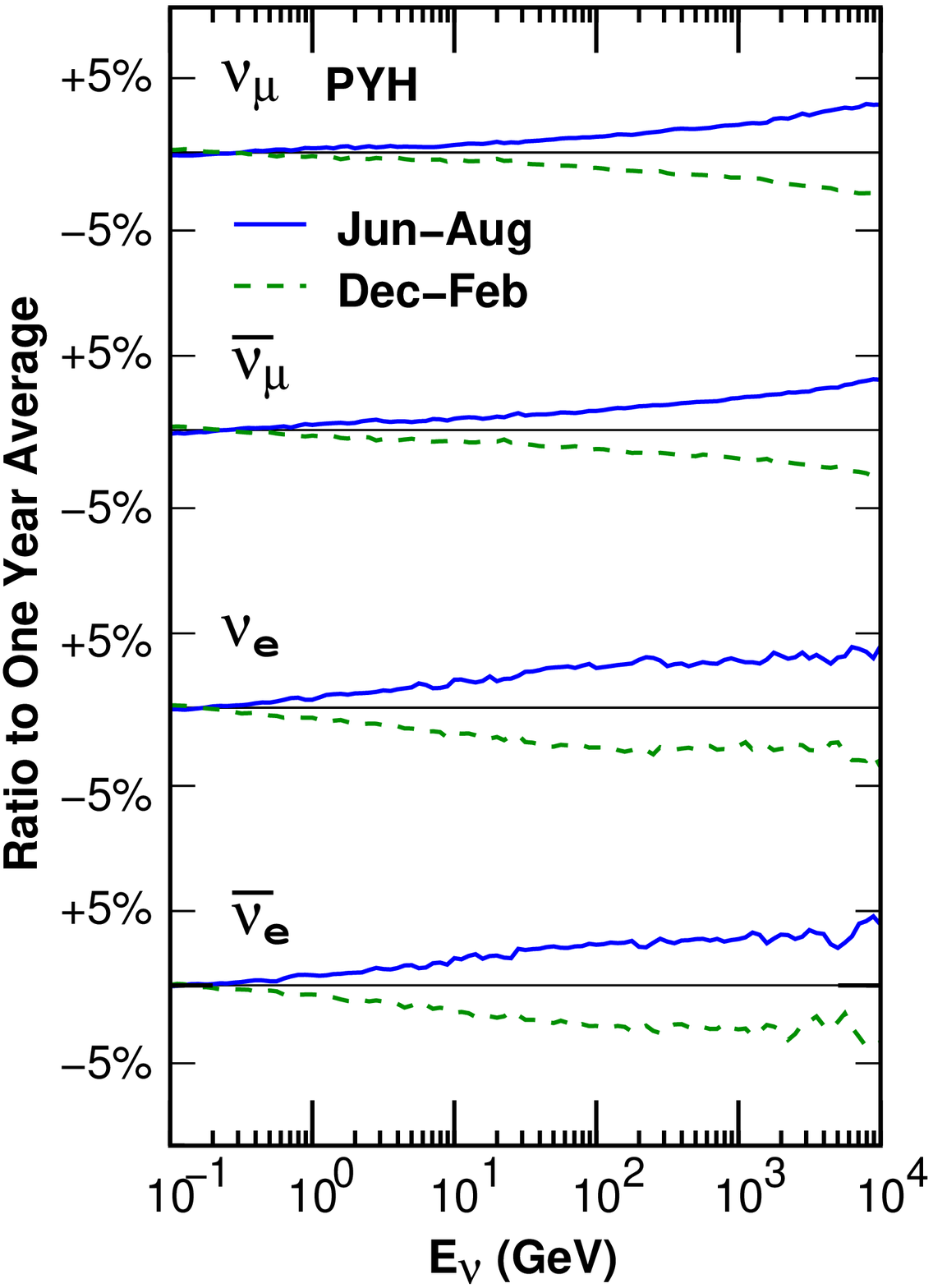}
  }
\caption{
  Ratios of the all-direction averaged flux in June -- August
  and in December --  February to that of the yearly average.
  For the SK site, we also plot the ratio for the calculation with
  the US-standard '76 atmospheric model to the yearly average in 
  dash-dot below 1 TeV. 
}
 \label{fig:r2ally}
 \end{figure*}

To study the seasonal variations,
we calculate the  ratio of seasonal fluxes  to the yearly flux average,
and show them in Fig.~\ref{fig:r2ally}.
Also the ratio of the flux calculated with the US-standard '76
atmospheric model to the yearly
flux average is shown in the panel for SK site below 1 TeV.
Note,
the calculation of the fluxes with the US-standard '76 atmospheric model
was carried out in the previous work~\cite{hkkm2011},
and the statistics of the Monte Carlo simulation is poorer than that in
the present work especially above 1~TeV.

At the mid-latitude region (SK site) and tropical region (INO site), 
the seasonal differences are small, and is difficult to see even 
in the ratio.
On the other hand, 
the seasonal variaston is large in the Polar region,
as the summer--winter difference reaches more than 10~\% at the South Pole  at 10~TeV,
and more than 5~\% at the Pyh\"asalmi mine.
At both sites, the atmospheric neutrino flux is higher
in summer (December -- February at the South Pole and 
June -- August at the Pyh\"asalmi mine).
This may be understood by the fact that the air density at
 higher altitudes ($\gtrsim$ 15~km) is higher in the summer at both sites
(see Fig.~\ref{fig:ratio2us}). 
We note the seasonal variation of the $\nu_e$ and $\bar\nu_e$ fluxes 
starts from $\lesssim$~10~GeV, where that of
the $\nu_\mu$ and $\bar\nu_\mu$ is still small.

The effect of the air density profile on the atmospheric neutrinos is 
generally discussed considering the relative probability
of pions to decay or to interact with air nuclei.
This effect works equally on all flavors of neutrino 
at high energies ($\gtrsim$ 100~GeV),
where  the probability of the  parent pions to interact becomes comparable 
to that to decay,
and  may explain the large seasonal variation above 1~TeV.
However, it does not explain the seasonal variation of the 
$\nu_e$ and $\bar\nu_e$ fluxes starting from $\lesssim$~10~GeV,
and the difference from that of the $\nu_\mu$ and $\bar\nu_\mu$ fluxes.

To explain these difference,
we need to consider the muon propagation in the atmosphere.
When the air shrinks down lower, the muons are created at 
lower altitudes, and the probability of muons to hit the ground before
decaying increases.
When the muons hit the rock or ice, they lose their energy 
quickly producing neutrinos with very low energies 
($\lesssim$~0.1~GeV) only.
Then the flux of neutrinos produced in the decay of muons decreases.
This mechanism results in the variation of neutrino flux
near the vertical directions at relatively high energies ($\gtrsim$~10~GeV).
It explain the variation of neutrino flux there,
and the difference of the $\nu_e$ and $\bar\nu_e$ fluxes 
and the $\nu_\mu$ and $\bar\nu_\mu$ fluxes,
as half of the $\nu_\mu$ and $\bar\nu_\mu$ are created in the 
pion decay directly.
%%%%%

There is another mechanism which has a large effect on the energy spectra 
of neutrinos at lower energies.
When the atmosphere shrinks lower,
the muons travel in denser air and lose more energy before the decay.
The larger energy loss of the muons causes a shift to the  energy 
spectra of the neutrinos produced in the muon decay towards the lower energy 
direction.
As the energy spectra of atmospheric neutrinos are steeply decreasing
at the energies $\gtrsim$ 0.1 GeV, 
the fluxes decrease in the denser air at those energies.
However, 
this mechanism is effective when most muons decay in the air
($E_\mu \lesssim$~GeV),
and for the neutrinos with  energies below a few GeV.
This mechanism is responsible to the seasonal variation of neutrino 
flux below 10 GeV.

The differences between the fluxes calculated 
with the US-standard '76 atmospheric model and yearly averaged one at the 
SK site are far less than 5~\%.
It is difficult to say that the differences are due to the
difference in the  atmospheric model, since there are some improvement in the 
calculation scheme and statistics of Monte Carlo simulation
after the previous work~\cite{hkkm2011}.


\subsection{\label{sec:rr} Flavor-ratio of the atmospheric neutrino flux}

\begin{figure*}[htb]
   \centering
       {
         \includegraphics[width=4cm]{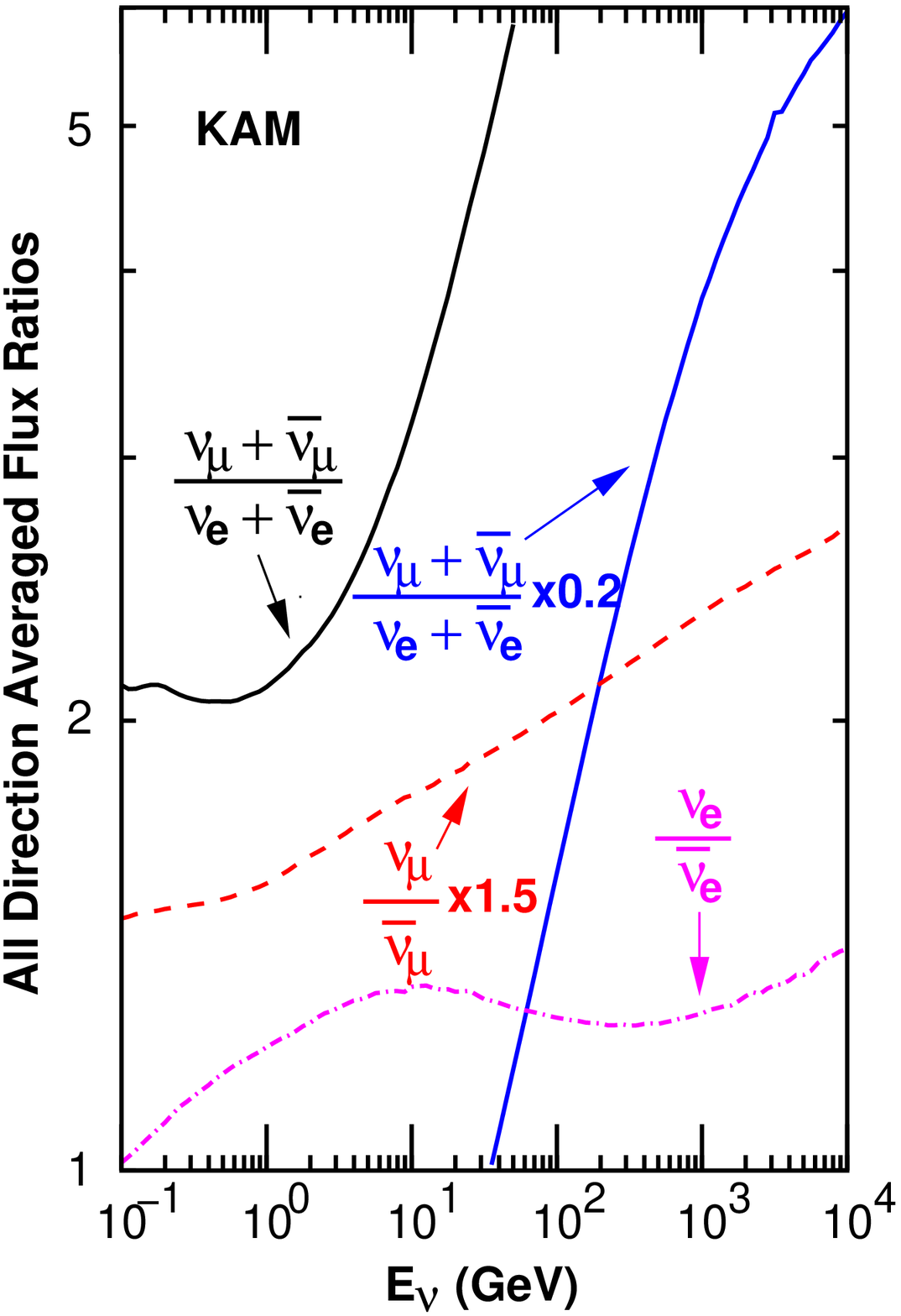} 
         \includegraphics[width=4cm]{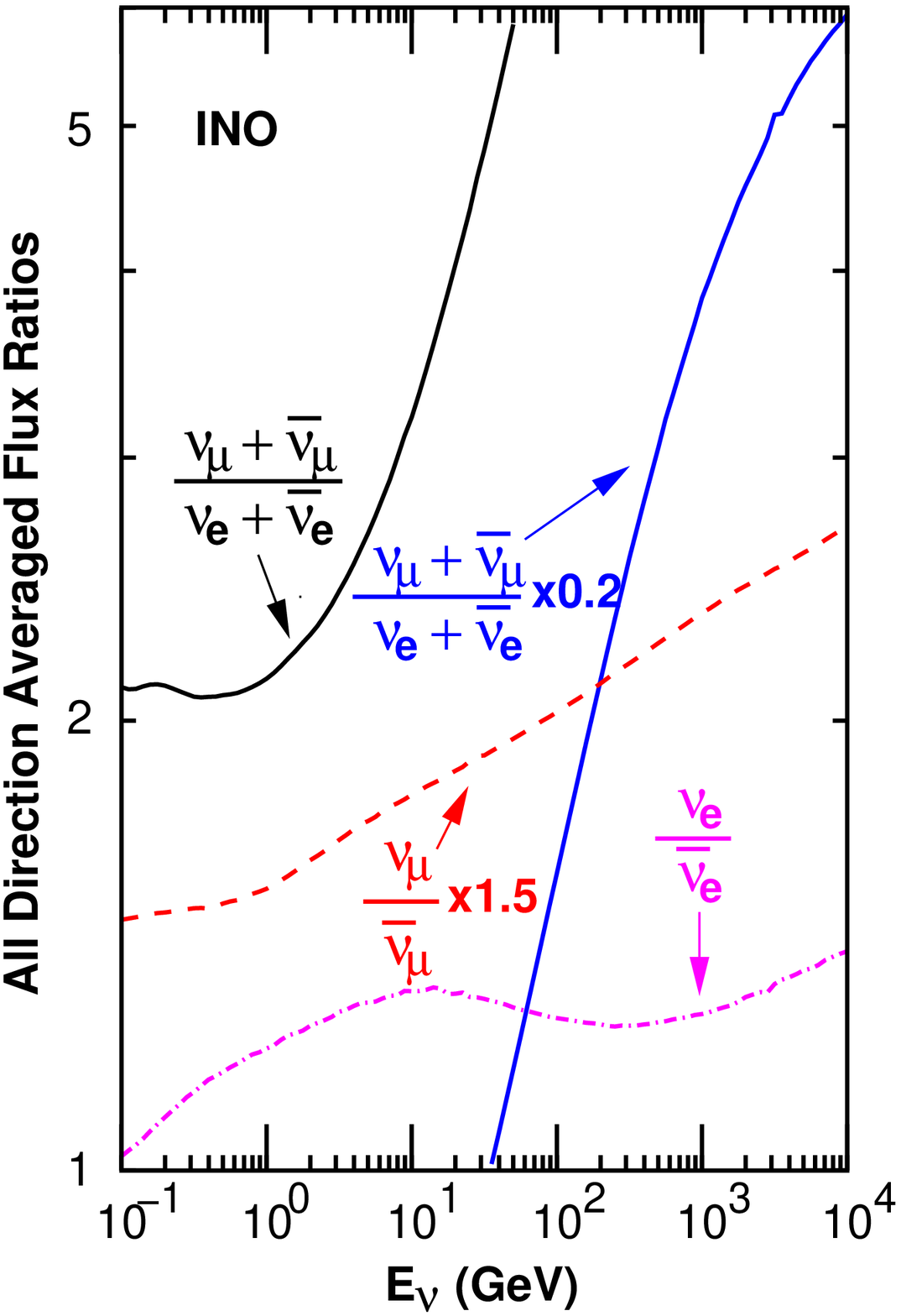}
         \includegraphics[width=4cm]{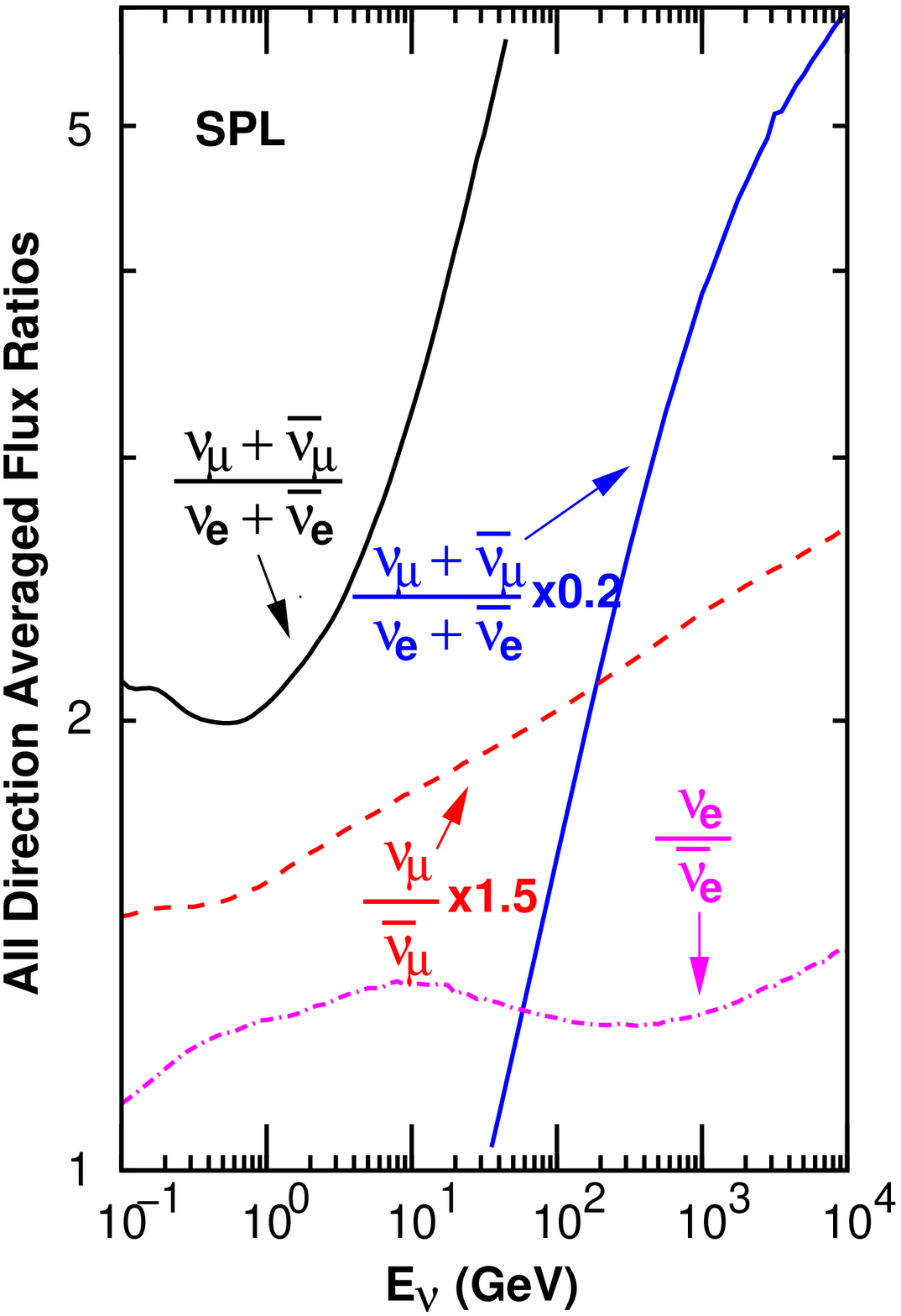}
         \includegraphics[width=4cm]{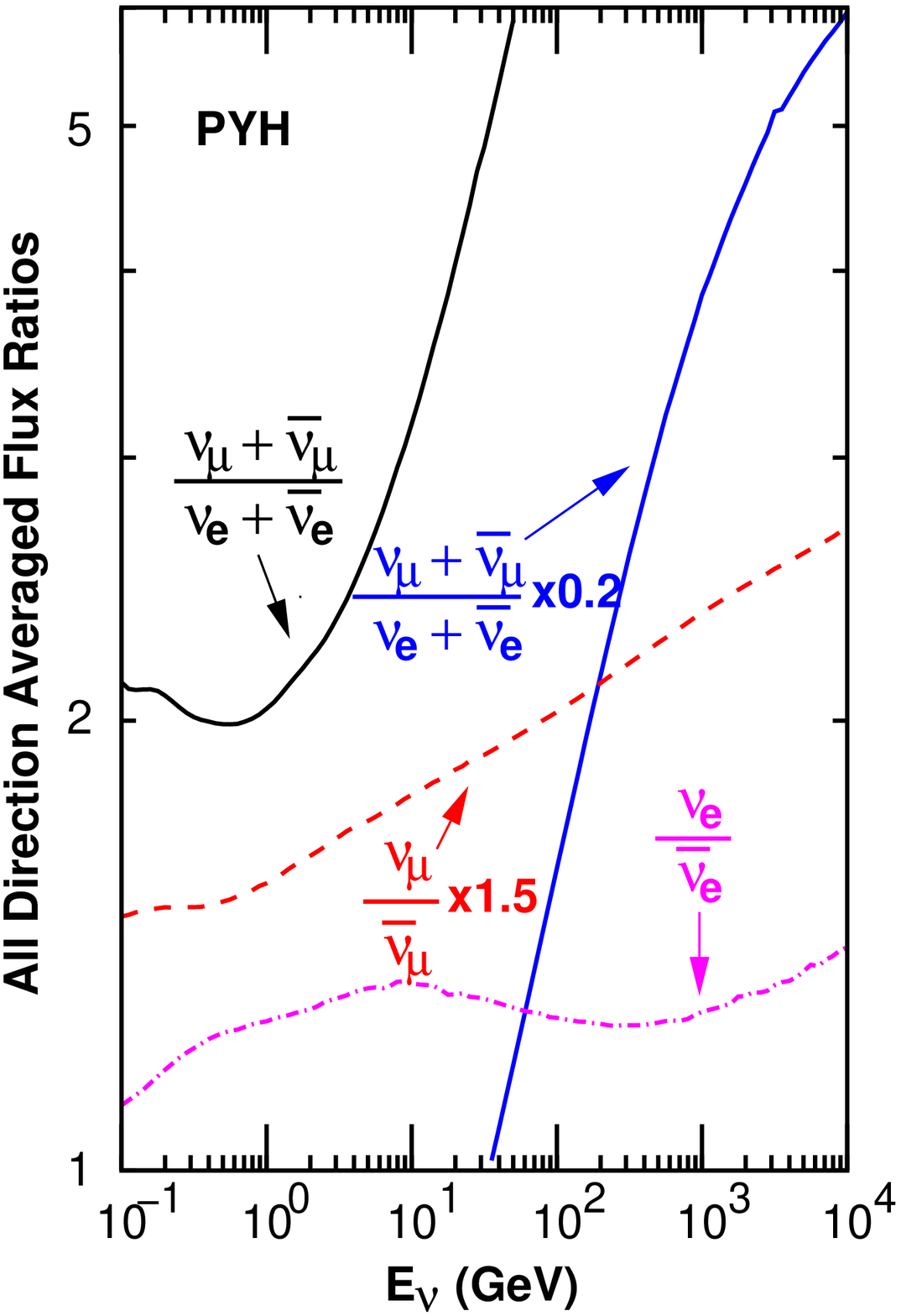}
       }
       \caption{
         Neutrino flavor-ratio calculated with the all-direction and one-year averaged
         atmospheric neutrino flux.
         KAM stands for the SK site, INO for the INO site, SPL for the South Pole, and 
         PYH for the Pyh\"asalmi mine.
       }
       \label{fig:nratio}
 \end{figure*}

In Fig.~\ref{fig:nratio}, we show the flavor-ratio defined as the flux ratio 
of different neutrino flavors such as,
$(\nu_\mu + \bar\nu_\mu)/(\nu_e+\bar\nu_e)$,
$(\nu_\mu / \bar\nu_\mu)$, and
$(\nu_e / \bar\nu_e)$
at the SK site, INO site, South Pole, 
and Pyh\"asalmi mine,
averaging over all the directions.
We find that the flavor-ratio is very similar to each other among these 
sites, confirming the stability of the flavor-ratio.
However, the flavor-ratio is an important quantity 
in the study of neutrino oscillations,
so we need to study the seasonal variations 
and position dependence's more precisely.

\begin{figure*}[htb]
   \centering
       {
         \includegraphics[width=4cm]{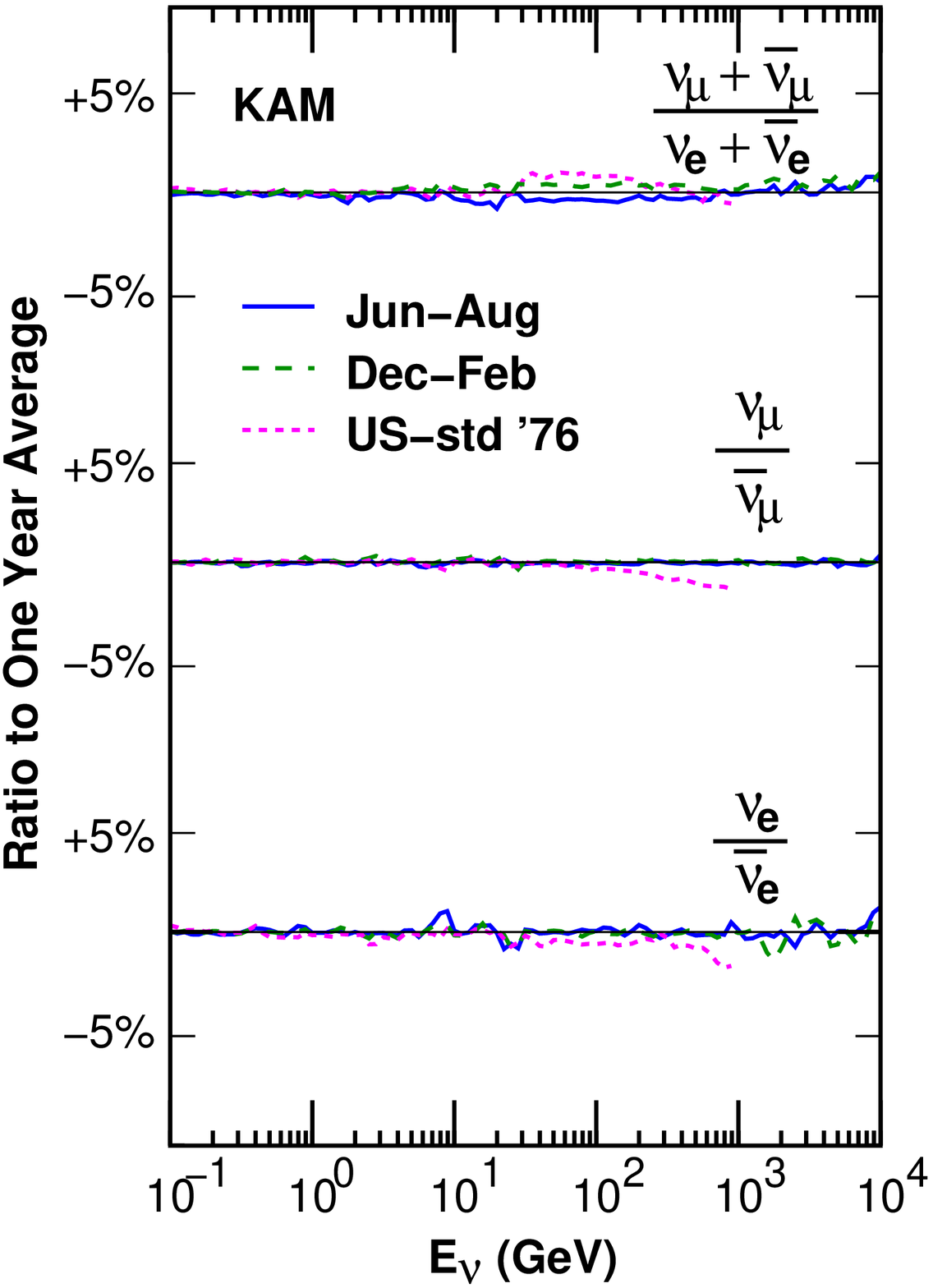} 
         \includegraphics[width=4cm]{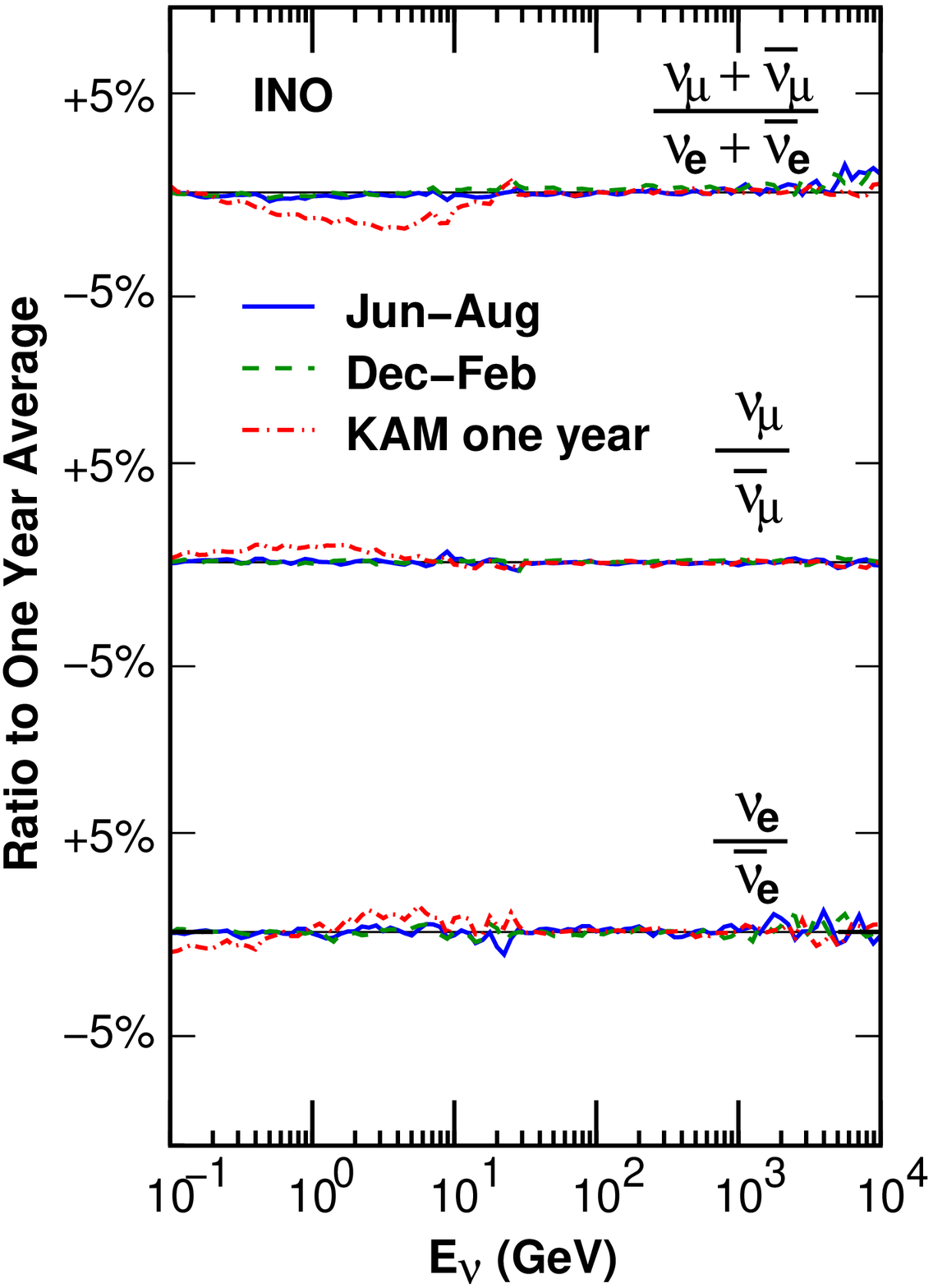}
         \includegraphics[width=4cm]{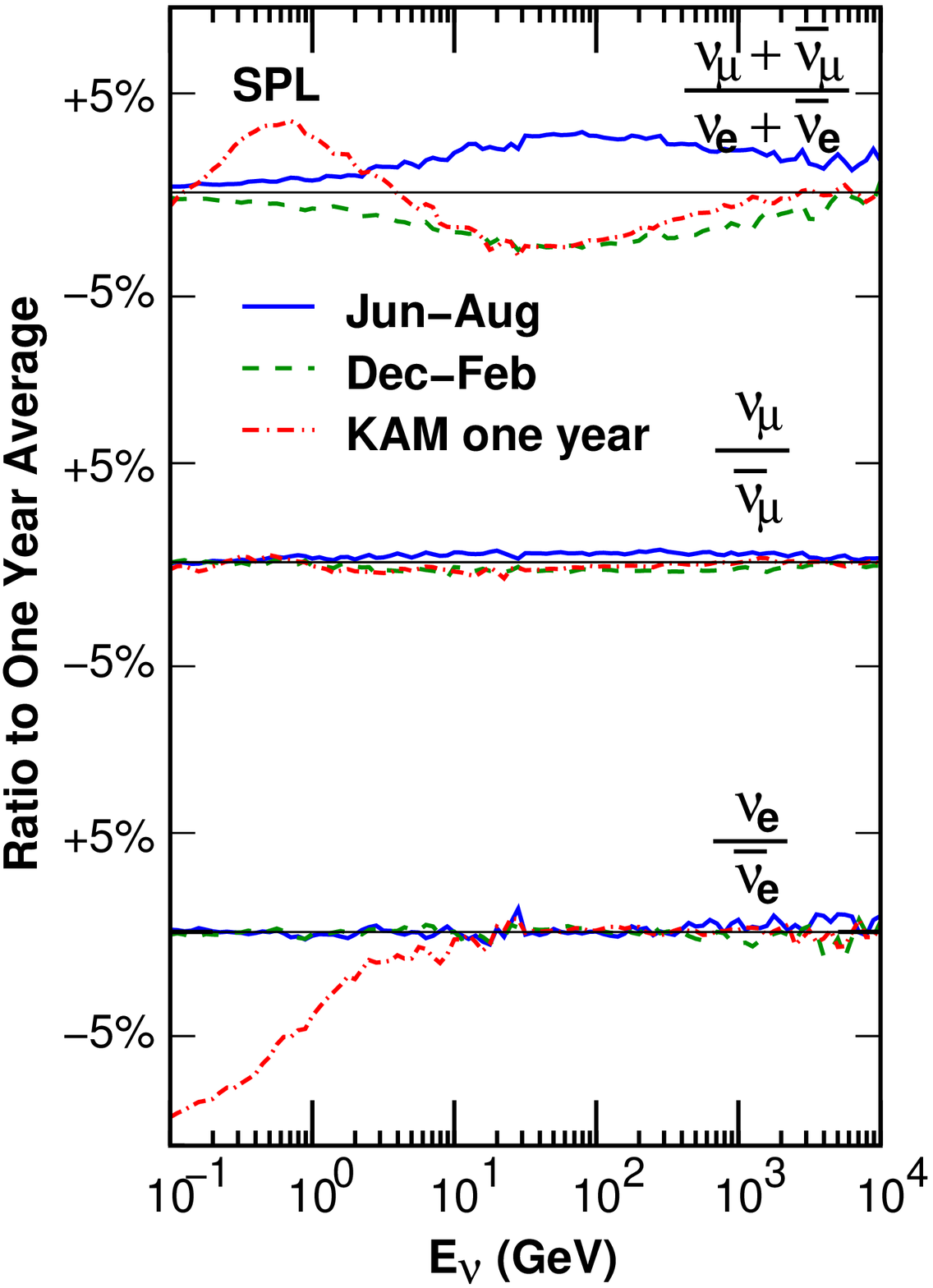}
         \includegraphics[width=4cm]{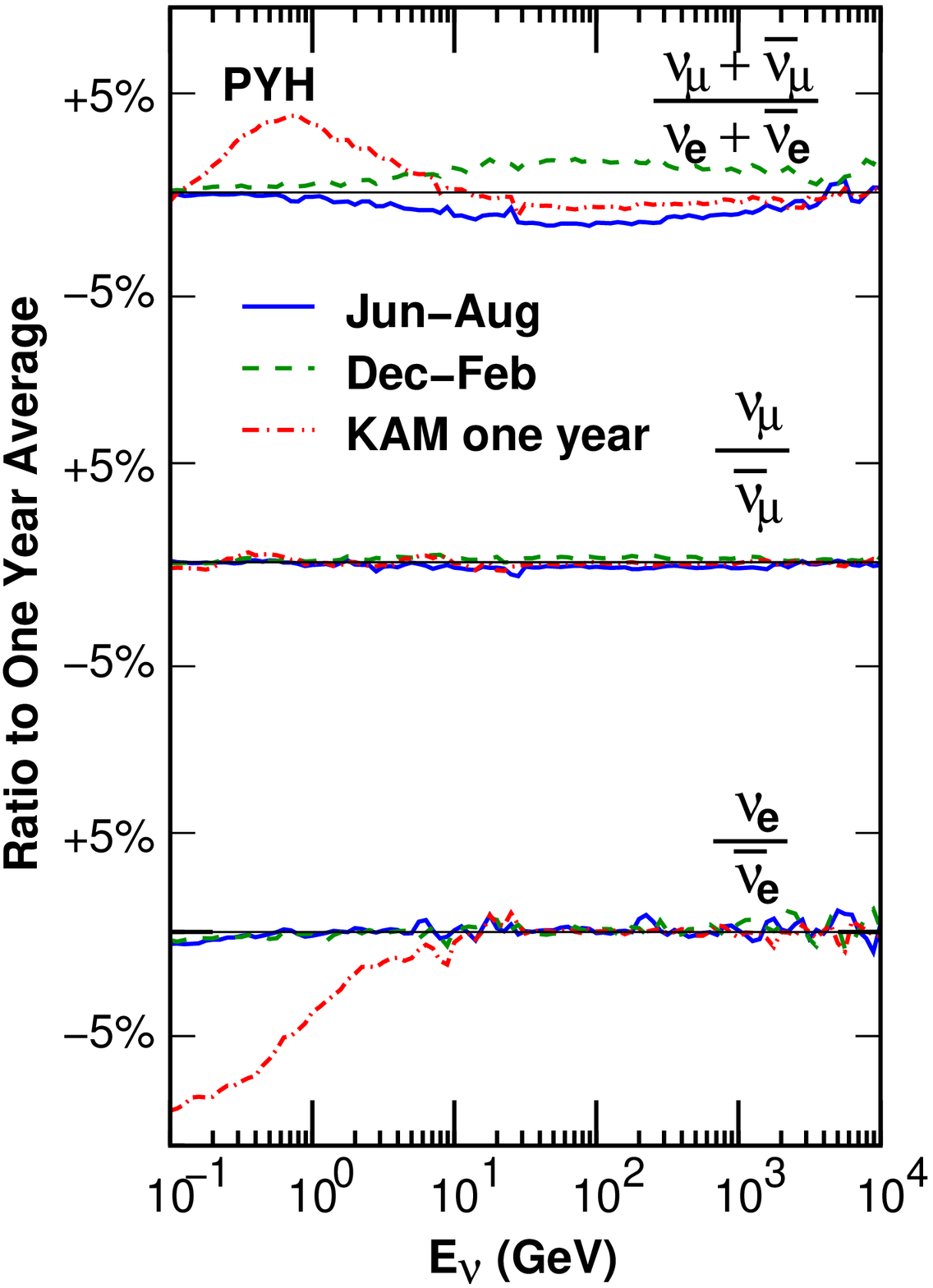}
       }
       \caption{
         Ratios of the flavor-ratio calculated with the all-direction and
         seasonally averaged flux to that calculated with all-direction
         and yealy averaged flux. 
         The solid lines are the flux ratio in June -- August, and dashed lines
         are in December --  February.
         KAM stands for the SK site, INO for the INO site, SPL for the South Pole, and 
         PYH for the Pyh\"asalmi mine.
         Taking the yealy averaged flavor-ratio at SK-site as the '`reference flavor-ratio'',
         the  ratio of '`reference flavor-ratio'' to the yearly averaged flavor-ratio
         at each site are plotted with dash-dot other than the panel for the SK-site.
         In the panel for the SK-site, the ratio of the flavor-ratio with
         the US-standard '76 atmospheric model to the yearly averaged flavor-ratio
         at SK-site is plotted with dash-dot.     
       }
       \label{fig:rr}
 \end{figure*}

To see the seasonal variation of the flavor-ratio,
we calculate the ratio of 
the flavor-ratio calculated with the seasonally averaged
fluxes to that calculated with the yearly averaged fluxes
and plot them in Fig.~\ref{fig:rr} for each site.
Also, to see the positional dependence on the earth,
we take the flavor-ratio calculated with the yearly averaged
fluxes at the SK site as the ``reference flavor-ratio''.
Then we calculate the ratio of the ``reference flavor-ratio''
to those calculated with yearly averaged fluxes at other sites,
and plotted them in their panels of Fig.~\ref{fig:rr}.
While in the panel of the SK site, we show the ratio of the flavor-ratio
calculated with the US-standard '76 atmospheric model
to that calculated with the yearly averaged fluxes at
the SK site (i.e. the ``reference flavor-ratio'').

At the sites in the Polar region (South Pole and Pyh\"asalmi mine),
the flavor-ratio,
$(\nu_\mu + \bar\nu_\mu)/(\nu_e+\bar\nu_e)$ 
shows a seasonal variation, high in summer and low in winter,
with the maximum of the amplitude at $\sim$~100~GeV.
This is considered to be due to the seasonal variation of 
the altitude of cosmic ray interactions.
Also the flavor-ratio
shows some differences from the SK site 
at the energy below a few GeV.
This is considered to be due to the lower air density at the 
neutrino production height  of 10$\sim$20 km a.s.l
(see section~\ref{sec:height}) in the Polar region.
The smaller muon energy loss causes a smaller shifts of the energy 
spectra of  the neutrinos produced in the muon decay .

Note, 
the air density at the South pole at the 10$\sim$20 km 
a.s.l. is much lower than at the Pyh\"asalmi mine in
Fig.~\ref{fig:ratio2us},
but the difference of the flavor-ratio,
$(\nu_\mu + \bar\nu_\mu)/(\nu_e+\bar\nu_e)$ at the South Pole 
from the SK site is similar to that at the Pyh\"asalmi mine.
This is considered to be due to the higher observation site
at the South Pole (2835m a.s.l.).
The shorter distance from the ground to the production height
 of muons reduces the neutrino fluxes produced by muon decay
at the South Pole.

In the $\nu_e/\bar\nu_e$ ratio, we also find a difference
from the SK site at the South Pole and Pyh\"asalmi mine,
below a few GeV.
This difference is considered to be due to the difference of 
cutoff rigidity.
The $\nu_e/\bar\nu_e$ ratio reflects the $\pi^+/\pi^-$
ratio of parent pions.
As the majority of primary cosmic rays are protons,
there is a $\pi^+$ excess generally.
Especially when the cutoff rigidity is low enough, 
the pion production of primary cosmic rays overwhelms that
of secondary cosmic rays,
and  $\pi^+/\pi^-$ and  $\nu_e/\bar\nu_e$ ratios are
high even at low energies.
However, when cutoff rigidity is high,
the pion production by secondary cosmic rays can not be
ignored, and the $\pi^+$ excess is diluted by the 
secondary neutron cosmic ray interactions.

In the comparison of neutrino flavor-ratio between the
SK site (mid-latitude region) and the INO site (tropical region),
we find a small difference in the
$(\nu_\mu + \bar\nu_\mu)/(\nu_e+\bar\nu_e)$ ratio
due to the difference of air density at $15$~km a.s.l.,
and the difference of the muon energy loss.
Other ratios are quite similar to each other.
The differences of the flavor-ratio with the US-standard '76 
atmosphere model are very small to that in the present calculation.

\subsection{\label{sec:zdep}Zenith angle variation of the atmospheric neutrino flux at 3.2~GeV}

\begin{figure*}[htb]
 \centering
  {
    \includegraphics[width=4cm]{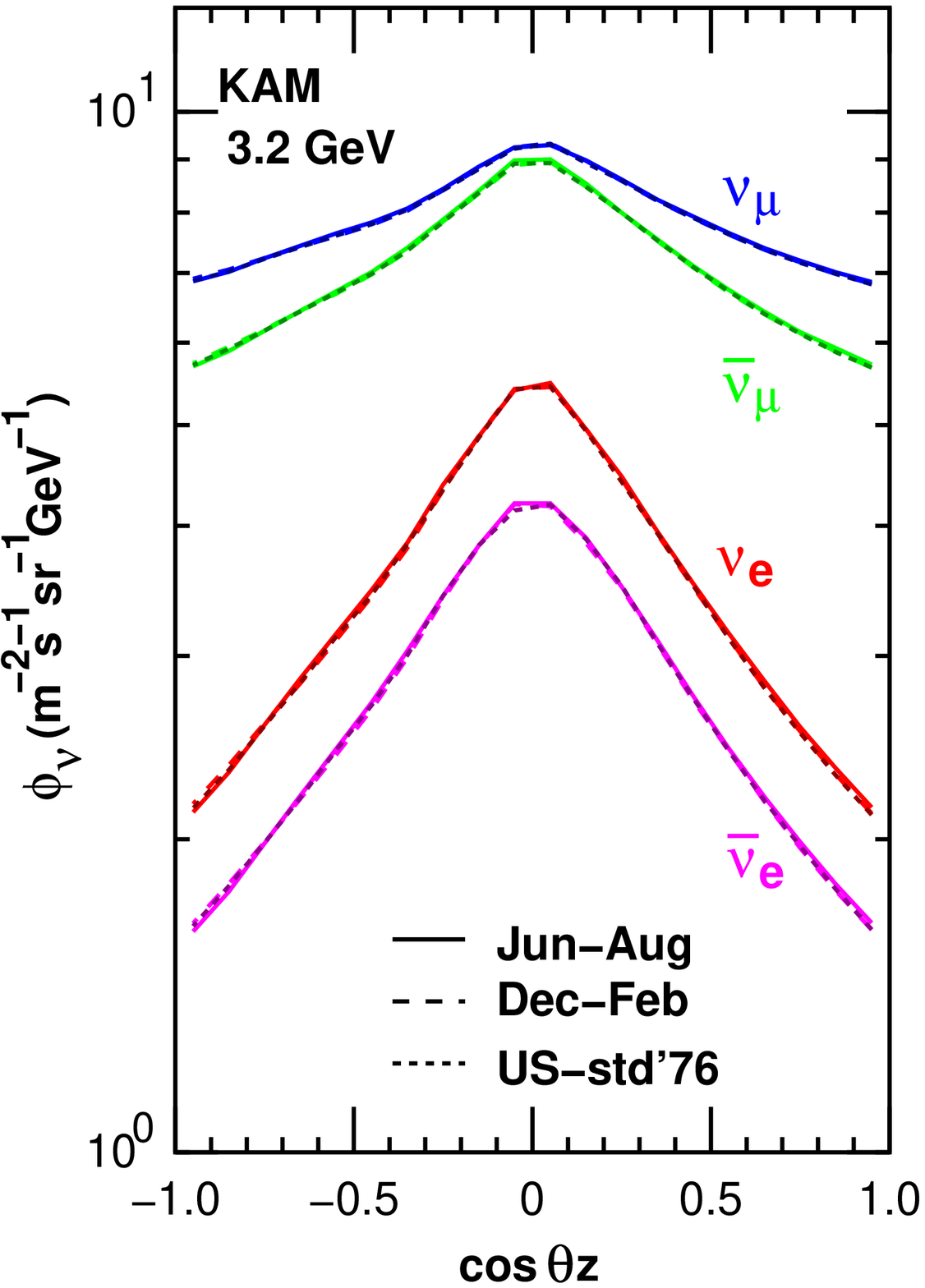} 
    \includegraphics[width=4cm]{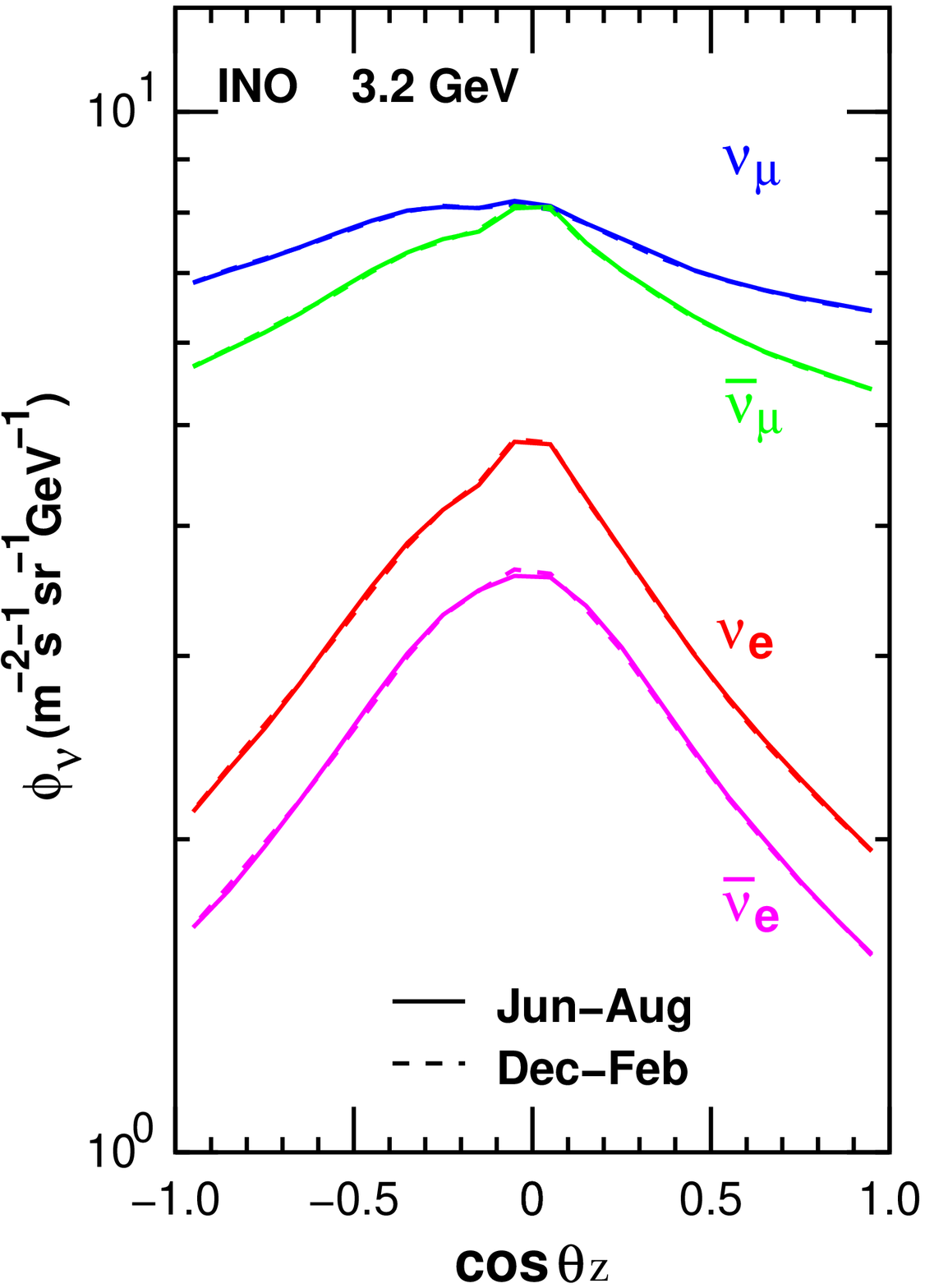}
    \includegraphics[width=4cm]{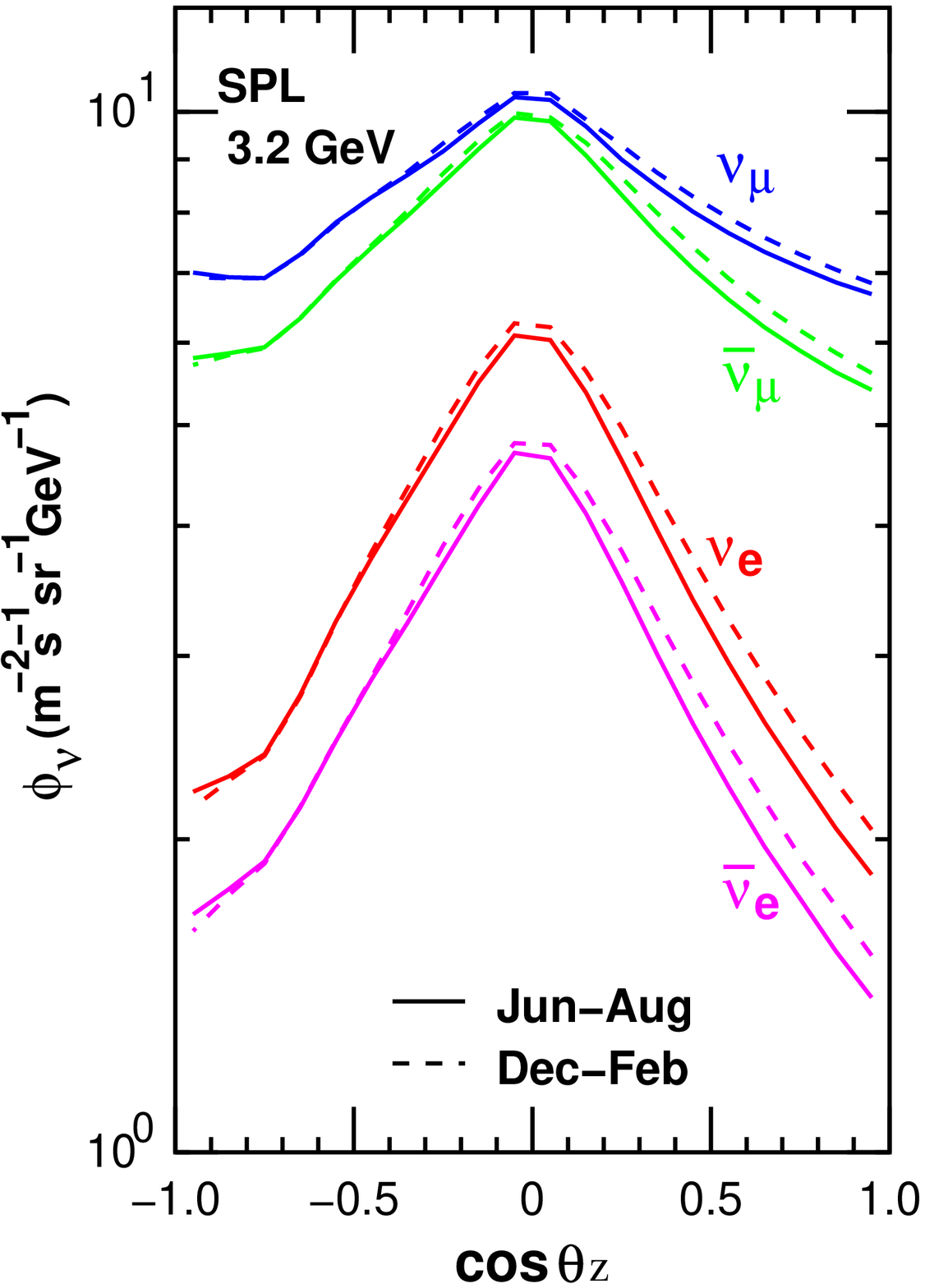}
    \includegraphics[width=4cm]{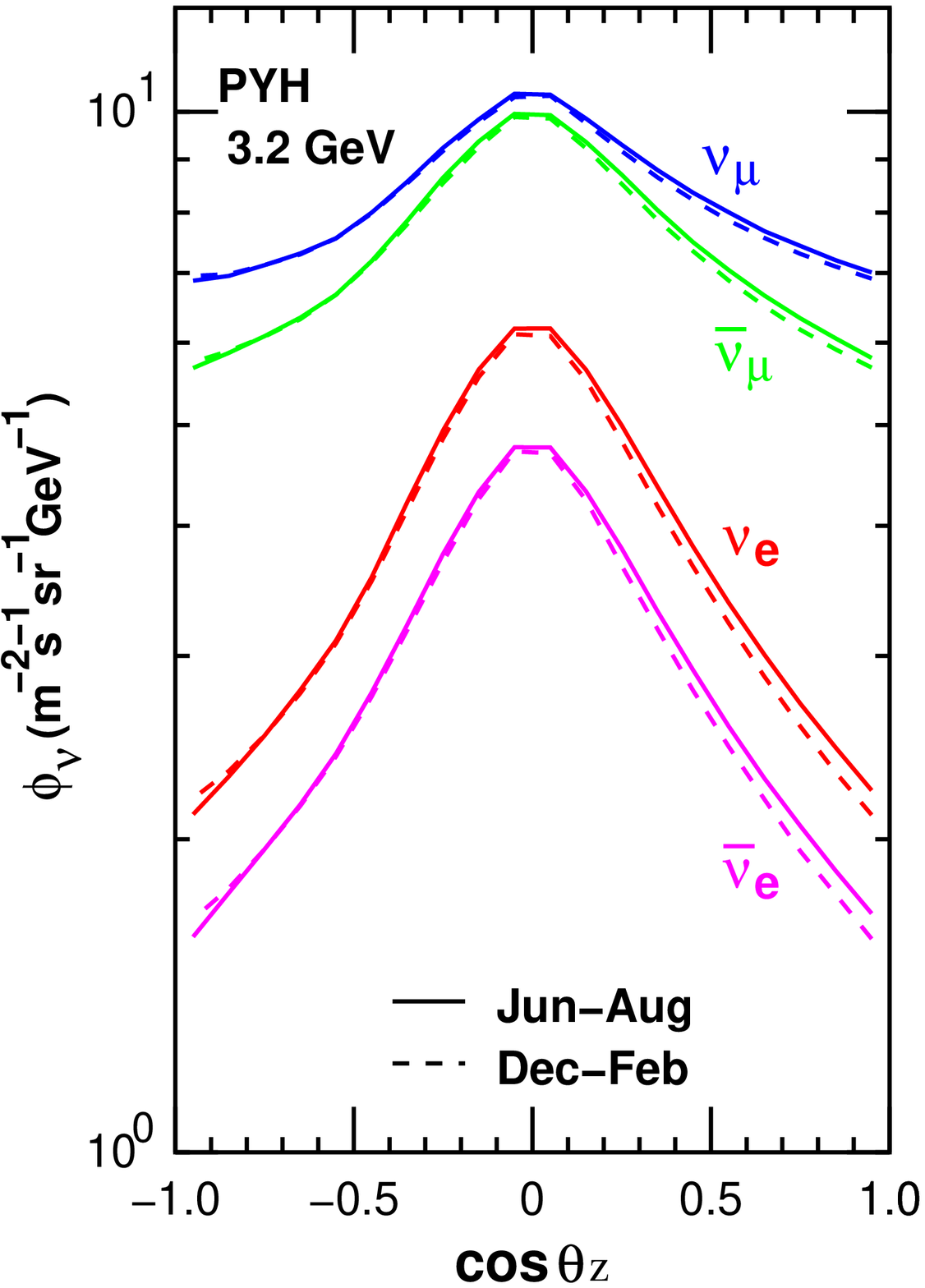}
  }
\caption{
  Arrival zenith angle dependence of atmospheric neutrino fluxes averaged
  over all azimuth angles at 3.2~GeV. 
  KAM stands for the SK site, INO for the INO site, SPL for the South Pole, 
  and PYH for the Pyh\"asalmi mine.
  The solid lines are the time average in June -- August, and dashed lines
  are in December --  February.
  For SK site, we also plot the calculation with the US-standard '76 atmospheric
  model using a dotted line (but is difficult to see separately).
  Note, $\theta_z$ stands for the zenith angle.
}
 \label{fig:zdep}
 \end{figure*}

In Fig.~\ref{fig:zdep}, we show the the zenith angle dependence 
of the atmospheric neutrino flux calculated at the 4 sites at 3.2~GeV. 
We also plot the atmospheric neutrino flux calculated with 
the US-standard '76 atmospheric model~\cite{hkkm2011} at the SK site,
although the differences are small and within the thickness of the line.
The seasonal variation
is very small at the SK and INO sites as in the all-direction average,
but 
is seen clearly in the down going neutrino fluxes
at the South Pole and Pyh\"asalmi mine even at this energy.
At both the sites the trend is the same as the
all-direction average at high energies, 
the atmospheric neutrino flux is large in summer
(June -- August at  Pyh\"asalmi mine and December --  February 
at the South Pole).

The amplitude of the seasonal variation is different among 
different neutrino flavors in Fig.~\ref{fig:zdep}.
It is $\sim$ 2.5\% for $\nu_\mu$ and $\bar\nu_\mu$, 
and $\sim$ 10 \% for $\nu_e$ and $\bar\nu_e$
for the vertically down going direction at the South Pole.
The muon energy loss in the atmosphere before the decay seems 
to be mainly responsible to this seasonal variations.

Leaving from the seasonal variation,
the atmospheric neutrino flux at near horizontal directions
are largely different from site to site.
The horizontal fluxes at the SK site are $\sim 10$~\% smaller than 
those at the South Pole or Pyh\"asalmi mine,
but $\sim 10$~\% larger than those at the INO site.
The differences in the atmospheric neutrino fluxes 
at near horizontal directions is mainly caused by 
the difference of rigidity cutoff at near horizontal directions.
The rigidity cutoff works most strongly for the neutrinos arriving
from near horizontal geomagnetic East directions.
At the SK site, we can almost ignore the rigidity cutoff at
near vertical directions, but it is still strong 
at near horizontal directions.
We can explain the difference between the SK site and the Polar region 
by the difference of rigidity cutoff, 
but can not explain the difference between INO and SK sites
only by the difference of rigidity cutoff.
We need to consider the muon bending, 
as the suppression for the $\nu_\mu$ and $\bar\nu_e$ is stronger 
than that for  $\bar\nu_\mu$ and $\nu_e$ at the INO site.
The suppression by the muon bending will be understood better 
with the azimuthal variation of atmospheric neutrinos described 
in subsection \ref{sec:adep}.

The fluxes at the South Pole and Pyh\"asalmi mine 
are a little smaller than the SK and INO sites at 
$\cos\theta_z \sim -0.7$, and they have some structure
there due to the rigidity cutoff and the muon bending 
at the places where the neutrinos are produced.
On the other hand, the rigidity cutoff and the muon bending
effects are weaker at the SK and INO sites at $\cos\theta_z \sim -0.7$ 
than the South Pole and Pyh\"asalmi mine.

Apart from the geomagnetic field, 
the vertically down going neutrino fluxes at the South Pole are
a little smaller than those at the Pyh\"asalmi mine due to the 
difference of the observation altitudes.
For the South Pole, 2835m a.s.l. is assumed as the observation site,
and sea level for Pyh\"asalmi mine.

\subsection{\label{sec:adep} Azimuthal variation of atmospheric neutrino flux}

 \begin{figure*}[htb]
   \centering
       {
         \includegraphics[width=4cm]{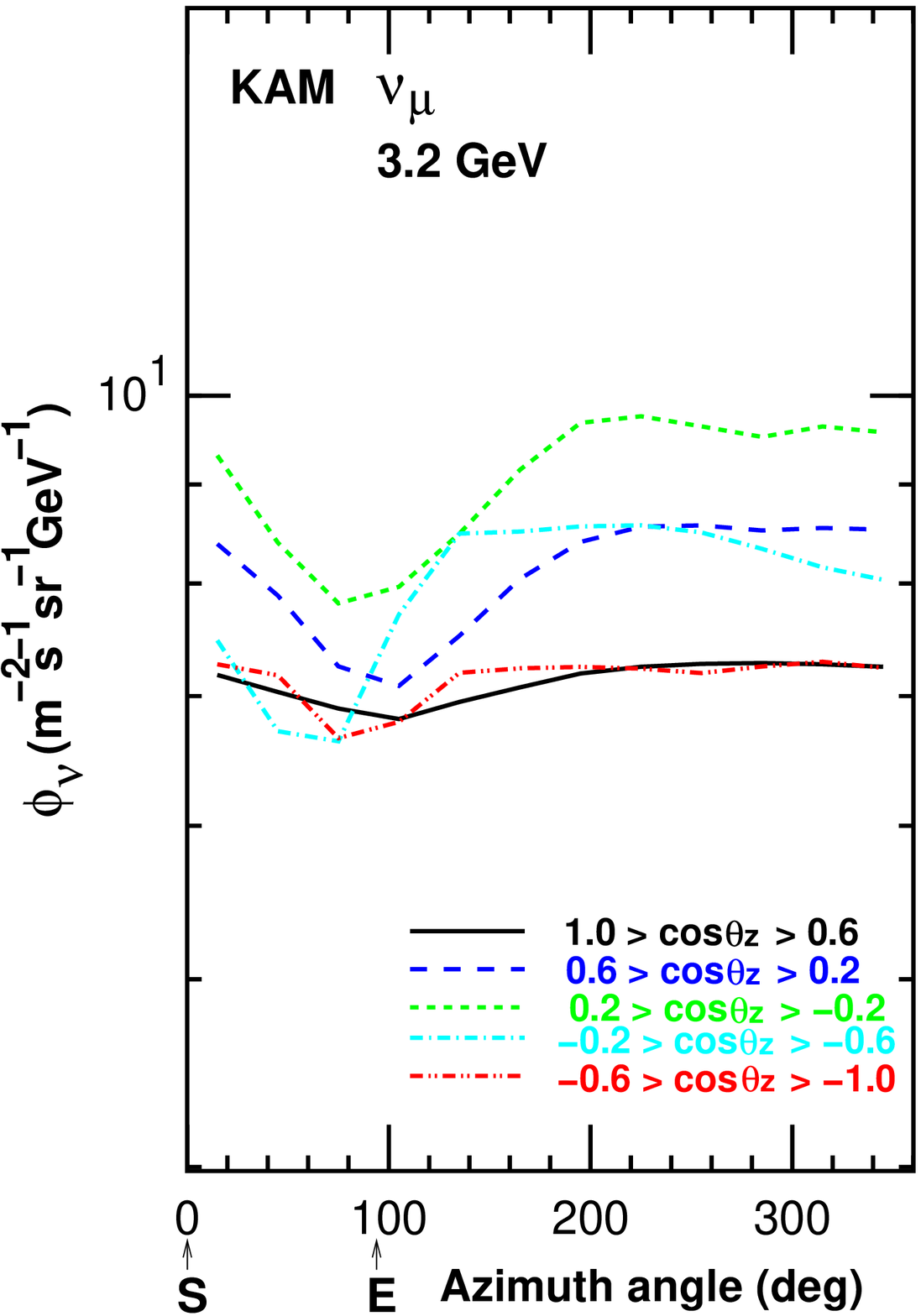}
         \includegraphics[width=4cm]{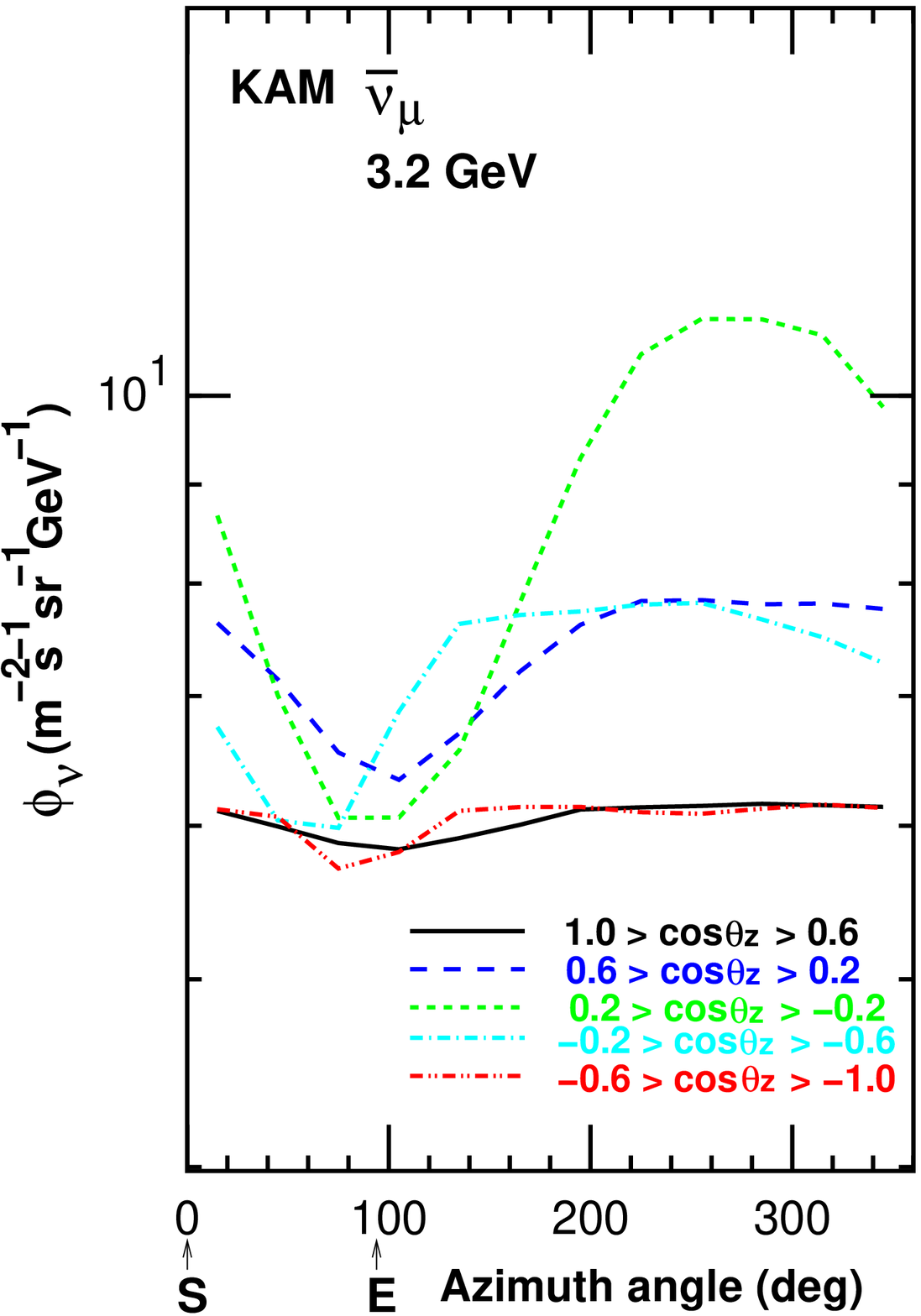}
         \includegraphics[width=4cm]{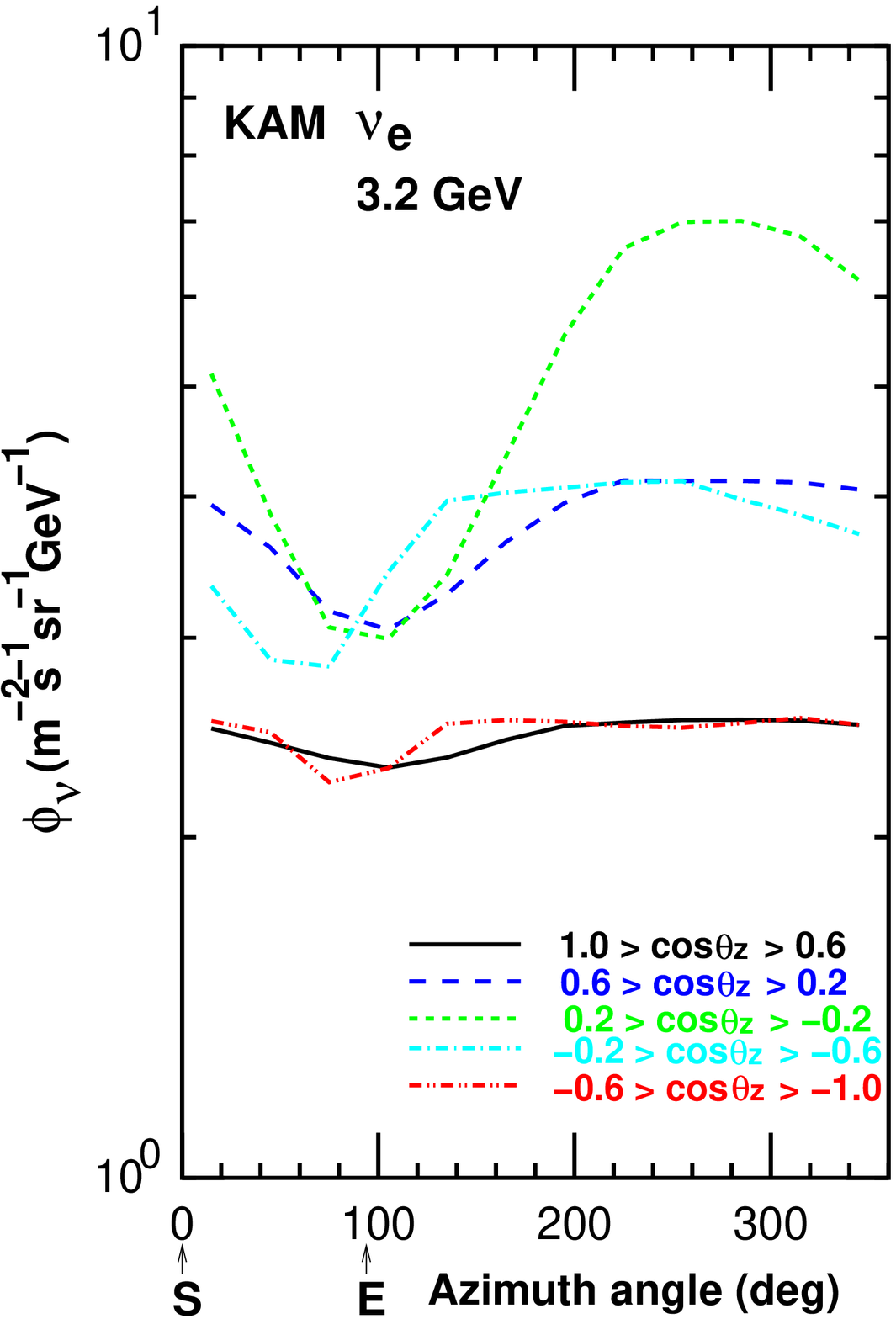} 
         \includegraphics[width=4cm]{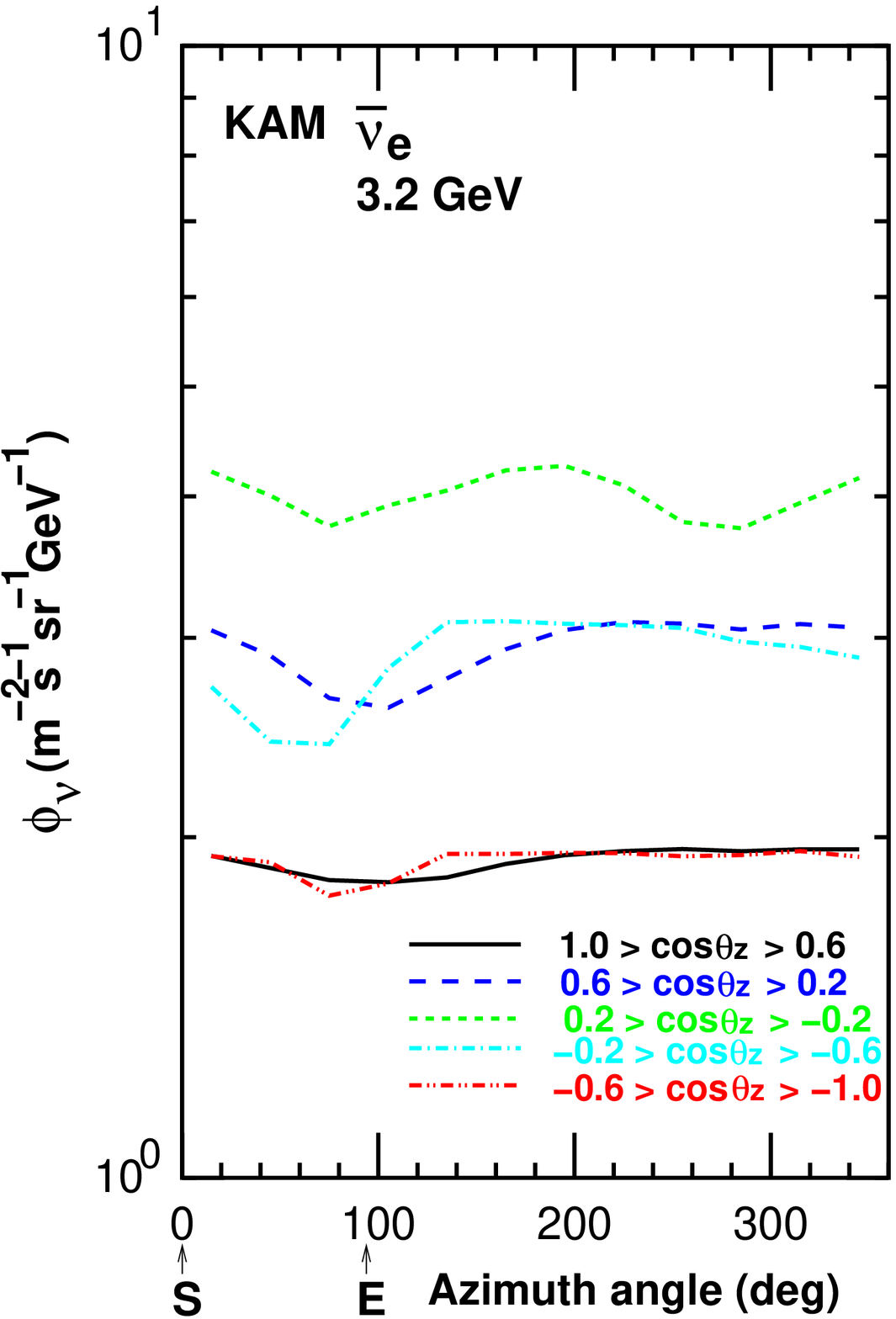}
       }
       \caption{
         Arrival azimuth angle dependence of the time averaged atmospheric neutrino flux 
         in 5 zenith angle bins,
         $1>\cos\theta_z>0.6$,
         $0.4>\cos\theta_z>0.2$,
         $0.2>\cos\theta_z>-0.2$,
         $-0.2>\cos\theta_z>-0.6$, and
         $-0.6>\cos\theta_z>-1$ at the SK site.
         The azimuth angle is measured counter-clockwise from South.
         Solid line shows the averaged fluxe in  $1>\cos\theta_z>0.6$,
         dashed line that in  $0.4>\cos\theta_z>0.2$,
         dotted line that in  $0.2>\cos\theta_z>-0.2$,
         dash dot that in $-0.2>\cos\theta_z>-0.6$, and
         dash 2dots aht in $-0.6>\cos\theta_z>-1$.
       }
       \label{fig:azim-kam}
 \end{figure*}

\begin{figure*}[htb]
   \centering
       {
         \includegraphics[width=4cm]{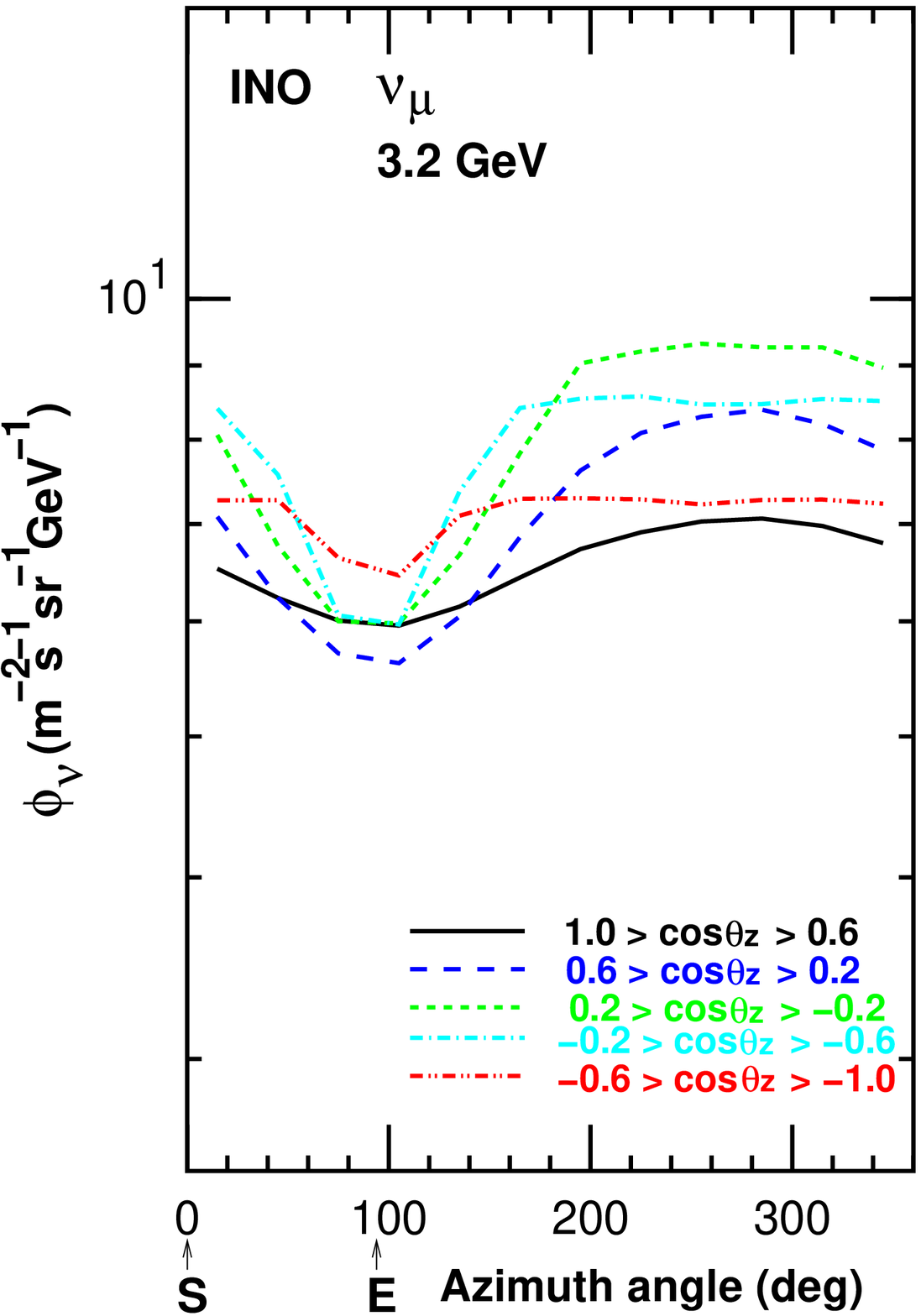}
         \includegraphics[width=4cm]{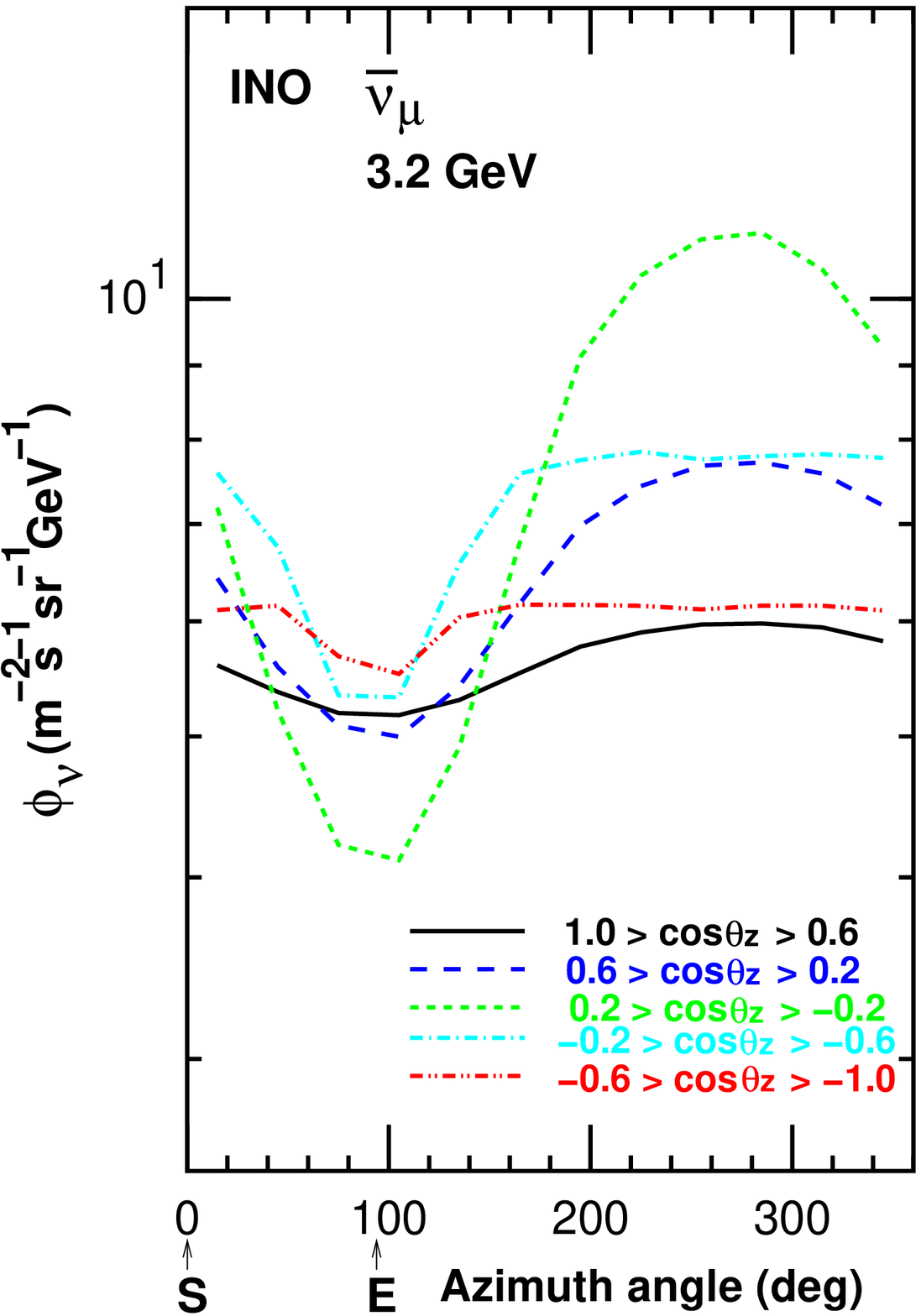}
         \includegraphics[width=4cm]{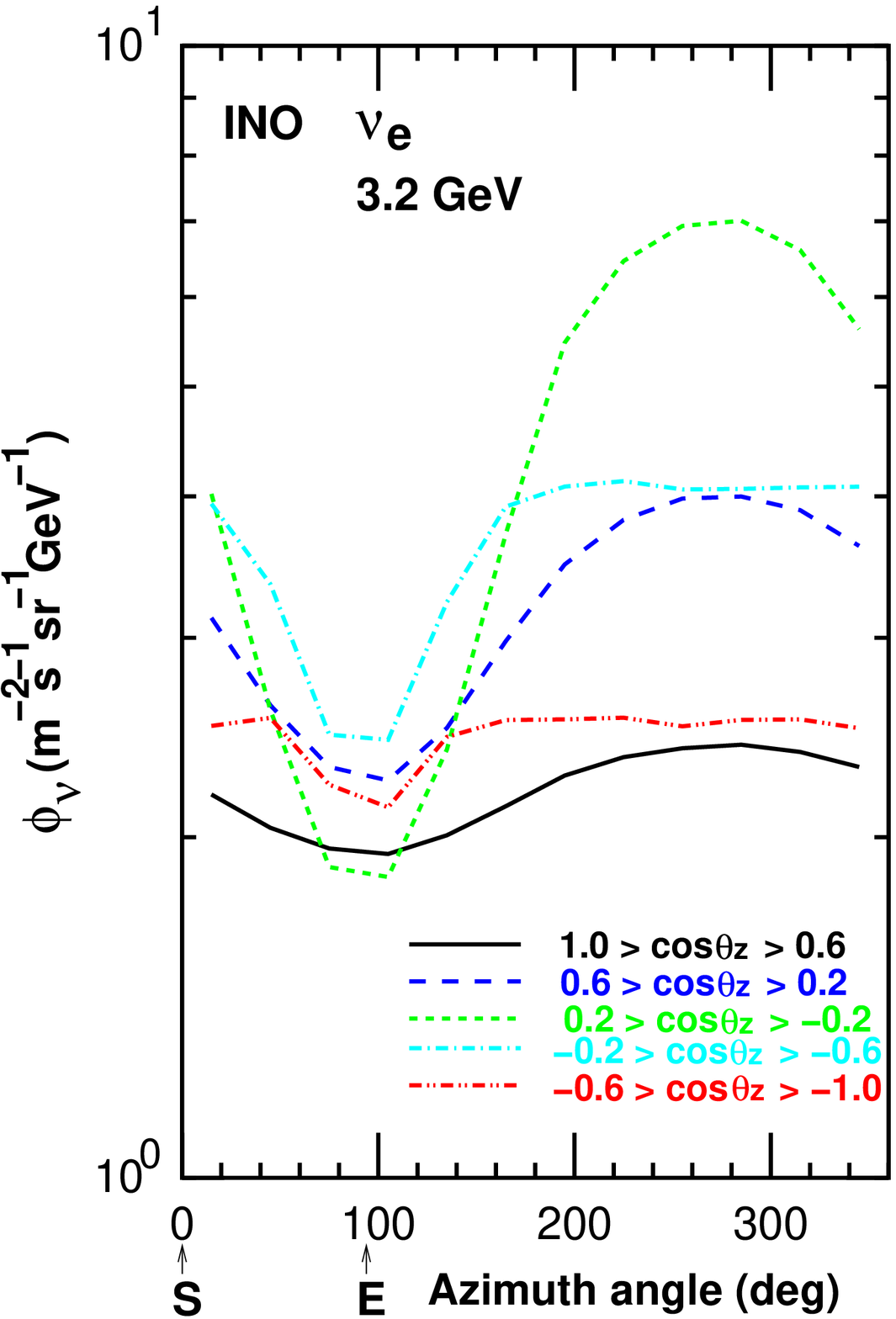} 
         \includegraphics[width=4cm]{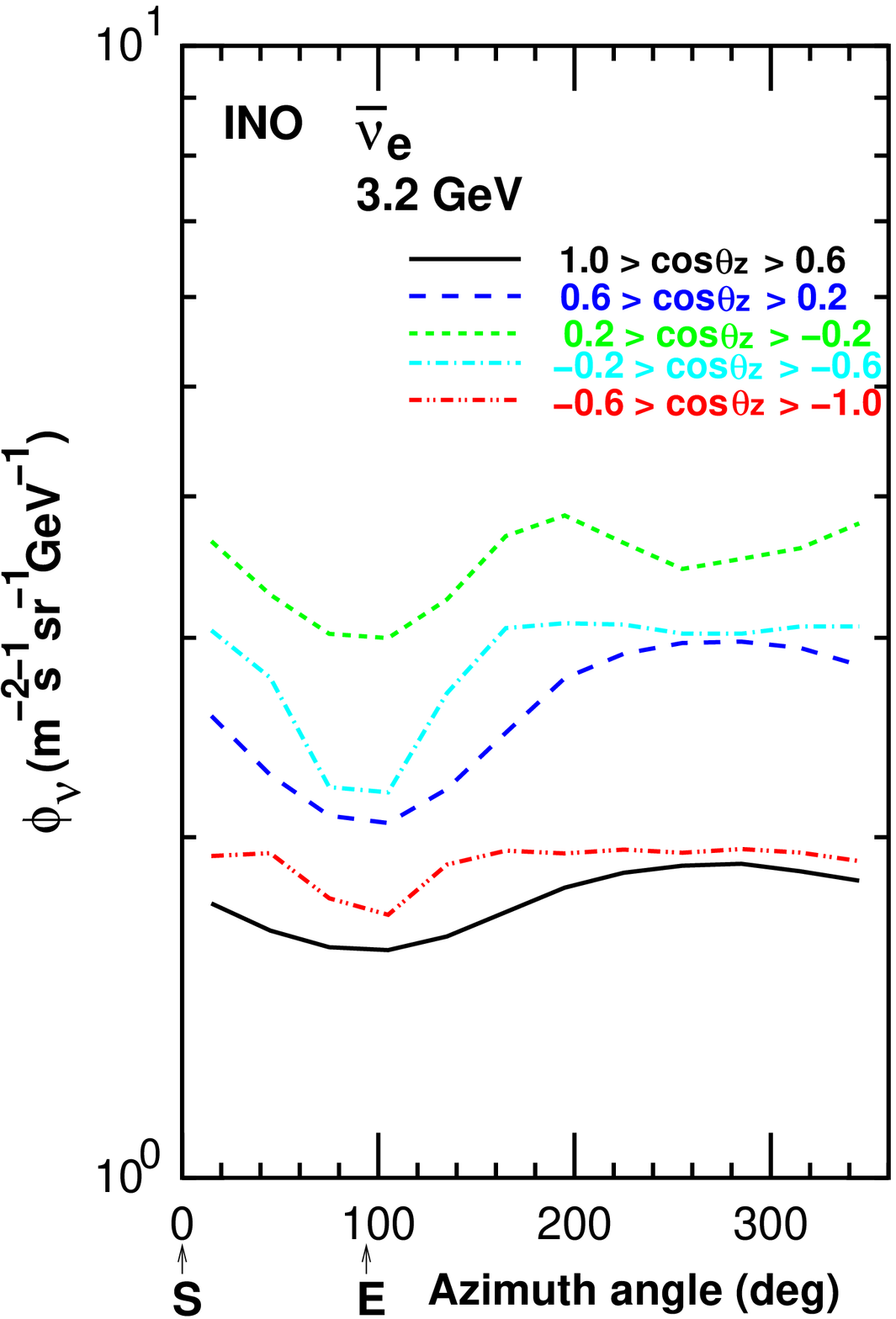}
       }
       \caption{
         Azimuth angle dependence of the time averaged atmospheric neutrino flux 
         in 5 zenith angle bins,
         $1>\cos\theta_z>0.6$,
         $0.4>\cos\theta_z>0.2$,
         $0.2>\cos\theta_z>-0.2$,
         $-0.2>\cos\theta_z>-0.6$, and
         $-0.6>\cos\theta_z>-1$ at the INO site.
         The azimuth angle is measured counter-clockwise from the South. 
         Solid line shows the averaged fluxe in  $1>\cos\theta_z>0.6$,
         dashed line that in  $0.4>\cos\theta_z>0.2$,
         dotted line that in  $0.2>\cos\theta_z>-0.2$,
         dash dot that in $-0.2>\cos\theta_z>-0.6$, and
         dash 2dots aht in $-0.6>\cos\theta_z>-1$.
       }
       \label{fig:azim-ino}
 \end{figure*}

\begin{figure*}[htb]
  \centering
      {
        \includegraphics[width=4cm]{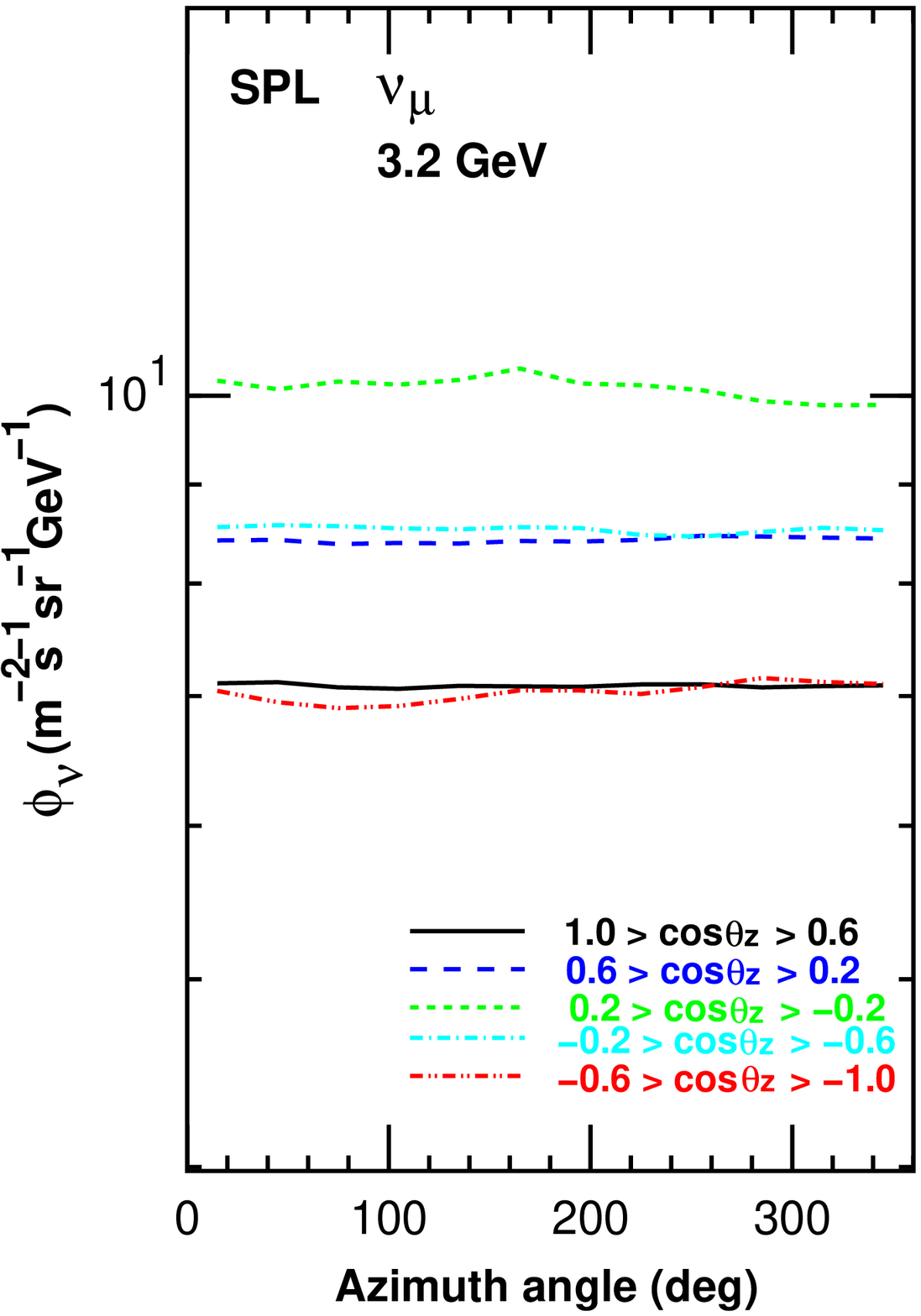}
        \includegraphics[width=4cm]{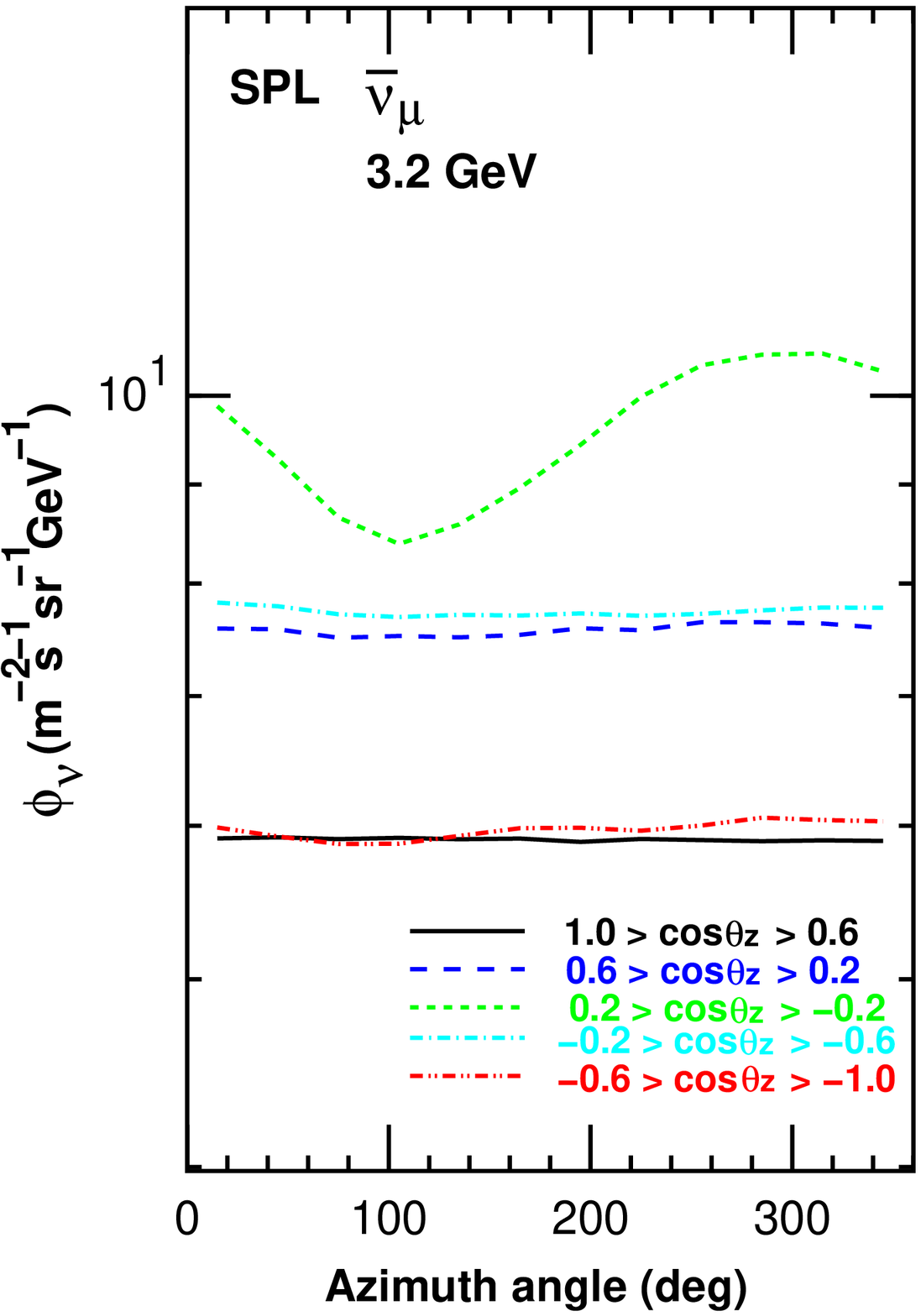}
        \includegraphics[width=4cm]{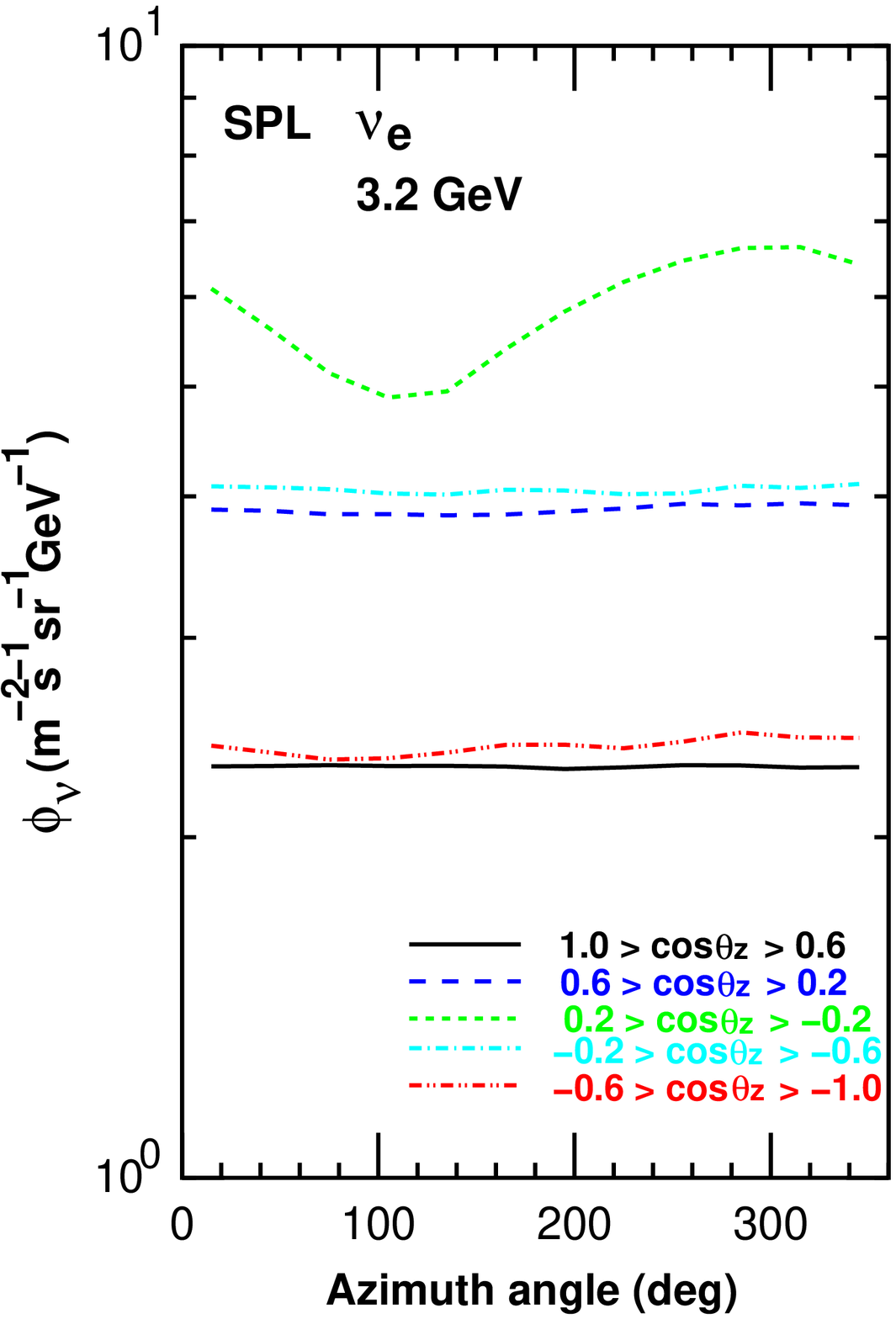} 
        \includegraphics[width=4cm]{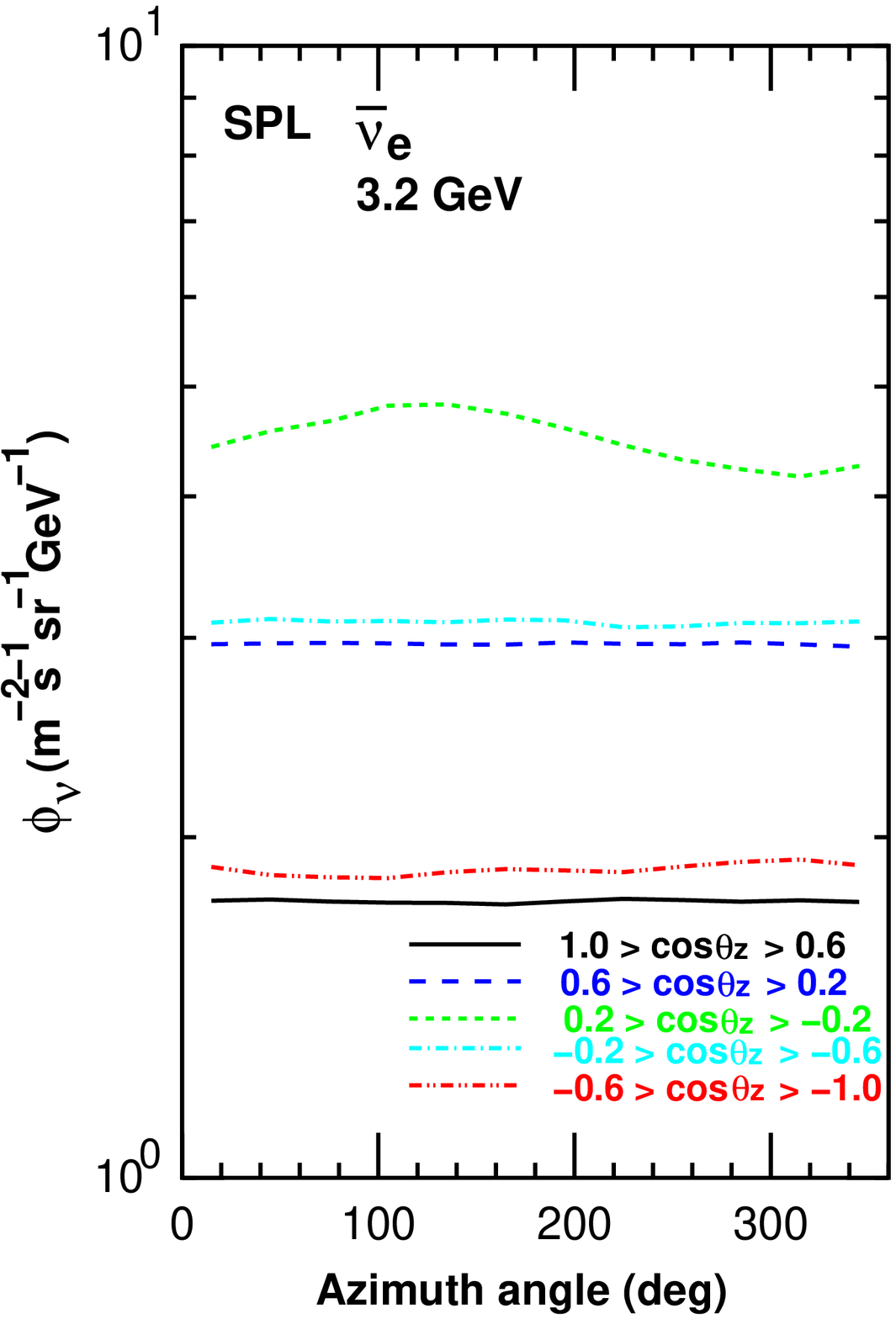}
      }
      \caption{
        Azimuth angle dependence of the time averaged atmospheric neutrino flux 
        in 5 zenith angle bins,
        $1>\cos\theta_z>0.6$,
        $0.4>\cos\theta_z>0.2$,
        $0.2>\cos\theta_z>-0.2$,
        $-0.2>\cos\theta_z>-0.6$, and
        $-0.6>\cos\theta_z>-1$ at the South Pole.
        The azimuth angle is measured counter-clockwise from the meridian 
        line of 180 degree in longitude.
        Solid line shows the averaged fluxe in  $1>\cos\theta_z>0.6$,
        dashed line that in  $0.4>\cos\theta_z>0.2$,
        dotted line that in  $0.2>\cos\theta_z>-0.2$,
        dash dot that in $-0.2>\cos\theta_z>-0.6$, and
        dash 2dots aht in $-0.6>\cos\theta_z>-1$.
      }
      \label{fig:azim-spl}
\end{figure*}

\begin{figure*}[htb]
   \centering
       {
         \includegraphics[width=4cm]{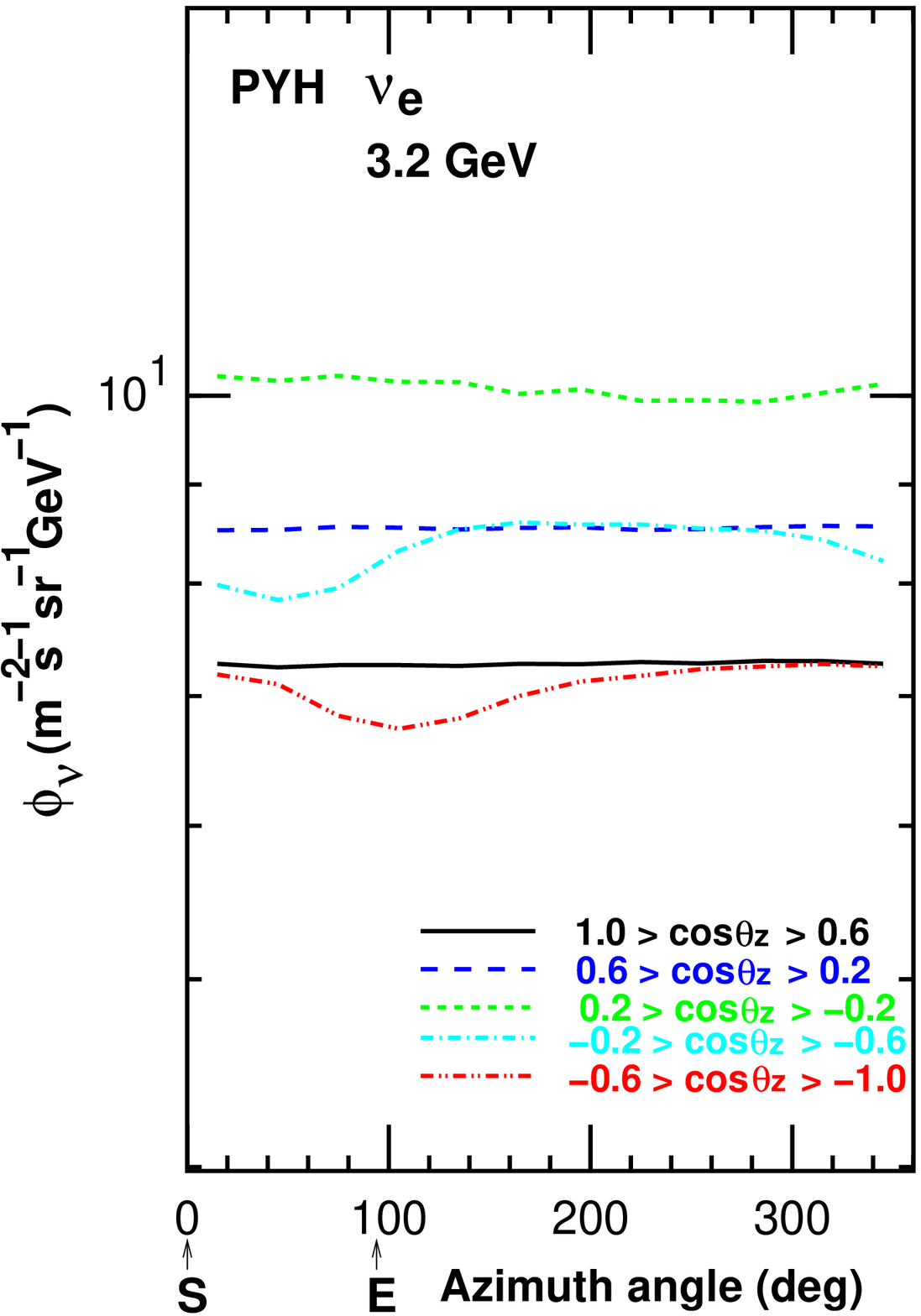}
         \includegraphics[width=4cm]{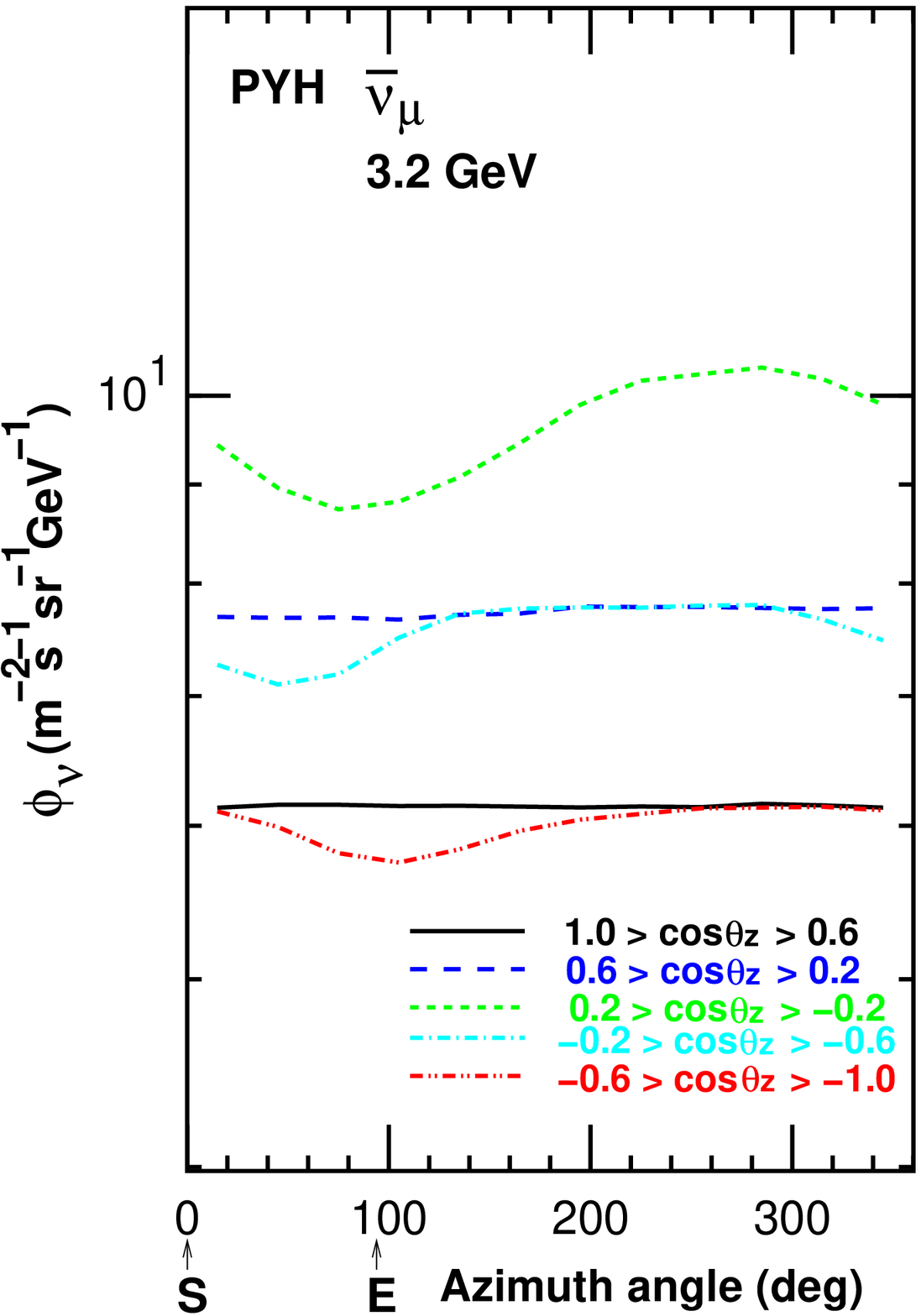}
         \includegraphics[width=4cm]{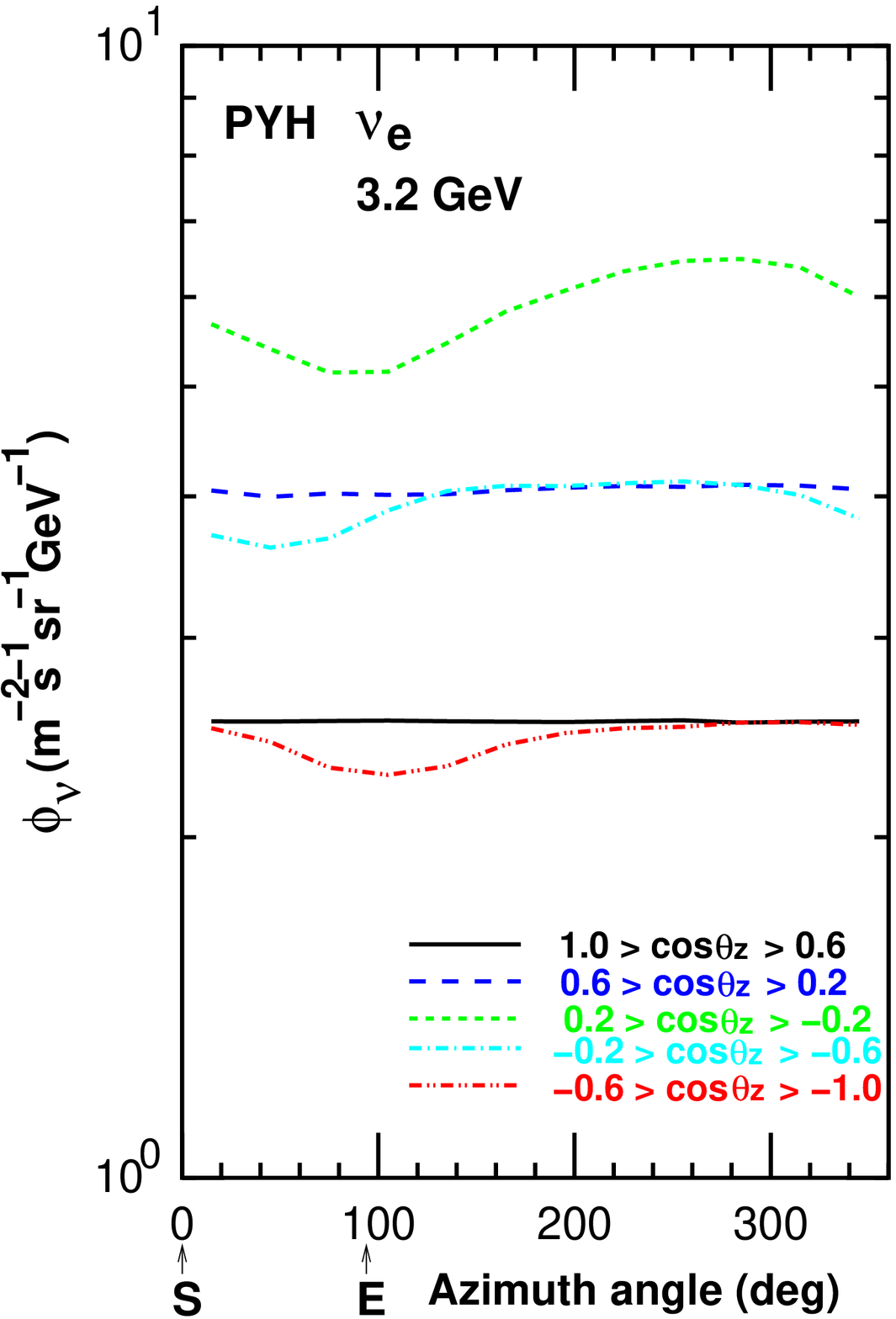} 
         \includegraphics[width=4cm]{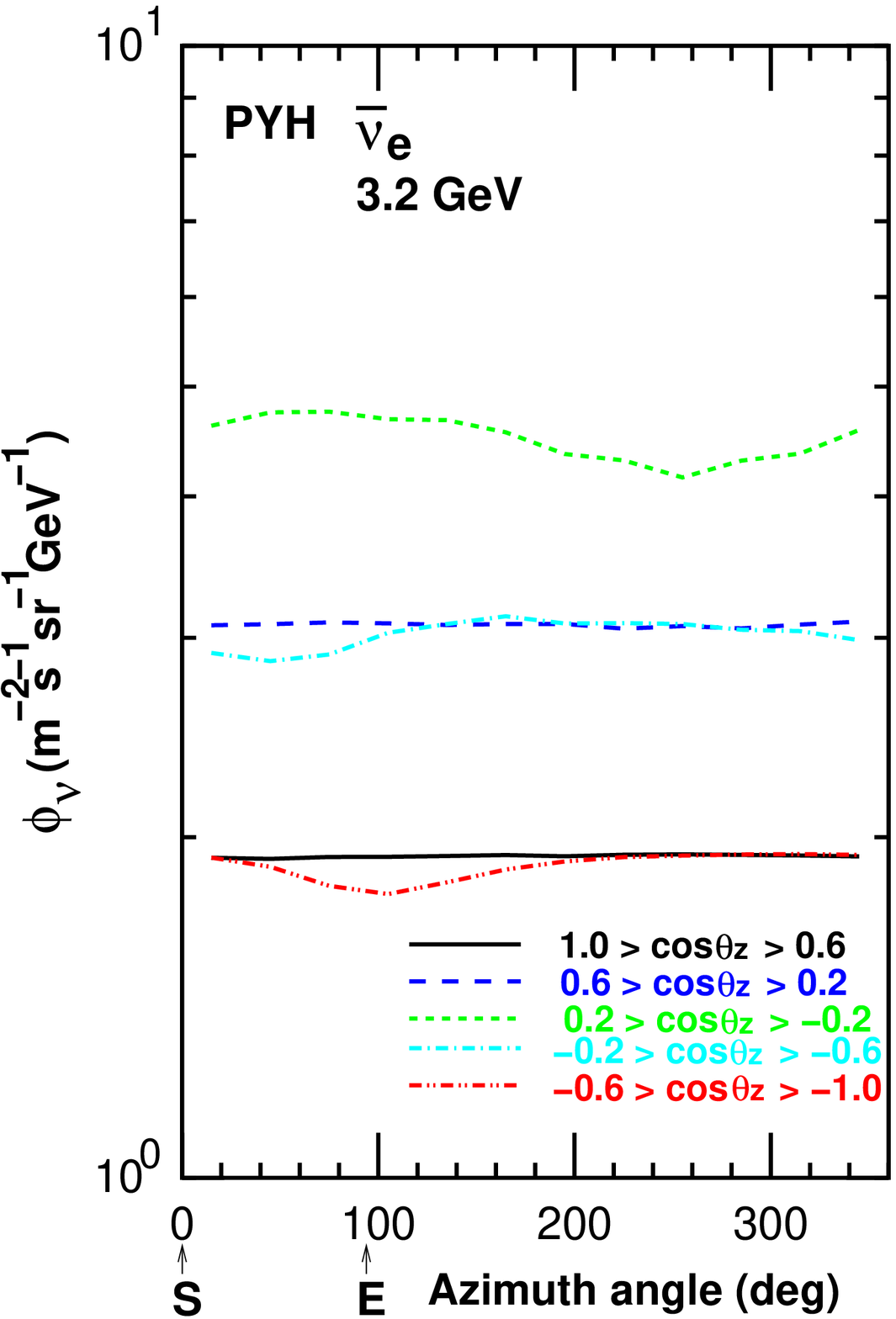}
       }
       \caption{
         Azimuth angle dependence of the time averaged atmospheric neutrino flux 
         in 5 zenith angle bins,
         $1>\cos\theta_z>0.6$,
         $0.4>\cos\theta_z>0.2$,
         $0.2>\cos\theta_z>-0.2$,
         $-0.2>\cos\theta_z>-0.6$, and
         $-0.6>\cos\theta_z>-1$ at the Pyh\"asalmi mine.
         The azimuth angle is measured counter-clockwise from the South. 
         Solid line shows the averaged fluxe in  $1>\cos\theta_z>0.6$,
         dashed line that in  $0.4>\cos\theta_z>0.2$,
         dotted line that in  $0.2>\cos\theta_z>-0.2$,
         dash dot that in $-0.2>\cos\theta_z>-0.6$, and
         dash 2dots aht in $-0.6>\cos\theta_z>-1$.
       }
       \label{fig:azim-pyh}
 \end{figure*}

Next,
we study the azimuthal variation of the neutrino fluxes 
in 5 zenith angle bins;
$1>\cos\theta_z>0.6$,
$0.4>\cos\theta_z>0.2$,
$0.2>\cos\theta_z>-0.2$,
$-0.2>\cos\theta_z>-0.6$, and
$-0.6>\cos\theta_z>-1$ at 3.2~GeV,
averaging over a year.
In Fig.~\ref{fig:azim-kam}, 
we show the azimuthal variation 
of atmospheric neutrino flux at the SK site, 
in Fig.~\ref{fig:azim-ino}, those at the INO site,
in Fig.~\ref{fig:azim-spl}, those at the South Pole  
and in Fig.~\ref{fig:azim-pyh}, those at the Pyh\"asalmi mine. 

In these figures we observe two kinds of effects from the geomagnetic field.
One is the rigidity cutoff, and the other is the muon bending. 
The geomagnetic field directed to the North filters the low energy 
cosmic rays from the East directions, since the cosmic rays 
generally carry a positive charge.
The rigidity cutoff reduces 
all flavors of neutrino from the East at the same rate.

On the other hand,
the effect of muon bending depends on the muon charge.
The positive muon and its decay products are affected by the geomagnetic
field in the same way as the rigidity cutoff,
and the negative muon and its decay products are affected in the
opposite way.
Then, muon bending reduces the $\bar\nu_\mu$ and $\nu_e$ fluxes from the
East, but reduces the 
$\nu_\mu$ and $\bar\nu_e$ fluxes from the West.

In Fig.~\ref{fig:azim-kam}, 
we find small azimuthal variations of atmospheric neutrino 
fluxes at near vertically down going directions
($1>\cos\theta_z>0.6$) at the SK site.
As the variation shape is almost the same among all neutrino flavors, 
this variation is considered to be due to the rigidity cutoff, 
but the effect is already small at the SK site for near vertical 
directions at 3.2~GeV.

At near horizontal directions ($0.2>\cos\theta_z>-0.2$),
$\bar\nu_\mu$ and $\nu_e$ fluxes have large sinusoidal azimuthal variations,
but $\nu_\mu$ and $\bar\nu_e$ fluxes have small but more complicated
azimuthal variations. 
This is because the rigidity cutoff and the muon bending work
in the same directions for $\bar\nu_\mu$ and $\nu_e$ fluxes,
and in the opposite directions for $\nu_\mu$ and $\bar\nu_e$ fluxes. 
The dip at $\sim  90^\circ$ commonly seen in all neutrino flavors is 
considered to be due to the rigidity cutoff, 
and the dip at $\sim 270^\circ$ seen in the $\nu_\mu$ and $\bar\nu_e$ fluxes
is due to muon bending.
 
In Fig.~\ref{fig:azim-ino}, 
we find larger azimuthal variations of atmospheric neutrino 
fluxes at near vertically down going directions
($1>\cos\theta_z>0.6$) at the INO site compared to the SK site.
As the variation shape is similar among all neutrino flavors,
this is considered as an effect of the rigidity cutoff,
and is still strong at the INO site for vertical down going 
directions at 3.2~GeV.
The azimuthal variation at near horizontal directions 
($0.2>\cos\theta_z>-0.2$) are also larger than the SK site 
for each neutrino flavor, due to the larger horizontal component
of the geomagnetic field.

We note that the muon bending effect is seen in the azimuthal averaged 
flux plot (Fig.~\ref{fig:zdep}) at the INO site at near horizontal
directions.
The muon bending suppresses the $\nu_\mu$ and $\bar\nu_e$ 
fluxes from the West, but enhances them from the East.
However, the rigidity cutoff works strongly for the East directions,
canceling the enhancement from the East,
and the muon bending is seen as a suppression of the 
$\nu_\mu$ and $\bar\nu_e$ fluxes in Fig.~\ref{fig:zdep}.
On the other hand for $\bar\nu_\mu$ and $\nu_e$, 
the muon bending enhances the 
fluxes from the West, but suppresses them from the East.
As the rigidity cutoff works weakly for the West directions,
the muon bending is seen as an enhancement of the 
$\bar\nu_\mu$ and $\nu_e$ 
fluxes in Fig.~\ref{fig:zdep}.
The same mechanism works at other sites, but the amplitudes of the
suppression and enhancement are small, and it is not seen 
clearly even at the SK site (KAM).

In Fig.~\ref{fig:azim-spl}, 
we find that there are almost no azimuthal variations of the atmospheric 
neutrino flux except for the near horizontal directions
($0.2>\cos\theta_z>-0.2$) at the South Pole,
since the geomagnetic field is almost vertical at the South Pole.
At near horizontal directions we find the azimuthal variations
of the $\bar\nu_\mu$ and $\nu_e$, having the minimum at $\sim$ 90$^\circ$
and the maximum at $\sim$ 270$^\circ$.
Also there are slightly smaller variations of $\nu_\mu$ and 
$\bar\nu_e$, having the minimum at  $\sim$ 90$^\circ$
and the maximum at $\sim$ 270$^\circ$.
These features represent the fact that the rigidity 
cutoff works very weakly at the South Pole, and 
the muon bending works mainly with the residual 
horizontal component of the geomagnetic field.
We note that the azimuthal variation for 
upward going directions ($\cos\theta_z<-0.2$) is also very small
at the South pole.
This is probably because the South Pole is close to the geomagnetic
South pole in the dipole approximation of the geomagnetic field.

In Fig.~\ref{fig:azim-pyh}, 
we find the features of azimuthal variations of down going and 
near horizontal atmospheric neutrino flux ($\cos\theta_z > -0.2$) 
at the Pyh\"asalmi mine are almost the same as those at the South Pole.
We note that the horizontal component of the geomagnetic field at 
the Pyh\"asalmi 
mine ($B_h \sim 13000$ nT) is even smaller than that of the South Pole
($B_h \sim 16000$ nT).
However, as the Pyh\"asalmi mine sits at a distant position from the
geomagnetic North Pole, the rigidity
cutoff and muon bending affects the azimuthal variations of 
the upward going neutrinos.

\section{\label{sec:height} production height of atmospheric neutrino}

\begin{figure*}[htb]
  \centering
      {
        \includegraphics[width=4cm]{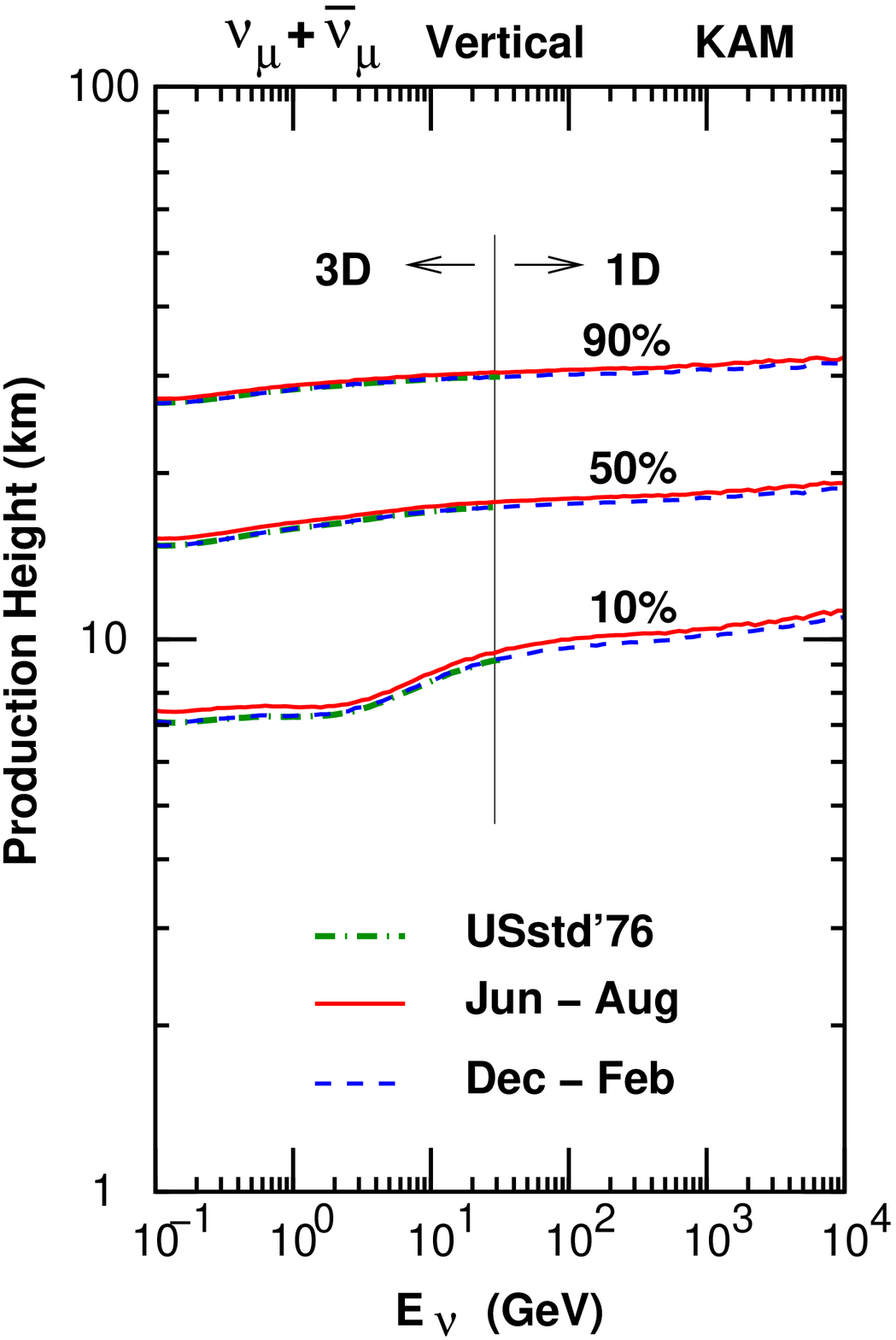}
        \includegraphics[width=4cm]{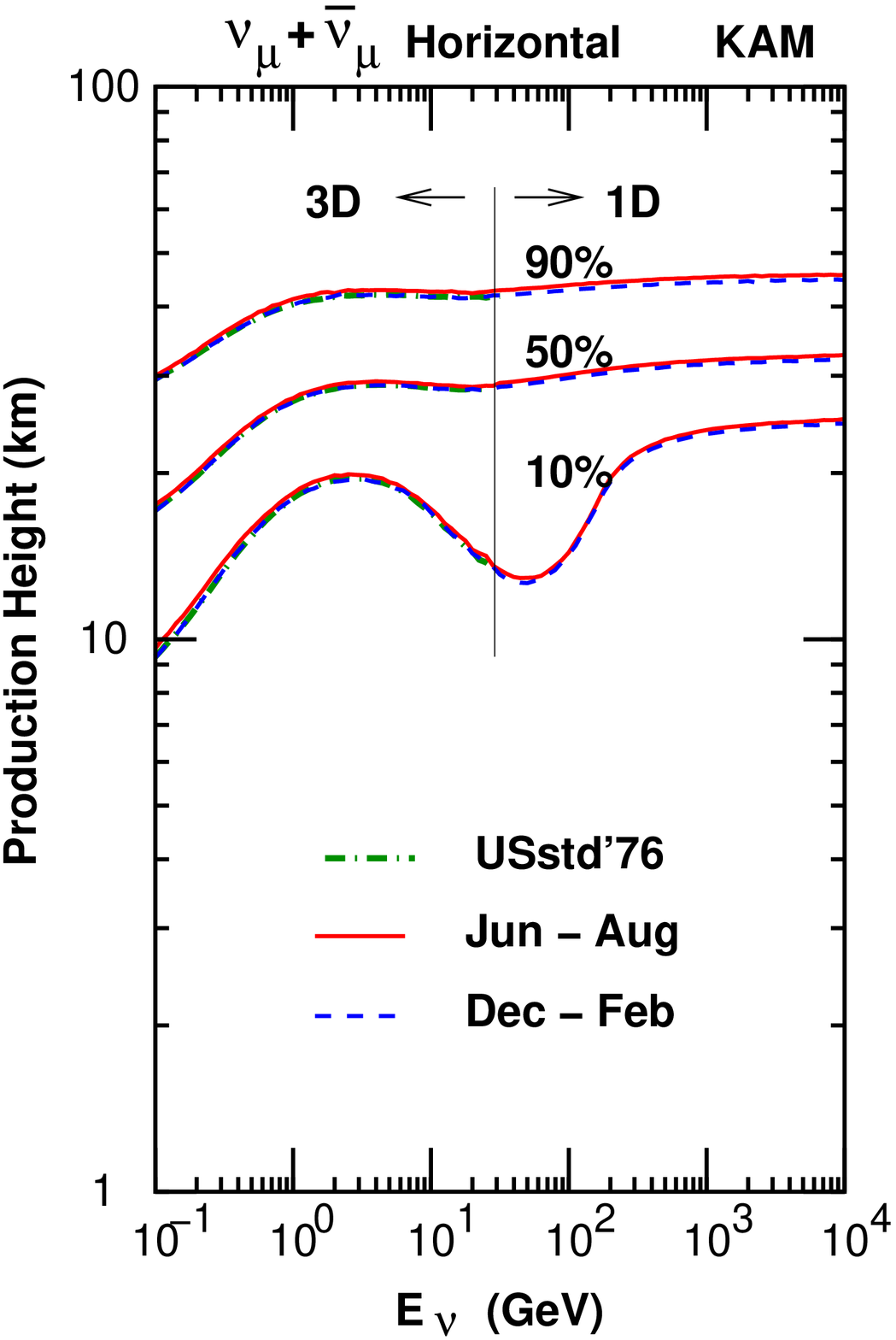}
        \includegraphics[width=4cm]{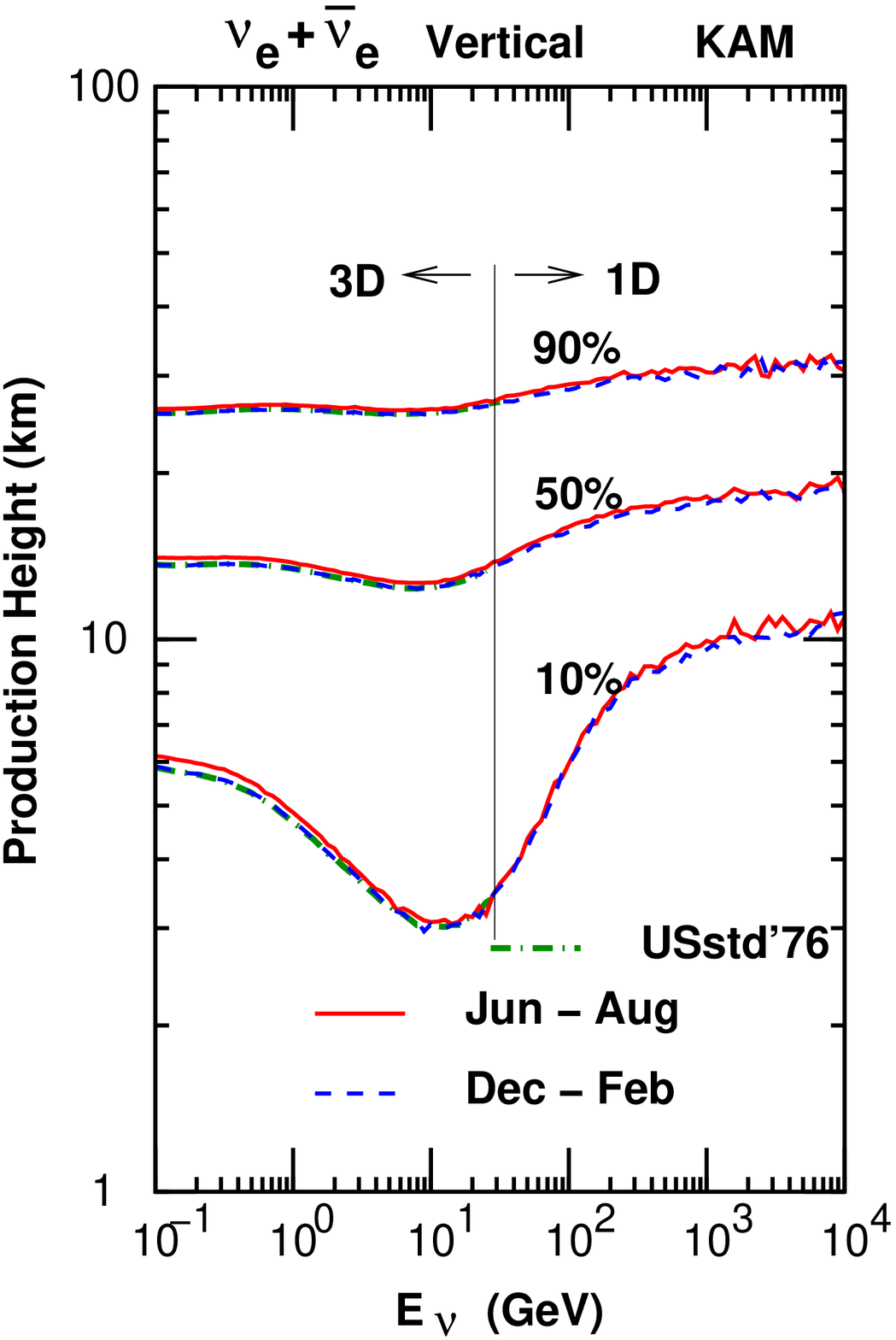}
        \includegraphics[width=4cm]{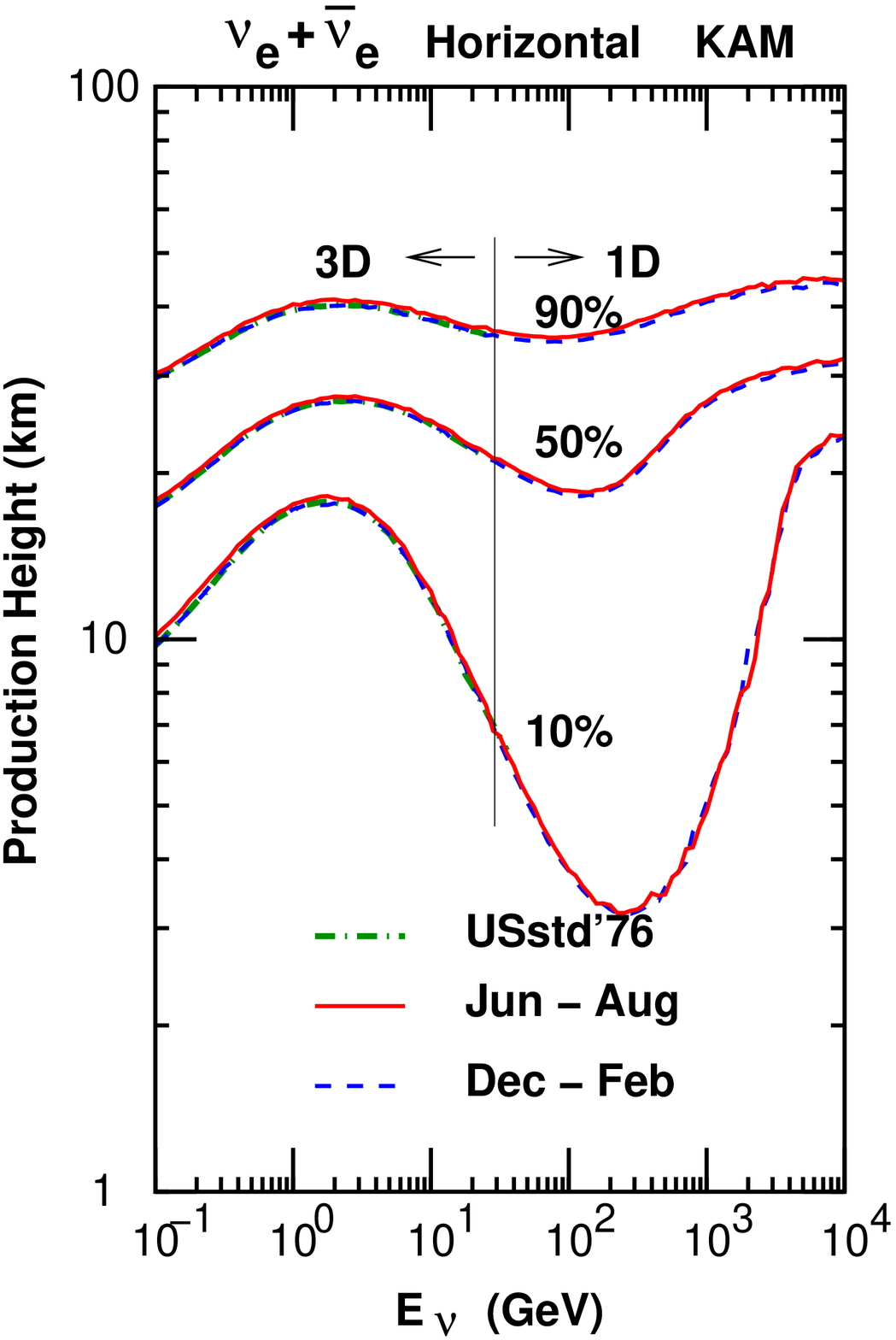}
      }
      \caption{
        Production height of atmospheric neutrinos for 
        $\nu_\mu$ + $\bar\nu_\mu$ and $\nu_e$ + $\bar\nu_e$ 
        going to vertically down and horizontal directions,
        summing all azimuth directions at the SK site.
        The height that cumulative distribution reaches 10~\%,
        50~\%, and 90~\% are shown as the function of neutrino energy.
        The time average values in June -- August are shown as solid lines
        and in December -- February as Dashed line.
        The same values with the US-standard '76 atmospheric model
        are also shown in dash dot in the 3D region. 
      }
      \label{fig:height-kam}
\end{figure*}

\begin{figure*}[htb]
  \centering
      {
        \includegraphics[width=4cm]{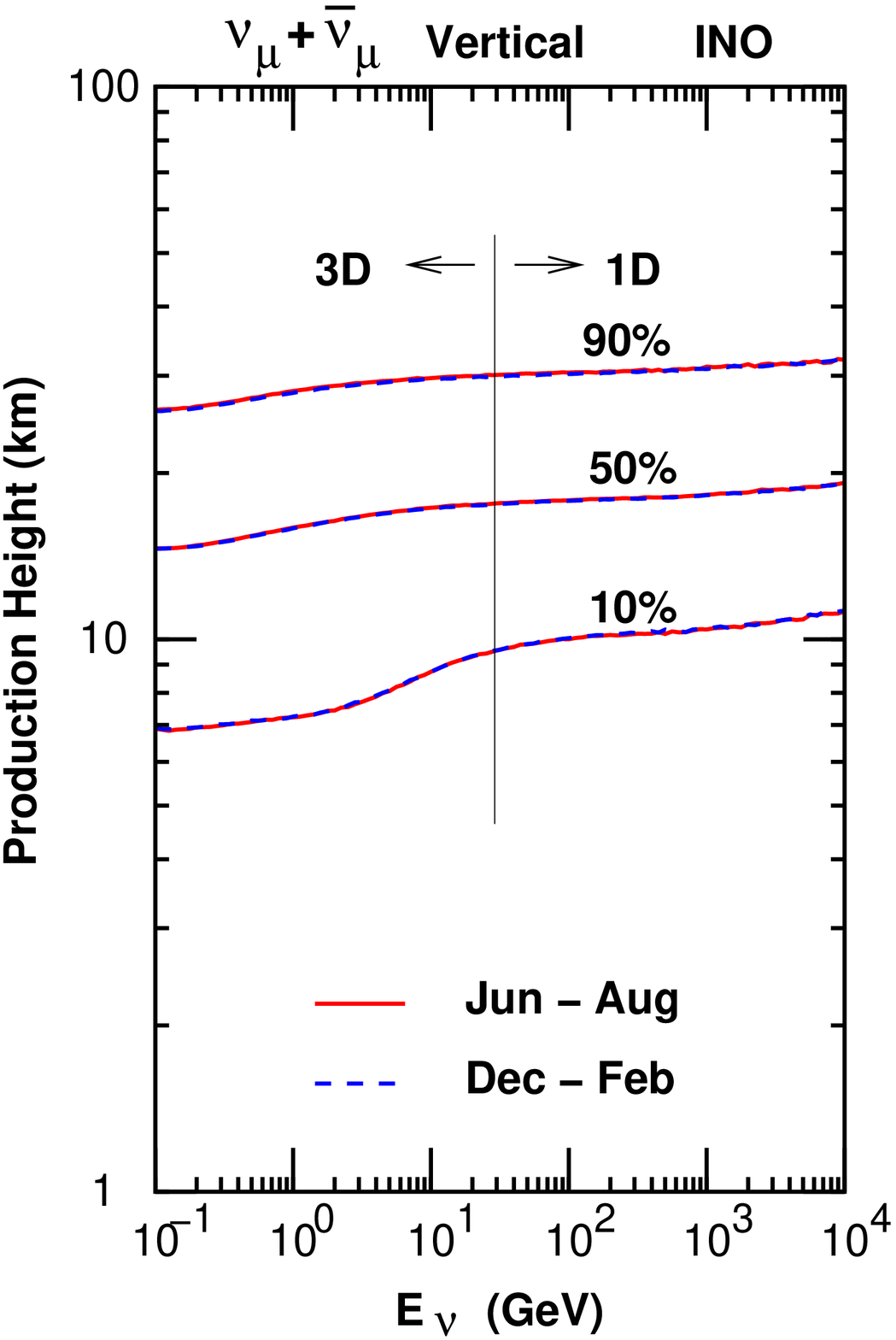}
        \includegraphics[width=4cm]{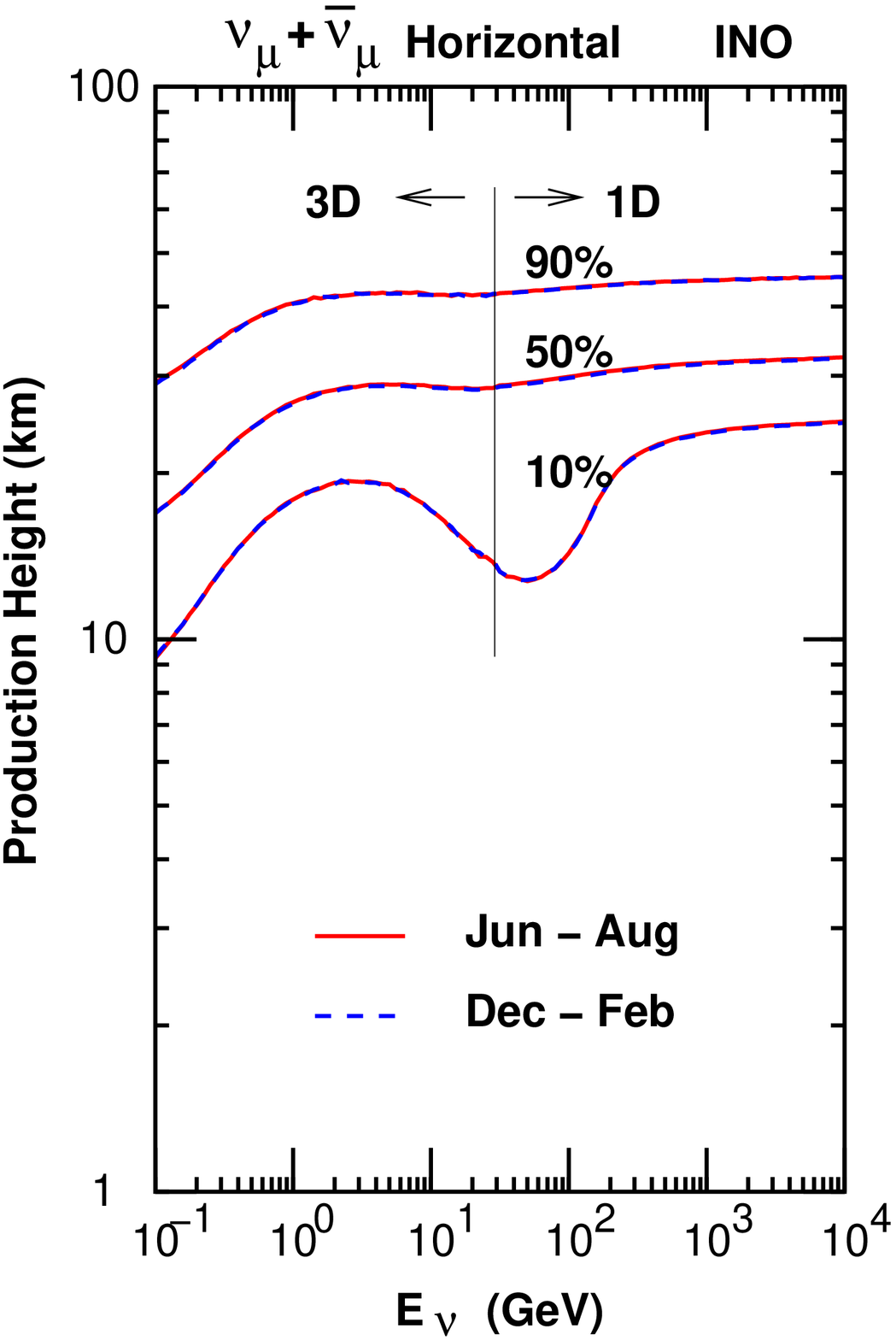}
        \includegraphics[width=4cm]{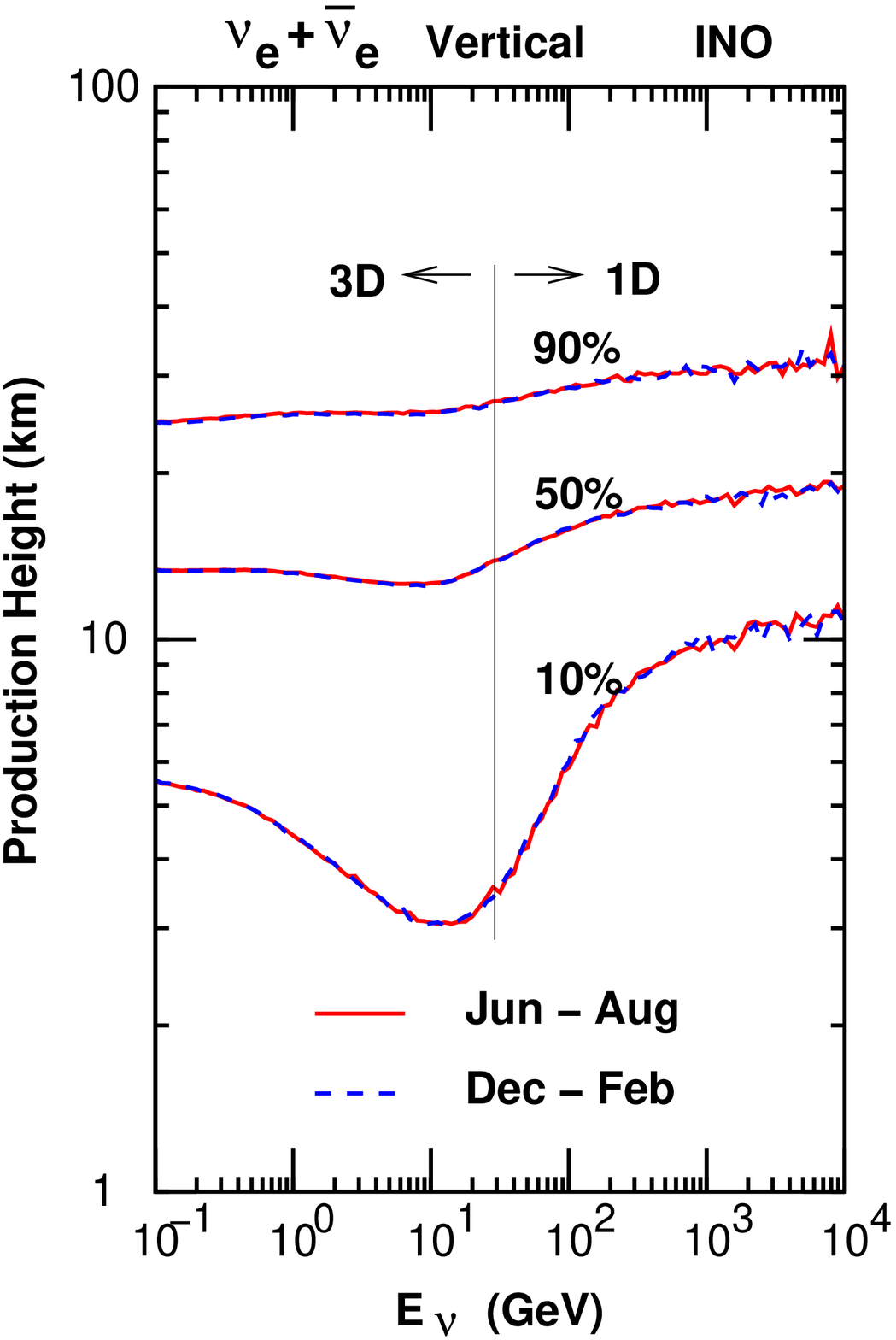}
        \includegraphics[width=4cm]{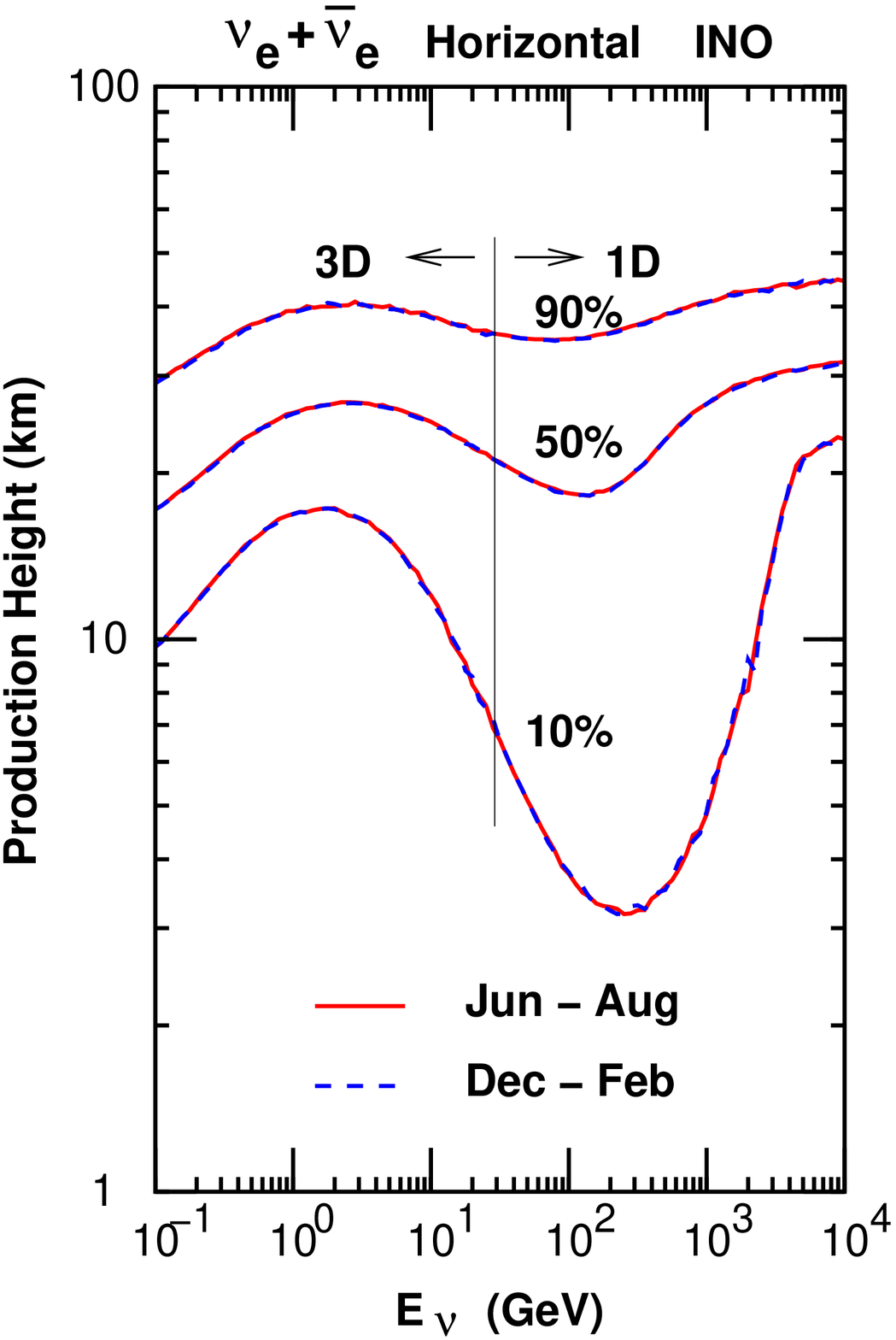}
      }
      \caption{
        Production height of atmospheric neutrinos for 
        $\nu_\mu$ + $\bar\nu_\mu$ and $\nu_e$ + $\bar\nu_e$ 
        going to vertical down and horizontal directions,
        summing all azimuth directions at the INO site.
        The height that cumulative distribution reaches 10~\%,
        50~\%, and 90~\% are shown as a function of neutrino energy.
        The time average values in June -- August are shown as solid lines
        and in December -- February as Dashed line.
      }
      \label{fig:height-ino}
\end{figure*}

\begin{figure*}[htb]
  \centering
      {
        \includegraphics[width=4cm]{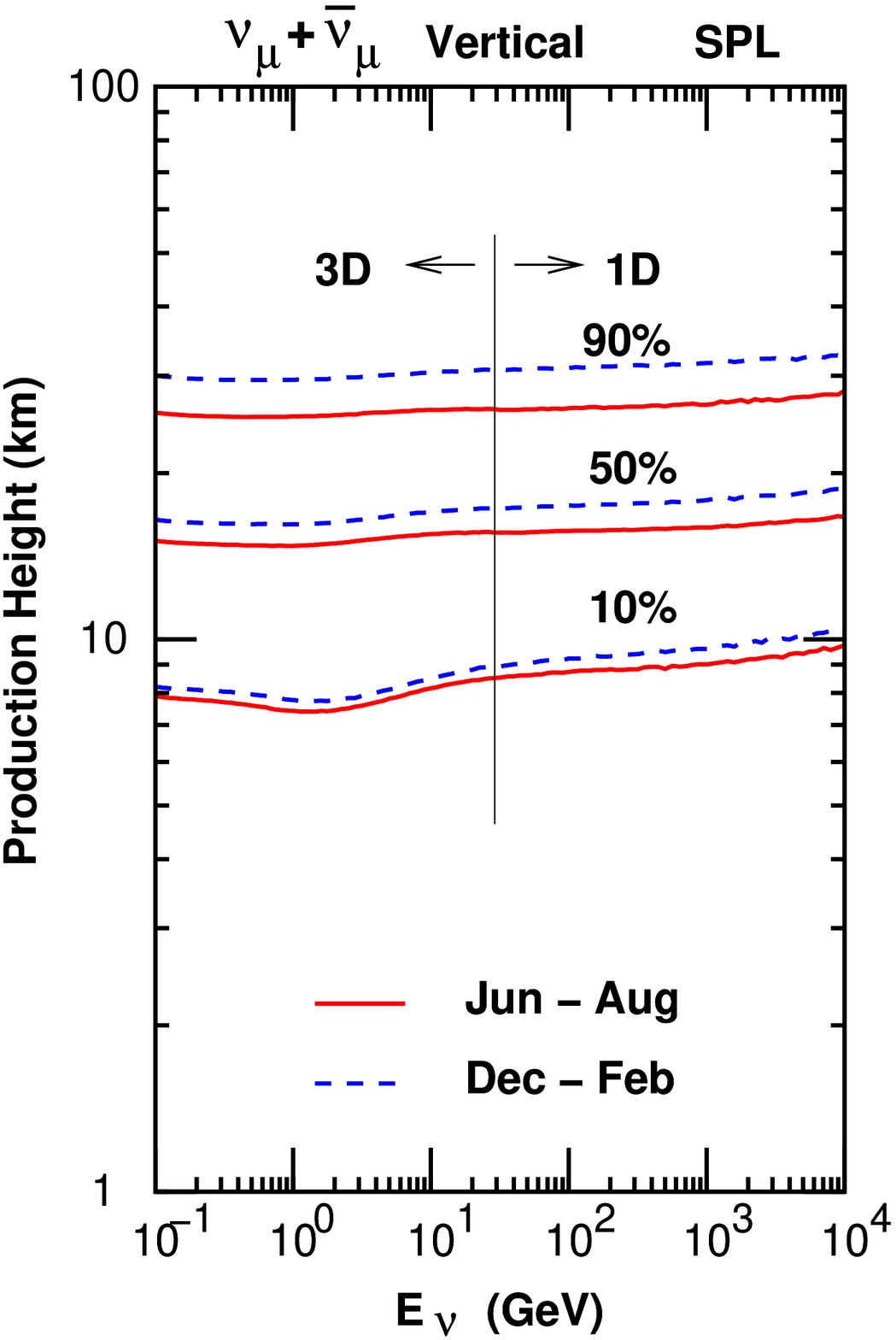}
        \includegraphics[width=4cm]{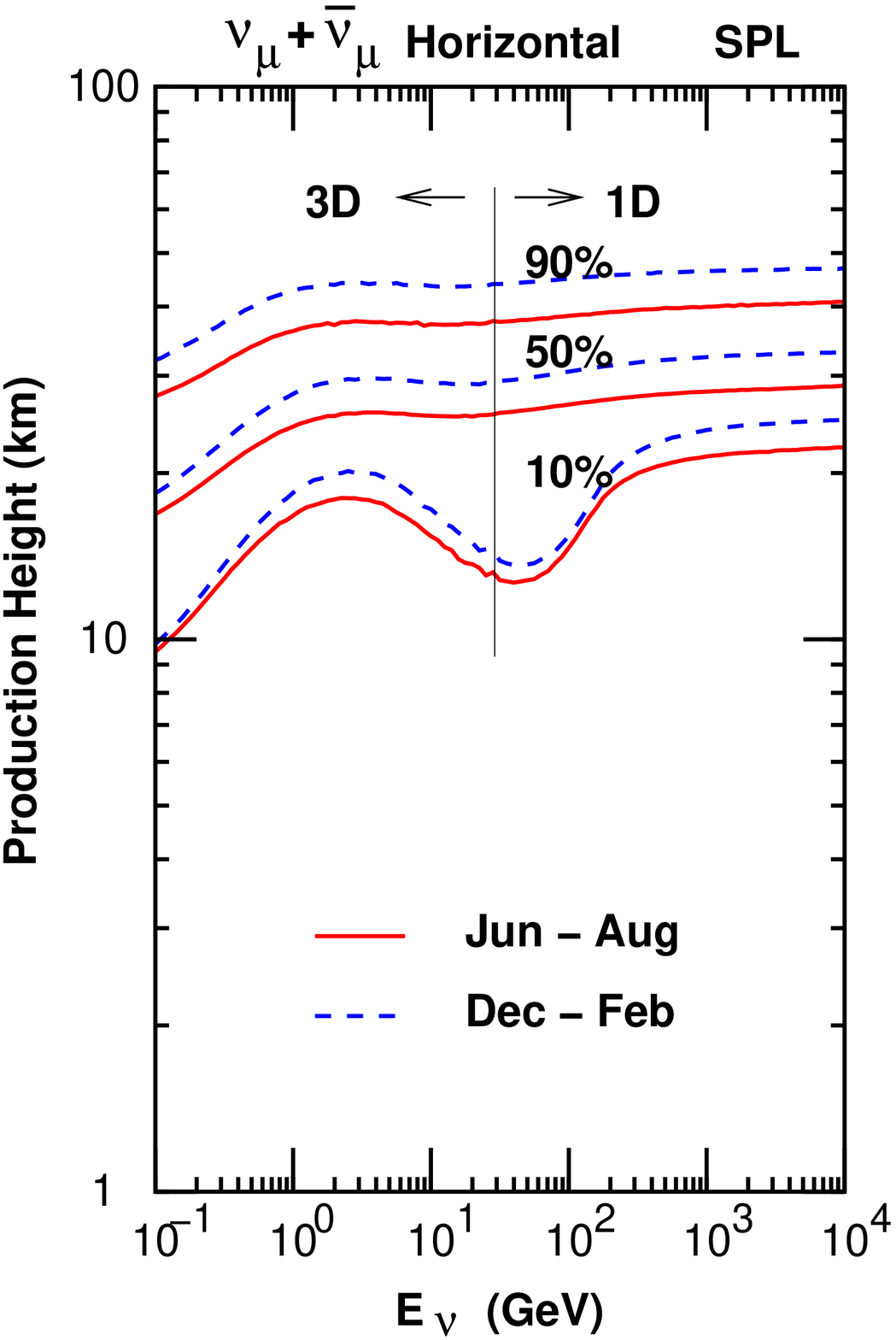}
        \includegraphics[width=4cm]{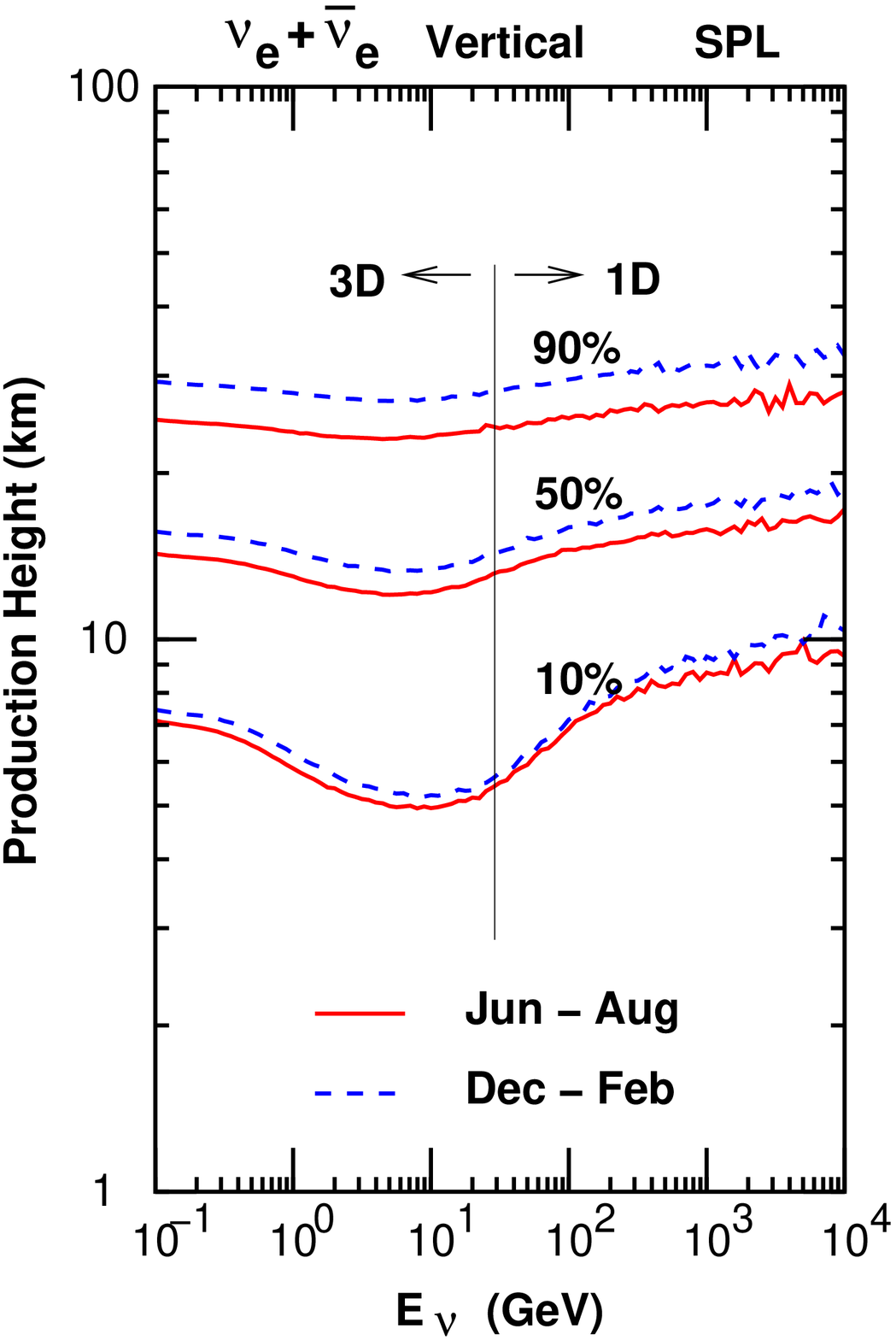}
        \includegraphics[width=4cm]{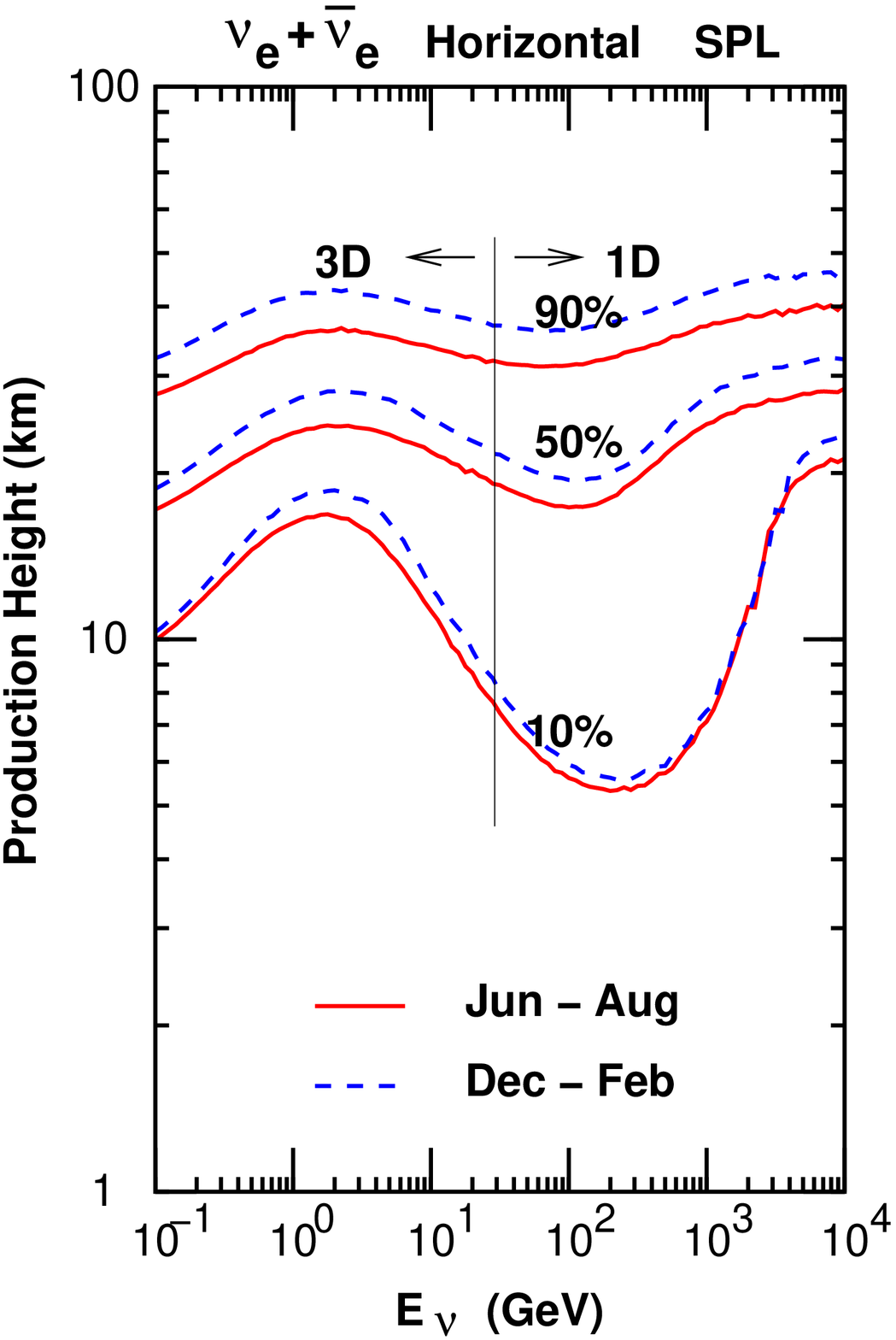}
      }
      \caption{
        Production height of atmospheric neutrinos for 
        $\nu_\mu$ + $\bar\nu_\mu$ and $\nu_e$ + $\bar\nu_e$ 
        going to vertically down and horizontal directions,
        summing all azimuth directions at the South Pole.
        The height that cumulative distribution reaches 10~\%,
        50~\%, and 90~\% are shown as a function of neutrino energy.
        The time average values in June -- August are shown as solid lines
        and in December -- February as Dashed line.
      }
      \label{fig:height-pyh}
\end{figure*}

\begin{figure*}[htb]
  \centering
      {
        \includegraphics[width=4cm]{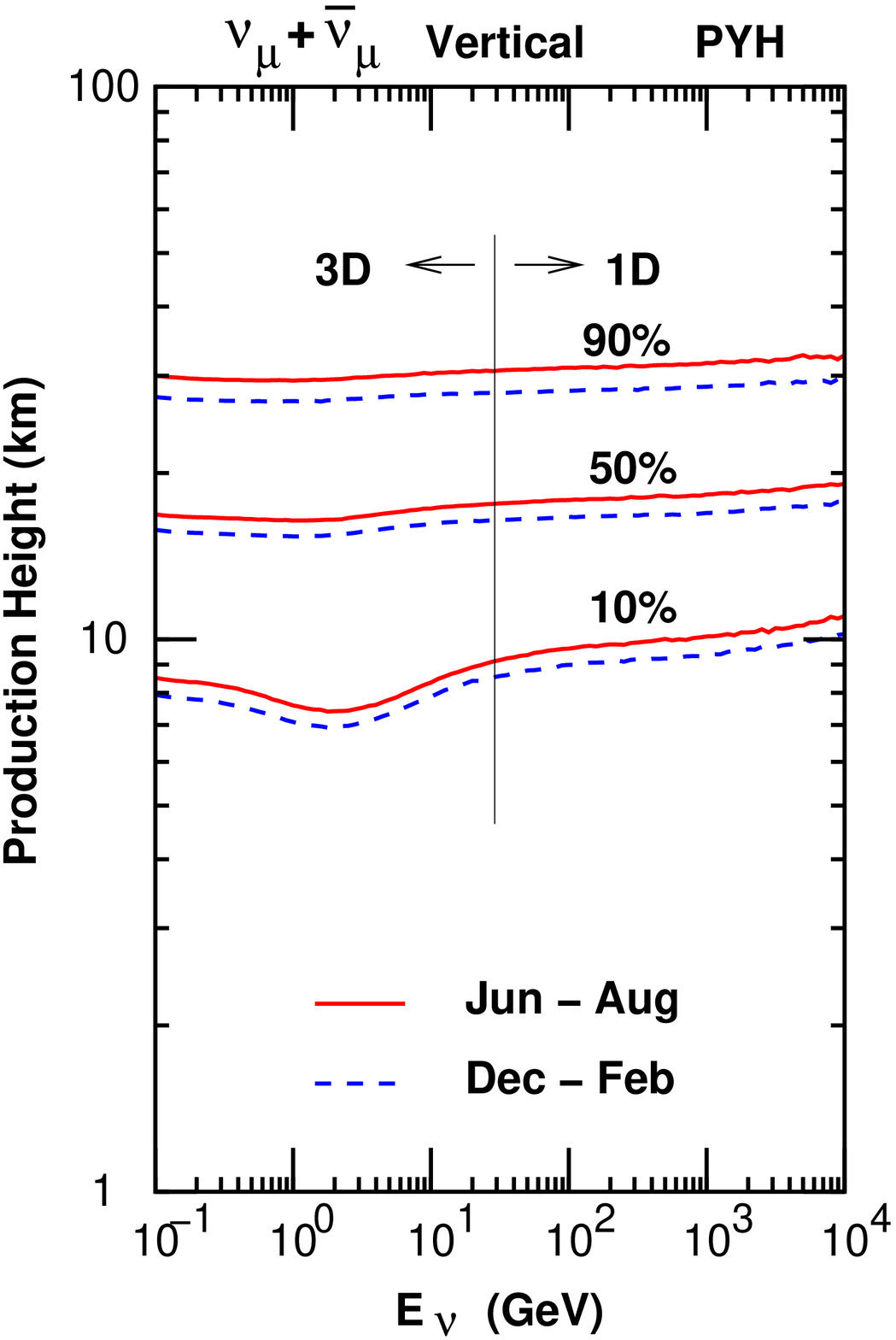}
        \includegraphics[width=4cm]{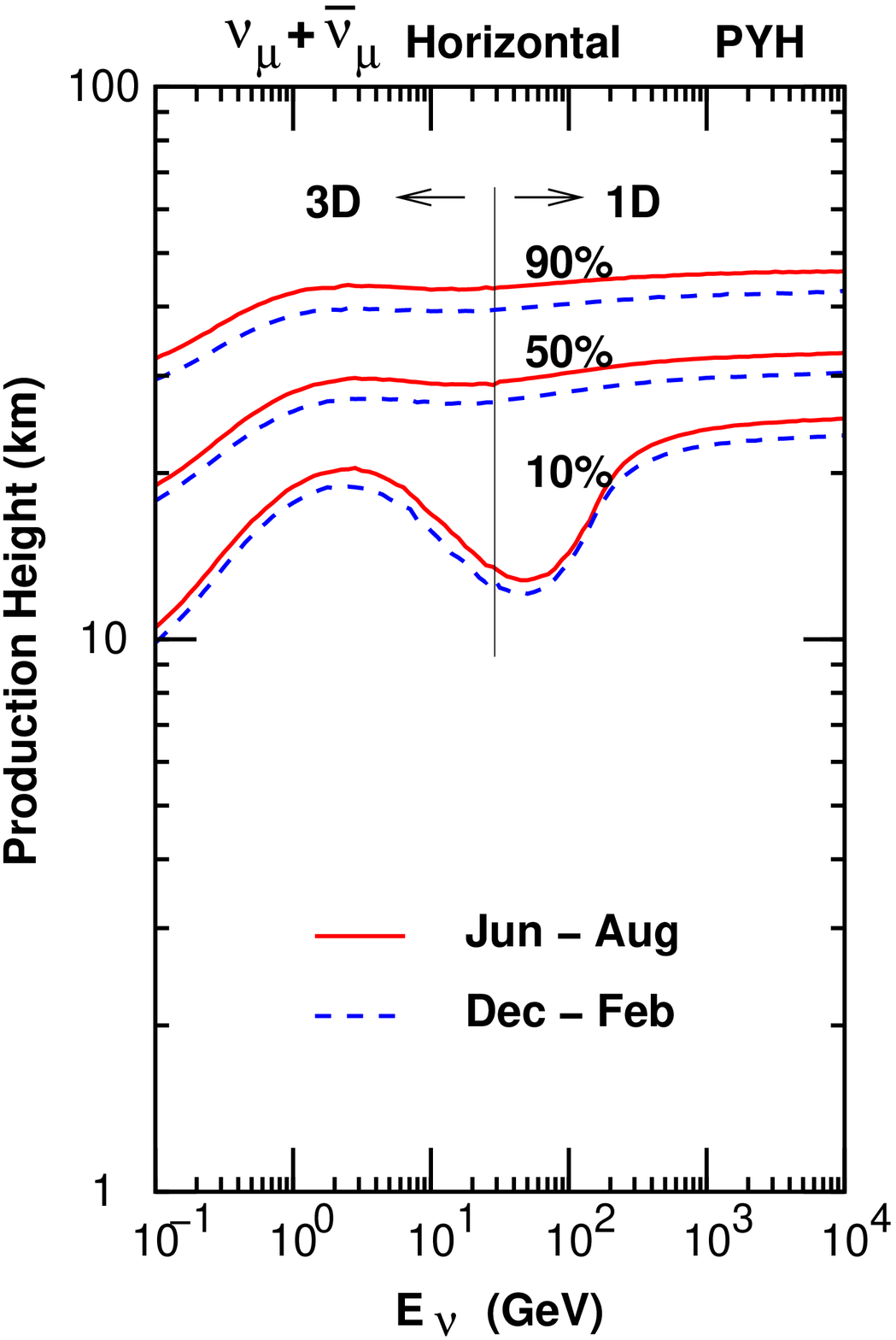}
        \includegraphics[width=4cm]{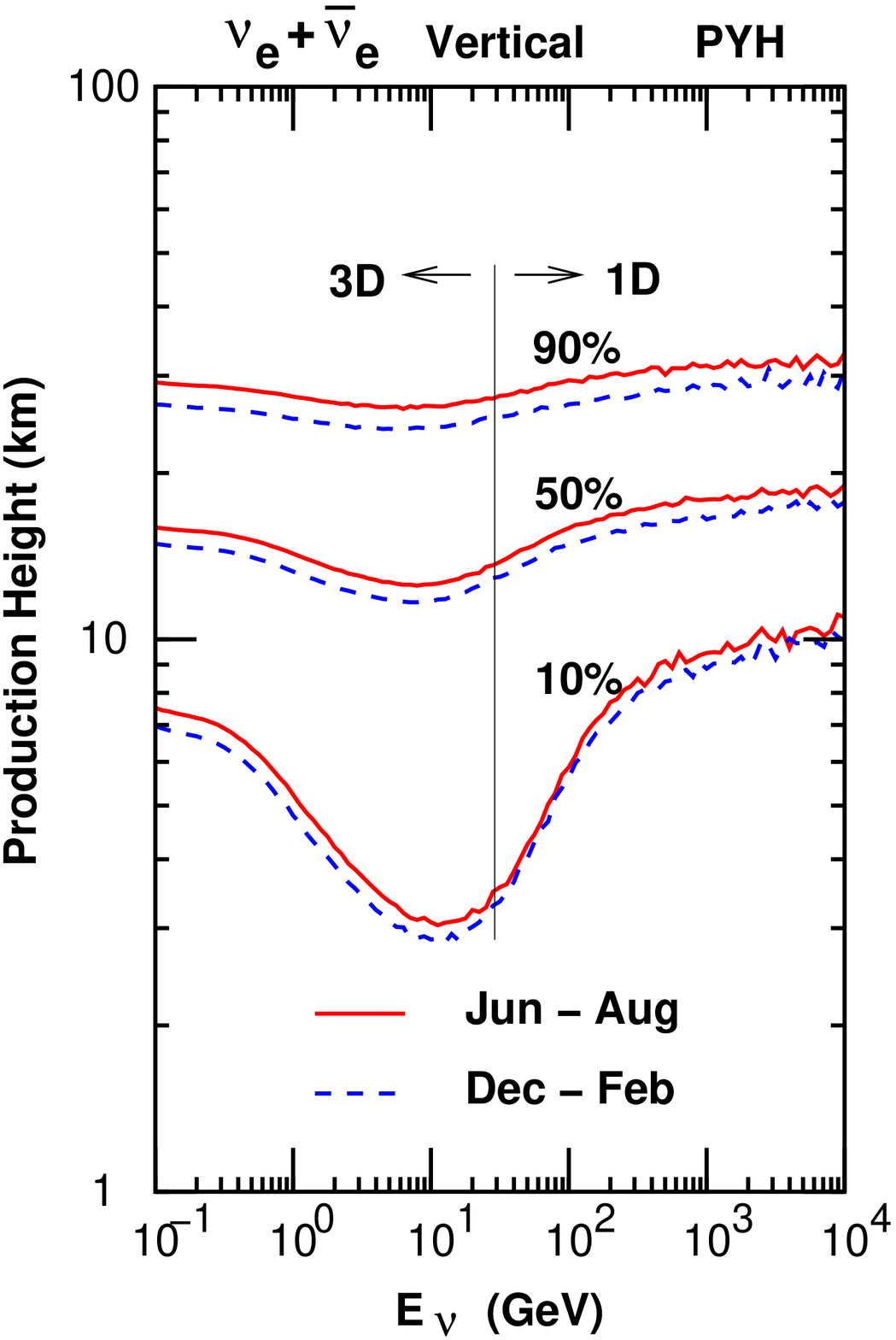}
        \includegraphics[width=4cm]{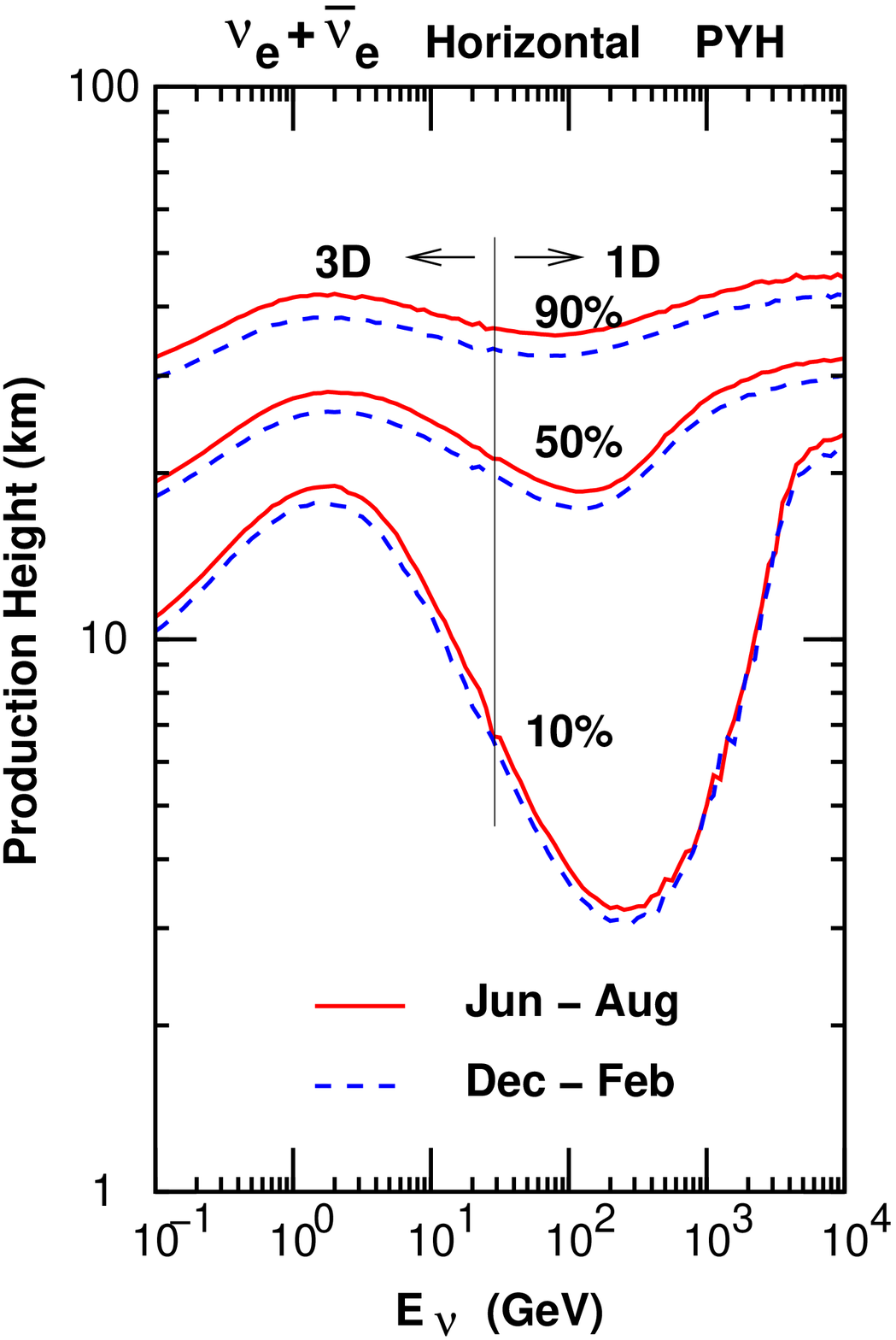}
      }
      \caption{
        Production height of atmospheric neutrinos for 
        $\nu_\mu$ + $\bar\nu_\mu$ and $\nu_e$ + $\bar\nu_e$ 
        going to vertically down and horizontal directions,
        summing all azimuth directions at the Pyh\"asalmi mine.
        The height that cumulative distribution reaches 10~\%,
        50~\%, and 90~\% are shown as a function of neutrino energy.
        The time average values in June -- August are shown as solid lines
        and in December -- February as Dashed line.
      }
      \label{fig:height-spl}
\end{figure*}

Here we study the production heights of atmospheric neutrinos
calculated with the NRLMSISE-00 atmospheric model.
The production height is an important parameter for analyzing the 
neutrino oscillations using atmospheric neutrinos.
As the production heights are distributed in a wide range
from sea level to $\sim$ 100~km a.s.l.,
we examine the cumulative distribution,
and plot the height where the cumulative distribution of production 
height reaches  10~\%, 50~\%, and 90~\% 
as a function of neutrino energy for vertical down going 
neutrinos ($\cos\theta_z > 0.9$) and horizontally going neutrinos
($0.1 > \cos\theta_z > 0$)
in Figs~\ref{fig:height-kam},~\ref{fig:height-ino},~\ref{fig:height-spl},
and ~\ref{fig:height-pyh},
In those figures, the heights shown by the lines of 10~\%, 50~\%,
and 90~\% indicate 10~\%, 50~\%, and 90~\% of neutrinos are created
below those height, respectively.
for the SK site, INO site, South Pole, and Pyh\"asalmi mine respectively. 
To make the distributions in the figures,
we combine particle and anti-particle neutrinos, and average
over all azimuth angles.
In the figures for the SK site,
we also plot the same flavor-ratio with the US-standard '76 
atmospheric model below 32~GeV,
calculated in the previous work~\cite{hkkm2011}.

The production heights are largely different between 
$\nu_\mu$ + $\bar\nu_\mu$ and $\nu_e$ + $\bar\nu_e$,
and also between vertical and horizontal directions.
The large energy dependence of production height of the 
$\nu_e$ and $\bar\nu_e$ are explained by the fact that they
are mainly produced by muon decay in the energies 
$\lesssim$ 100~GeV for near vertical directions, 
and $\lesssim$ 1~TeV for near horizontal directions.
The muons fly long distances before decays, and make neutrinos
at a lower altitude. 
As the muon flight length increases with the muon energy, 
the production height of the $\nu_e$ and $\bar\nu_e$ 
becomes lower as long as the muon is their major source.
However, above 100 GeV for vertical directions and 1 TeV for horizontal
directions, Kaon decay becomes the main source,
then the production height becomes high again.

Muon decay is also a source of $\nu_\mu$ and $\bar\nu_\mu$.
A similar variation to  $\nu_e$ and $\bar\nu_e$ is also seen 
in $\nu_\mu$ and $\bar\nu_\mu$, but with a smaller amplitude.
The fraction of $\nu_\mu$ and $\bar\nu_\mu$ created in 
muon decay is not so large, roughly a half at low energies 
and decreasing with the energy as muons tend to hit the ground.
The effect of muon flight in the production height is smaller
on $\nu_\mu$ and $\bar\nu_\mu$ than that 
on $\nu_e$ and $\bar\nu_e$.

Note, 
above 100 GeV for vertical directions and 1 TeV for horizontal
directions, 
Kaon decay becomes the main source of all flavors of neutrinos.
Then the production height becomes high, 
close to the production height of Kaons,
or the hadronic interaction zone of cosmic rays.
We would also like to note that the production height calculated in
3D-scheme is smoothly connected to that calculated in the 1D-scheme
as well as the flux.

At the SK and INO sites, we observe small difference between the averages
of June -- August and December -- February.
Especially at the INO site, the difference is almost invisible in the figures.
The difference between the calculations with the NRLMSISE-00 and 
US-standard '76 atmospheric models at the SK site is also small.
On the other hand, we find a large seasonal variation of 
production height at the South Pole. 
The variation of the median (cumulative distribution of 50~\%)
height extend to $\sim$~20\% variation at near horizontal 
directions,
and the variation of the amplitude is similar between 
$\nu_\mu$ + $\bar\nu_\mu$ and $\nu_e$ + $\bar\nu_e$.
At the Pyh\"asalmi mine, we also find a seasonal variation of 
$\sim$~10~\% for the median.

The energy dependence of the production height is similar 
among the different sites.
However, the of 10~\% line of the cumulative 
distribution for $\nu_e$ + $\bar\nu_e$ 
at the South Pole is a little higher than that at other 
sites due to the altitude of the sites,
as we assumed the observation site is 2835m a.s.l. for the 
South Pole, and sea level for other sites.

\begin{figure*}[htb]
  \centering
      {
        \includegraphics[width=6cm]{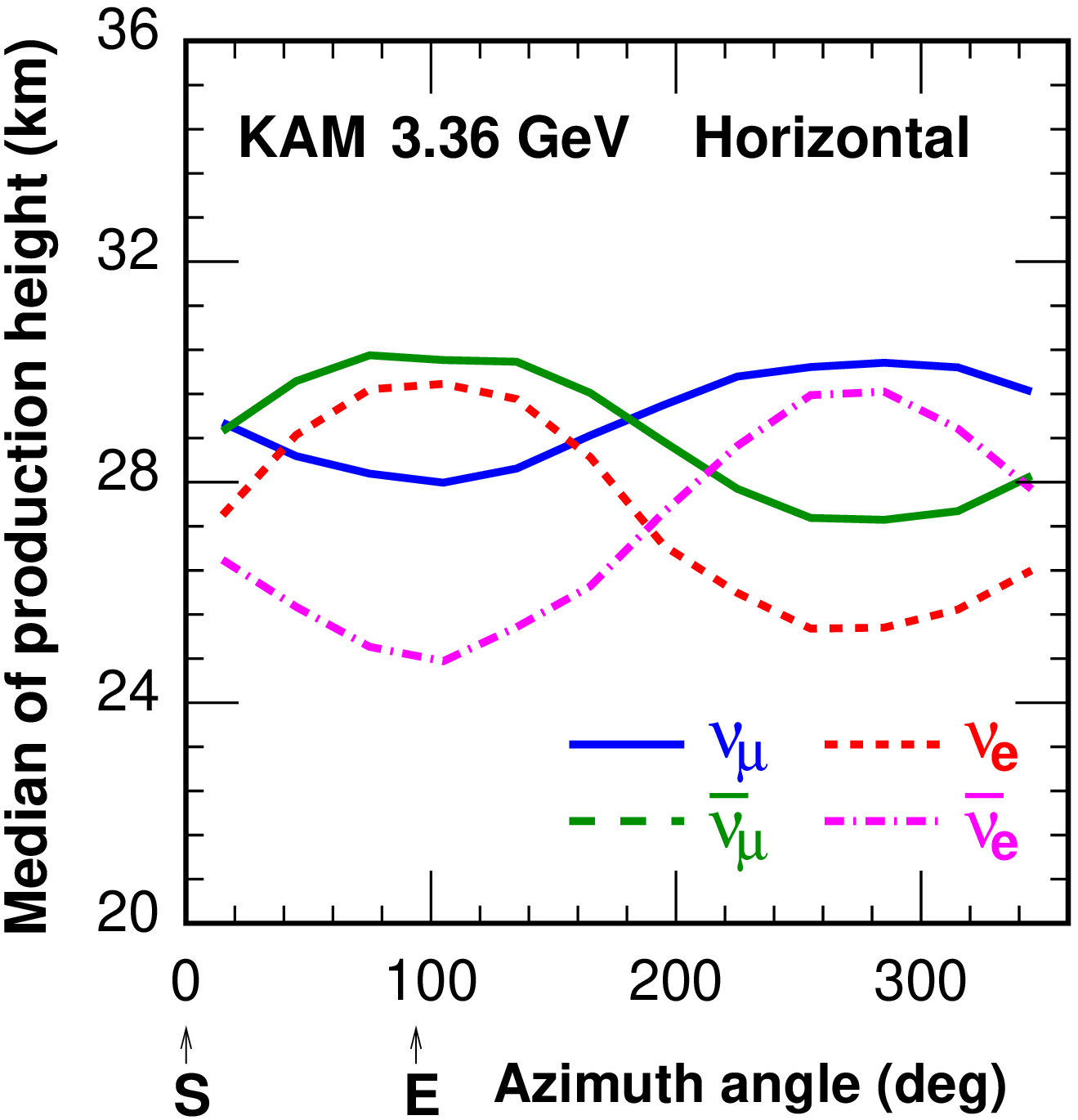} 
        \hspace{5mm}
        \includegraphics[width=6cm]{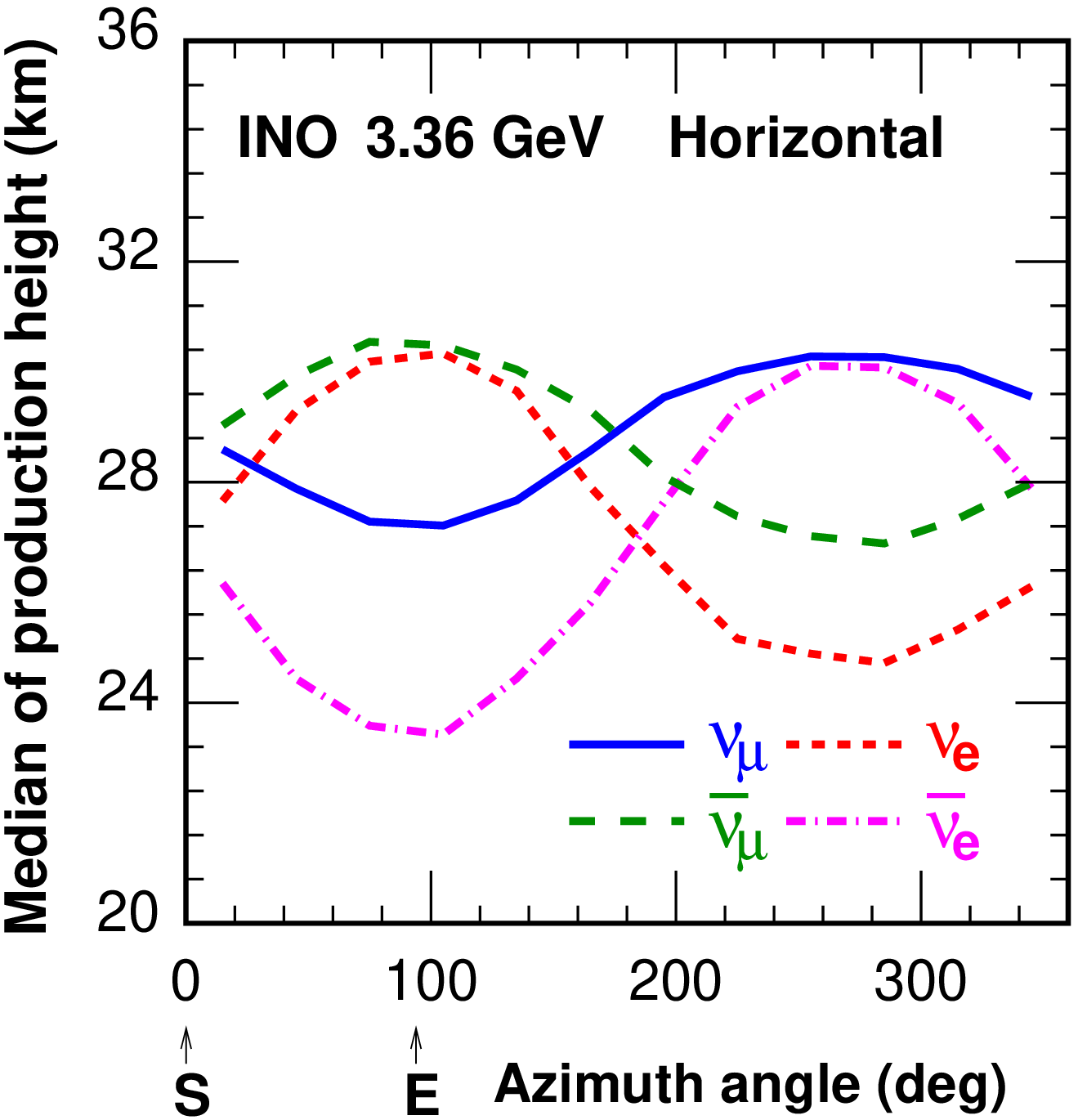} 
        \\
        \includegraphics[width=6cm]{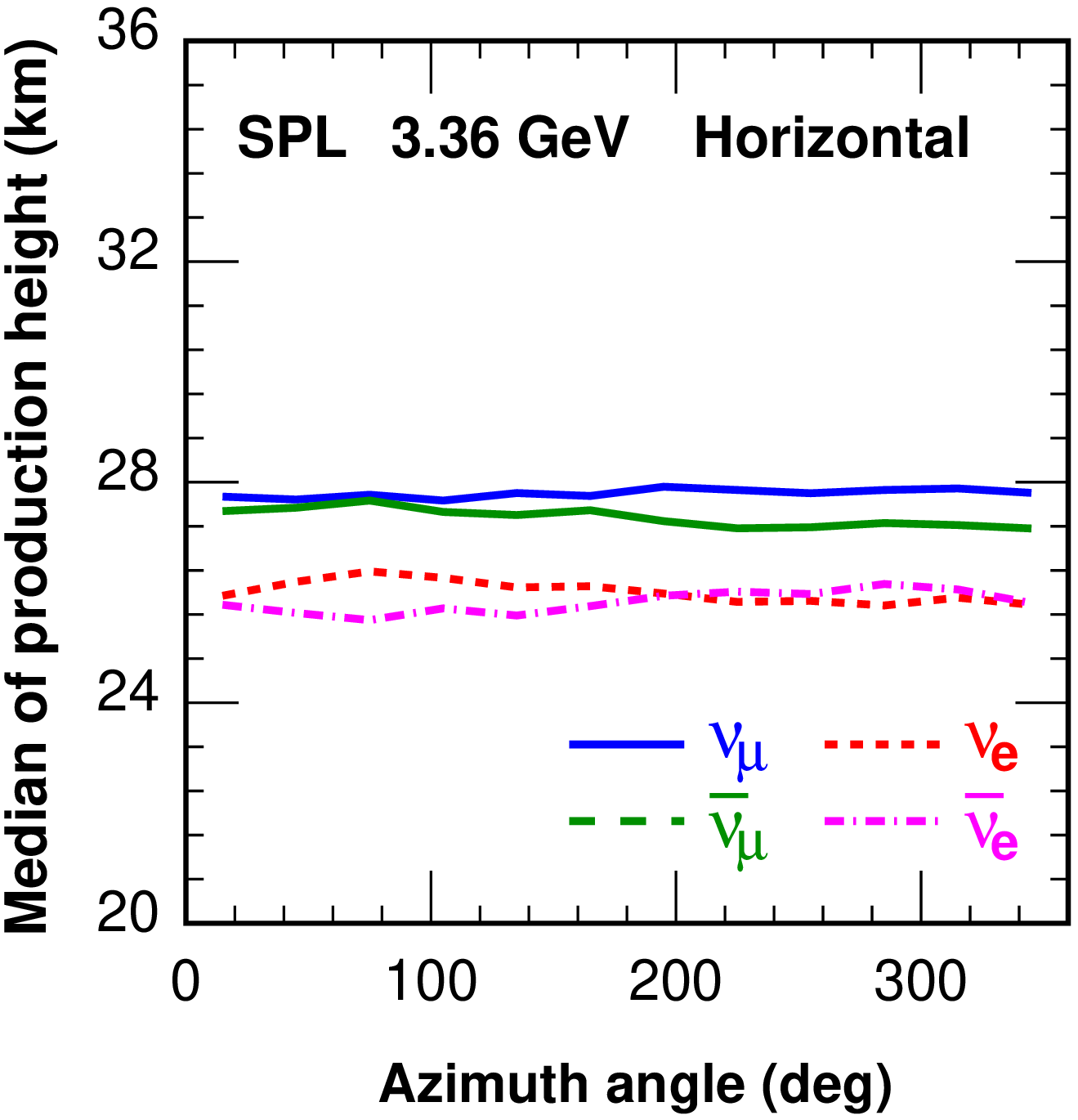} 
        \hspace{5mm}
        \includegraphics[width=6cm]{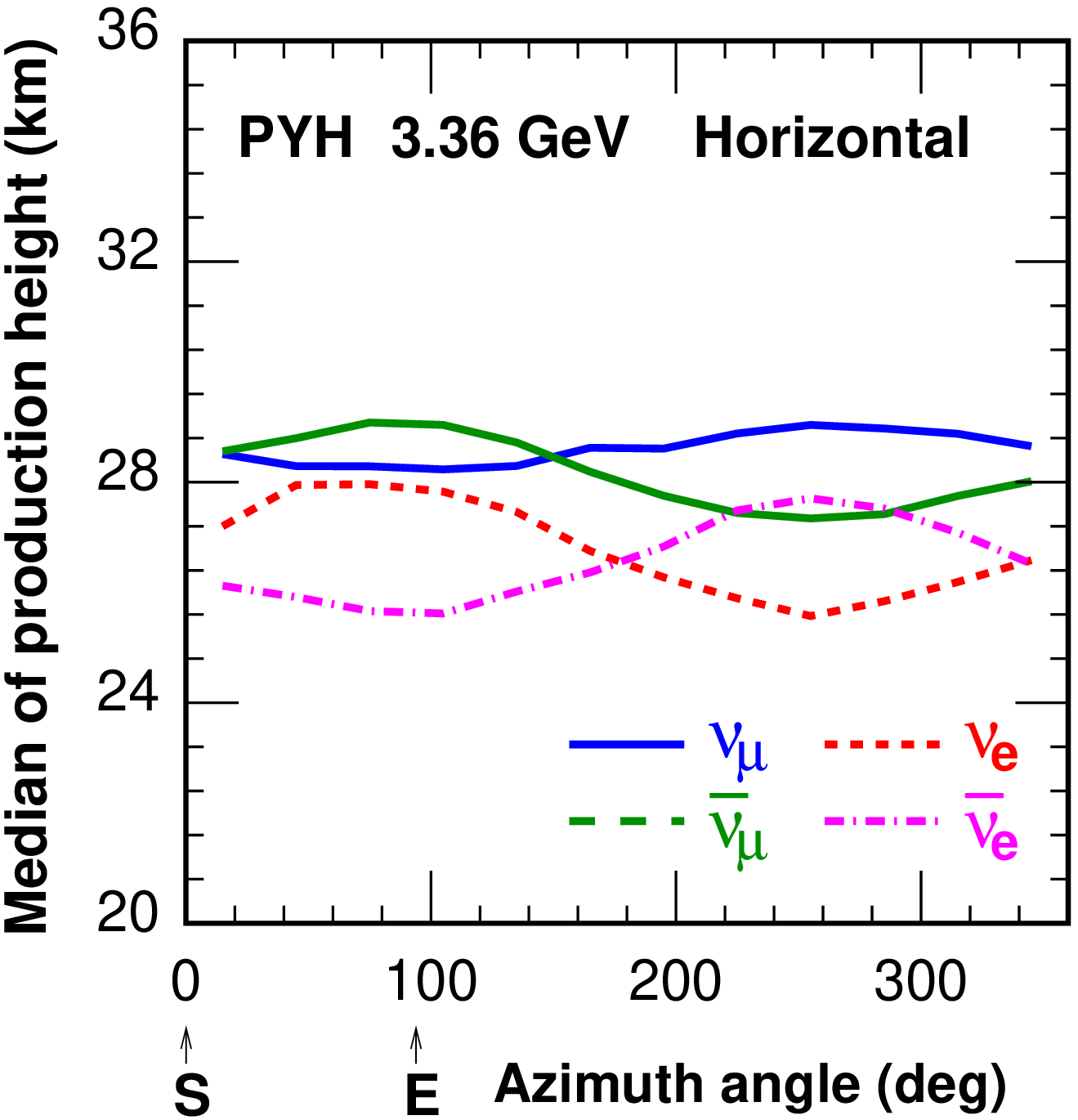} 
      }
      \caption{
        Azimuthal variation of production height of atmospheric
        neutrinos
        at the SK site, INO site, South Pole and Pyh\"asalmi mine for 
        near horizontal directions.
        The height that cumulative distribution reaches 50~\% is shown 
        as the function of azimuth for neutrino in the energy range of
        $3.16 < E_\nu < 3.55$~GeV.
      }
      \label{fig:height-azim}
\end{figure*}

In Fig.~\ref{fig:height-azim} we plotted the azimuthal variation 
of the median of the production height 
at near horizontal directions for the atmospheric 
neutrinos at  $E_\nu = 3.36$~GeV ($3.16 < E_\nu < 3.55$~GeV),
summing the production height distribution over a year.
We find there are sinusoidal variations with azimuth angle
for all flavor neutrinos, 
but in an opposite direction to each other among the neutrinos 
and  anti-neutrinos, at the SK and INO site.
The azimuth variation of production height at the Pyh\"asalmi mine 
is smaller than that at the SK and INO sites, but the shape is 
similar.
The azimuthal variation of production height at the South Pole
is small.
These sinusoidal features at the SK and INO sites
could be understood if we consider 
that the production 
height is mainly controlled by the horizontal component of 
the geomagnetic field
and the effect of the rigidity cutoff is small.

\section{\label{summary}summary and discussion}

We have extended our calculation of atmospheric neutrino flux to 
the tropical (INO site)
and polar regions (South Pole and Pyh\"asalmi mine).
In this extension, we have updated the atmospheric model from 
the US-standard '76 model to the NRLMSISE-00 model.
When we compare the two atmospheric models, we find the
NRLMSISE-00 atmospheric model almost agrees with the US-standard '76 
atmospheric model in the tropical and mid-latitude region. However, 
they disagree with each other largely in the polar region.
Also the NRLMSISE-00 atmospheric model suggests a large seasonal variation
in the polar region.
The atmospheric neutrino flux calculated with the NRLMSISE-00 
model is compared with
that calculated with the US-standard '76 model at the 
SK site.

Adding to the air density profile, there is a large difference of 
the geomagnetic field configuration between equatorial 
and polar regions.
It is well known that the influence of geomagnetic field on the
atmospheric neutrino is large. 
The extension in this paper is also the study of the atmospheric 
neutrino flux in the two extremes in the IGRF geomagnetic field model.
Note, our calculation so far was limited to the sites in mid-latitude 
region.

As expected, the calculated atmospheric neutrino flux at the equatorial 
(tropical) site (INO) suffers a strong effect from the horizontal component
of the geomagnetic field.
A large reduction of the neutrino flux is still seen at 3.2~GeV 
for down going directions due to the rigidity cutoff.
The muon bending causes a large azimuthal variation of neutrino fluxes.
reducing $\nu_\mu$ and $\bar\nu_e$ fluxes 
and enhancing a little $\bar\nu_\mu$ and $\nu_e$ flues 
at horizontal directions.

On the other hand, the large seasonal variation of air density 
profile at the polar regions (South Pole and Pyh\"asalmi mine) causes
a large seasonal variation of atmospheric neutrino flux. 
We had expected that the seasonal dependence is large at higher 
energies since the $\pi$ decay and interaction ratio changes with
the air density.
But at the South Pole, a $\sim$ 10~\% variation is seen even at 3.2~GeV
in the $\nu_e$ and $\bar\nu_e$ fluxes for vertically down going directions.
The fluxes of the $\nu_\mu$ and $\bar\nu_\mu$ also show variations, 
but smaller than those of the $\nu_e$ and $\bar\nu_e$.
This is considered to be due to the change of 
muon energy loss rate in the atmosphere according to the 
change of the air density by the seasons.

We also studied the production height of atmospheric neutrinos 
with the NRLMSISE-00 atmospheric model.
For the mid-latitude region (SK site), they are very close to the 
production height calculated with the US-standard '76 atmospheric model.
In the polar region, we find a seasonal variation of production 
height,
but in the mid-latitude and tropical regions, 
the seasonal variation of production height is small, and almost
invisible.
However, 
we find a large azimuthal variation of production height due to
the muon bending 
in the mid-latitude and tropical regions.

Here,
we would like to make a short comment on the calculation error
of the atmospheric neutrino flux, and on the recent observations
of the cosmic rays.
First note that this work is still within the calculation scheme
established in Ref~\cite{hkkms2006} which is based on the comparison
of atmospheric muon observation and calculation.
Therefore, the estimation made in Ref~\cite{hkkms2006} is
still valid in this work. The total error is a little lower than
10~\% in the energy region 1--10~GeV.
The error increases outside of this energy region 
due to the small number of available muon observation data
at the lower energies, and due to the uncertainty of Kaon production
at higher energies.

The recent observations of the primary cosmic
rays~\cite{atic05,cream,pamela2011,AMS02p},
suggest that we need to modify the interaction model again
to reconstruct the observed atmospheric muon fluxes with them.
In a preliminary work~\cite{honda:quynhon},
we find it is possible to modify the interaction model in such a way.
and the atmospheric neutrino flux calculated with that 
is very close to the present calculation in 1--10~GeV 
and well within the error estimated in Ref~\cite{hkkms2006}.
Now would be the time to study the interaction model with the
updated primary cosmic ray spectra model for the calculation
of atmospheric neutrino flux.

\begin{acknowledgments}
We are greatly appreciative to J.~Nishimura and A.~Okada
for their helpful discussions and comments through this paper.
We are grateful to E. Richard for discussions.
We also thank the ICRR of the University of Tokyo, 
especially for the use of the computer system.

%This study was supported by Grants-in-Aid, KAKENHI(12047206), from
%the Ministry of Education, Culture, Sport, Science and Technology 
%(MEXT) in Japan.
\end{acknowledgments}

 \bibliography{nflx-msise}

\end{document}